\documentclass[11pt]{article}
\usepackage{amsmath}
\usepackage{amsfonts,amssymb,latexsym}

\begin{document}

\newcommand{\nl}{\nonumber\\}
\newcommand{\nnl}{\nl[6mm]}
\newcommand{\nle}{\nl[-2.5mm]\\[-2.5mm]}
\newcommand{\nlb}[1]{\nl[-2.0mm]\label{#1}\\[-2.0mm]}

\renewcommand{\leq}{\leqslant}
\renewcommand{\geq}{\geqslant}

\renewcommand{\theequation}{\thesection.\arabic{equation}}
\let\ssection=\section
\renewcommand{\section}{\setcounter{equation}{0}\ssection}

\newcommand{\be}{\bes}
\newcommand{\ee}{\ees}
\newcommand{\bes}{\begin{eqnarray}}
\newcommand{\ees}{\end{eqnarray}}
\newcommand{\eens}{\nonumber\end{eqnarray}}
\newcommand{\barr}{\begin{array}}
\newcommand{\earr}{\end{array}}

\renewcommand{\/}{\over}
\renewcommand{\d}{\partial}
\newcommand{\dd}[1]{\delta/\delta {#1}}
\newcommand{\ddt}{{d\/dt}}

\newcommand{\no}[1]{{\,:\kern-0.7mm #1\kern-1.2mm:\,}}

\newcommand{\half}{{1\/2}}
\newcommand{\bra}[1]{\big{\langle}#1\big{|}}
\newcommand{\ket}[1]{\big{|}#1\big{\rangle}}
\newcommand{\bracket}[2]{\big{\langle}#1\big{|}#2\big{\rangle}}

\newcommand{\da}{\d_\alpha}
\newcommand{\db}{\d_\beta}
\newcommand{\dc}{\d_\gamma}

\newcommand{\ab}{{\alpha\beta}}
\newcommand{\ba}{{\beta\alpha}}
\newcommand{\ca}{{\gamma\alpha}}
\newcommand{\bc}{{\beta\gamma}}

\newcommand{\ua}{u^\alpha}
\newcommand{\ub}{u^\beta}
\newcommand{\uc}{u^\gamma}

\newcommand{\ra}{r^\alpha}
\newcommand{\rb}{r^\beta}
\newcommand{\rc}{r^\gamma}

\newcommand{\tpi}{\tilde\pi}

\newcommand{\va}{v^\alpha}
\newcommand{\vb}{v^\beta}

\newcommand{\fa}{\phi^\alpha}
\newcommand{\fb}{\phi^\beta}
\newcommand{\fc}{\phi^\gamma}
\newcommand{\pa}{\pi_\alpha}
\newcommand{\pb}{\pi_\beta}
\newcommand{\Ea}{\EE_\alpha}
\newcommand{\Eb}{\EE_\beta}

\newcommand{\fxM}{\phi_{;M}}

\newcommand{\fs}{\phi^*}
\newcommand{\ps}{\pi^*}
\newcommand{\bfi}{\bar\phi}
\newcommand{\bfs}{\bar\phi^*}
\newcommand{\bpi}{\bar\pi}
\newcommand{\bps}{\bar\pi^*}

\newcommand{\fsa}{\phi^*_\alpha}
\newcommand{\fsb}{\phi^*_\beta}
\newcommand{\fsc}{\phi^*_\gamma}
\newcommand{\psa}{\pi_*^\alpha}
\newcommand{\psb}{\pi_*^\beta}
\newcommand{\fsi}{\phi^*_i}

\newcommand{\fw}{\phi^\w}
\newcommand{\pw}{\pi_\w}

\newcommand{\ii}{{(i)}}
\newcommand{\jj}{{(j)}}
\newcommand{\minus}{{(-)}}
\newcommand{\plus}{{(+)}}
\newcommand{\mm}{{(m)}}
\newcommand{\pp}{{(p)}}

\newcommand{\si}{\sigma}
\newcommand{\varsi}{\varsigma}
\newcommand{\eps}{\epsilon}
\newcommand{\vareps}{\varepsilon}
\newcommand{\dlt}{\delta}
\newcommand{\w}{\omega}
\newcommand{\ww}{\varpi}
\newcommand{\ups}{\upsilon}
\newcommand{\al}{\alpha}
\newcommand{\bt}{\beta}
\newcommand{\gm}{\gamma}
\newcommand{\ka}{\kappa}
\newcommand{\la}{\lambda}
\newcommand{\La}{\Lambda}

\newcommand{\mn}{{\mu\nu}}
\newcommand{\MN}{{MN}}
\newcommand{\ij}{{ij}}
\newcommand{\hm}{\eta_\mu\kern+0.5mm}
\newcommand{\hn}{\eta_\nu\kern+0.5mm}

\newcommand{\ihat}{\hat\imath}
\newcommand{\jhat}{\hat\jmath}
\newcommand{\one}{{\mathbf 1}}

\newcommand{\sw}{{1\/\sqrt{2\w}}}
\newcommand{\kw}{(k^2 - \w^2)}
\newcommand{\wk}{\w_\kk}
\newcommand{\swk}{{1\/\sqrt{2\wk}}}

\newcommand{\Qdag}{Q^\dagger}
\newcommand{\QM}{Q_{\MM}}
\newcommand{\dltM}{\dlt_{\MM}}
\newcommand{\adag}{a^\dagger}

\newcommand{\vect}{{\mathfrak{vect}}}
\newcommand{\map}{{\mathfrak{map}}}

\newcommand{\wt}{{\rm wt}\, }
\newcommand{\im}{{\rm im}\, }
\newcommand{\sgn}{{\rm sgn}}
\newcommand{\e}{{\rm e}}
\renewcommand{\div}{{\rm div}}
\newcommand{\afn}{{\rm afn\,}}
\newcommand{\til}{{\tilde{\ }}}

\newcommand{\ku}{k_\mu \dot q^\mu}

\newcommand{\larroww}[1]{{\ \stackrel{#1}{\longleftarrow}\ }}
\newcommand{\llarroww}[1]{{\ \stackrel{#1}{\longleftarrow}\ }}
\newcommand{\sumwI}{\sum_{\w\in\KK}}

\newcommand{\sumall}[1]{\sum_{#1 = -\infty}^\infty }
\newcommand{\sumkpos}{\sum_{k=0}^\infty }
\newcommand{\intk}{\int_{-\infty}^\infty dk\ }
\newcommand{\intkpos}{\int_{0}^\infty dk\ }
\newcommand{\intm}{\int_{-\infty}^\infty dm\ }
\newcommand{\intmpos}{\int_{0}^\infty dm\ }
\newcommand{\intdk}{\int d^4k\ }
\newcommand{\intkzpos}{\int_{k_0>0} d^4k\ }

\newcommand{\tr}[1]{{\rm tr}_{#1}\kern0.7mm}
\newcommand{\g}{{\mathfrak g}}
\newcommand{\m}{{\mathfrak m}}
\newcommand{\hh}{{\mathfrak h}}
\newcommand{\uu}{{\mathfrak u}}

\newcommand{\xx}{{\mathbf x}}
\newcommand{\kk}{{\mathbf k}}

\renewcommand{\L}{{\mathcal L}}
\newcommand{\J}{{\mathcal J}}
\newcommand{\R}{{\mathcal R}}
\newcommand{\B}{{\mathcal B}}
\newcommand{\QQ}{{\mathcal Q}}
\newcommand{\PP}{{\mathcal P}}
\newcommand{\HP}{{\mathcal {HP}}}
\newcommand{\EE}{{\mathcal E}}
\newcommand{\FF}{{\mathcal F}}
\newcommand{\II}{{\mathcal I}}
\renewcommand{\H}{{{\mathcal H}}}
\newcommand{\HH}{{{\mathfrak H}}}
\newcommand{\KK}{{\mathcal K}}
\newcommand{\MM}{{\mathcal M}}
\newcommand{\TT}{{\mathcal T}}
\newcommand{\W}{{\mathcal W}}

\newcommand{\M}{{\mathbf M}}
\newcommand{\N}{{\mathbf N}}

\newcommand{\cl}{{cl}}
\newcommand{\st}{{st}}
\newcommand{\op}{{op}}
\newcommand{\gh}{{gh}}
\newcommand{\TOT}{{TOT}}

\newcommand{\Mobs}{M_{obs}}
\newcommand{\viz}{{\em viz.}}

\newcommand{\RR}{{\mathbb R}}
\newcommand{\CC}{{\mathbb C}}
\newcommand{\ZZ}{{\mathbb Z}}
\newcommand{\NN}{{\mathbb N}}

\title{{Quantum Jet Theory I: Free fields}}

\author{T. A. Larsson \\
Vanadisv\"agen 29, S-113 23 Stockholm, Sweden\\
email: thomas.larsson@hdd.se}

\maketitle
\begin{abstract}
QJT (Quantum Jet Theory) is the quantum theory of jets, which can be
canonically identified with truncated Taylor series. Ultralocality
requires a novel quantization scheme, where dynamics is treated as a
constraint in the history phase space.
QJT differs from QFT since it involves a new datum: the expansion
point. This difference is substantial because it leads to new gauge
and diff anomalies, which are necessary to combine background
independence with locality. Physically, the new ingredient is that the
observer's trajectory is explicitly introduced and quantized together
with the fields.
In this paper the harmonic oscillator and free fields are treated
within QJT, correcting previous flaws. The standard Hilbert
space is recovered for the harmonic oscillator, but there are
interesting modifications already for the free scalar field, due to
quantization of the observer's trajectory.
Only free fields are treated in detail, but the complications when
interactions are introduced are briefly discussed. We also explain why
QJT is necessary to resolve the conceptual problems of quantum gravity.
\end{abstract}

\vskip 3 cm
PACS (2003): 03.65.Ca, 03.70.+k, 11.10.Ef.


\newpage

\section{Introduction}

MCCQ (manifestly covariant canonical quantization) is a novel
quantization scheme, first proposed in \cite{Lar04}. As the name
indicates, this is a canonical quantization method which maintains
manifest covariance at all levels. This is achieved through two
inventions:
\begin{enumerate}
\item
Dynamics is regarded as a constraint in the phase space of arbitrary
histories, which allows us to apply powerful cohomological methods.
\item
All fields are expanded in a Taylor series around an operator-valued
curve, which is naturally identified as the observer's trajectory in
spacetime. Hence all fields are replaced by their jets, which
motivates the name Quantum Jet Theory (QJT).
\end{enumerate}

The first invention is mainly technical. Manifest covariance greatly
simplifies the understanding of constrained systems, in particular
general relativity. Since the covariant constraint algebra is the
spacetime diffeomorphism algebra rather than the Dirac algebra, we can
profit from its projective representation theory developed in
\cite{Lar98,RM94}. The outstanding lesson from this theory is that in
order to construct quantum representations of the diffeomorphism
algebra, we must introduce and quantize the observer's trajectory.
In an ultralocal theory like QJT, dynamics must be formulated without
reference to nonlocal integrals; this is further discussed in
subsection \ref{ssec:ultra}. The history-orientated approach has this
crucial property.

In constrast, the second invention is physically important. When
replacing fields by their Taylor expansions, we introduce a new
datum: the expansion point (or rather the expansion curve). This makes
it possible to write down new extensions of the algebras of gauge
transformations and diffeomorphisms, generalizing affine and Virasoro
algebras to higher dimensions. The relevant cocycles are functionals of
the observer's trajectory, which explains why they cannot be formulated
within QFT proper, where this trajectory has not been introduced. Gauge
and diff anomalies in QFT were classified long ago \cite{Bon86}, leaving
no room for
new ones. The fact that we do find new anomalies shows that QJT is
substantially different from QFT. This should be a good thing, because we
know that QFT is incompatible with gravity, and QJT may lack this
defect. Nevertheless, QJT is close to QFT, in the same sense as a
Taylor series is close to the function to which it converges.
Therefore, we expect QJT to inherit the pragmatic successes of
QFT. The relation between QJT and QFT, and why the former are
necessary to formulate a consistent quantum theory of gravity, are further
discussed in section \ref{sec:discussion}.

Before we can have confidence in a new quantization method, we
must first verify that it reproduces the correct answers to
well-understood problems. MCCQ, as formulated in \cite{Lar04}, fails
this test already for the harmonic oscillator. There were two major
flaws in that paper: there are spurious states which are not removed in
cohomology, and no inner product was defined. The purpose of the present
paper is to correct these errors. We show here that the corrected
version of MCCQ reproduces the correct spectrum and inner product of the
harmonic oscillator, and more generally of any free theory, without any
overcounting of states. The ingredients missing in \cite{Lar04} are
additional constraints, which remove remaining antifields and
identify momenta and velocities.

Unfortunately, the additional constraints which eliminate the spurious
cohomology also break manifest covariance, and hence covariance is only
recovered in a weaker form. This is explained in subsection
\ref{ssec:mancov}. The BRST operator $Q=Q_D+\QM$ consists of two parts.
The momentum part $\QM$ violates covariance, but the dynamics part
$Q_D$ does not; it commutes with Poincar\'e transformations, and in the
general-covariant context with spacetime diffeomorphisms. Hence covariance
is maintained in the sense that $H^0(Q_D)$ carries a representation of
the covariance algebra. Such representations were constructed in
\cite{Lar04,Lar05a,Lar05b,Lar05c}; the main error in those papers was that
$\QM$ was not considered. However, $H^0(Q_D)$ is not a Hilbert space,
because there is no interesting inner product before the momentum
constraint has been imposed.

An important difference between QJT and QFT is that in the former, the
relevant covariance algebras, e.g. gauge or diffeomorphism algebras, act
in a well-defined manner at the quantum level. After quantization, the
history phase space $\HP$ and the classical cohomology
$H^0_\cl(Q_D)$ turn into linear spaces which carry projective,
lowest-energy representations of the relevant algebras. Indeed, it is
this property which dictates the form of QJT, since it grew out of the
representation theory of the extended diffeomorphism algebra
(multi-dimensional Virasoro algebra), developed in \cite{Lar98,RM94}.
As is well-known from low-dimensional models, in particular from lineal
gravity \cite{Jac95}, a detailed understanding
of how the gauge symmetries act on the quantum level is crucial for a
correct treatment. This is only possible within QJT.

History methods have previously been considered by Isham, Savvidou and
coworkers \cite{Ish95,Sav98}, and jet space formalism has been
employed by Sardanashvily \cite{Sard02}. There might be some overlap
between their works and	the present paper. However, they do not cover
the most important aspects of QJT, namely the identification of
the expansion point with the observer's position and the appearence of
new anomalies.

This paper is organized as follows.
Section 2 contains a list of notation.
The harmonic oscillator is treated in section 3, and it is verified
that we recover the correct Hilbert space with the correct inner
product.
Section 4 deals with the free particle, which is important in QJT because
it applies to the observer's trajectory.
In section 5 the method is extended to the free scalar field, and again
we recover the correct results. However, despite the Lorentz invariance
of the final results, we are forced to introduce a privileged vector
in the time direction, which we identify as the observer's 4-velocity.
To eliminate the last remnants of non-covariance, we must introduce
Quantum Jet Theory.
In section 6, QJT is applied to the harmonic oscillator, and the
correct results are again recovered.
In section 7, QJT is applied to the free scalar field. Unlike the
situation for the harmonic oscillator, we here see some deviations
from the QFT treatment. This is partly expected because QJT amounts to
a $p$-jet regularization of the fields, but some new phenomena survive
the $p\to\infty$ limit, especially reparametrization anomalies.
In section 8, we treat the free Maxwell field within QJT. The new
ingredients here are gauge symmetries, which are neatly
incorporated into the cohomological implementation of dynamics.
Although the treatment of interactions is deferred to a later paper,
we briefly indicate in section 9 how they modify the formalism.
The final section 10 contains a general discussion of the philosophical
issues raised by QJT, in particular regarding quantum gravity,
and directions for future work.

\section{List of notation}
\begin{tabular}{ll}
$\mu,\nu$ & Spacetime index \\
$\al, \bt, \gm$ & History index \\
$a, b, c$ & Gauge index \\
$\w, \ww, \ups$ & Solution index \\
$M = (M_1, ..., M_d), N$ & Multi-index \\
$$ &  \\
$x^\mu$ & Spacetime coordinate \\
$x = x^0$ & Time coordinate in QM \\
$t$ & Parameter along observer's trajectory \\
$k_\mu$ & Fourier variable matching $x^\mu$ \\
$k$ & Fourier variable matching $x$ \\
$m$ & Fourier variable matching $t$ \\
\end{tabular}

\begin{tabular}{ll}
$\phi(x), \phi(k)$ & Spacetime field (history) \\
$\pi(x), \pi(k)$ & History momentum \\
$\fs(x), \fs(k)$ & Dynamics antifield \\
$\ps(x), \ps(k)$ & Dynamics antifield momentum \\
$$ &  \\
$q^\mu(t)$ & Observer's trajectory (free particle) \\
$p_\mu(t)$ & Observer's momentum \\
$q^\mu_*(t), p^*(t)$ & Associated antifields \\
$\phi_M(t), \phi_M(m)$ & Jet corresponding to $\phi(x)$ \\
$\pi_M(t), \pi_M(m)$ & Jet momentum \\
$\fs_M(t), \fs_M(m)$ & Antijet (dynamics) \\
$\ps_M(t), \ps_M(m)$ & Antijet momentum (dynamics) \\
$\bfi_M(t), \bfi_M(m)$ & Antijet (time) \\
$\bpi_M(t), \bpi_M(m)$ & Antijet momentum (time) \\
$\bfs_M(t), \bfs_M(m)$ & Antijet (dynamics+time) \\
$\bps_M(t), \bps_M(m)$ & Antijet momentum (dynamics+time) \\
$$ &  \\
$\EE(x), \EE_k$ & Dynamics constraint \\
$\MM(x), \MM_k$ & Momentum constraint \\
$\EE_M(t), \EE_M(m)$ & Dynamics constraint jet \\
$\MM_M(t), \MM_M(m)$ & Momentum constraint jet \\
$\TT_M(t), \TT_M(m)$ & Time constraint jet \\
$\R_M(t), \R_M(m)$ & Dynamics+time constraint jet \\
$Q,Q_\TOT$ & Total BRST operator \\
$Q_D,Q_T,\QM...$ & Partial BRST operators \\
$\H$ & Hamiltonian \\
$\HP$ & History phase space \\
$\HP^*$ & Extended history phase space \\
$\PP$ & Physical phase space \\
$H^0_\cl(Q) = C(\PP)$ & Classical cohomology \\
$\HH = H^0_\st(Q)$ & Hilbert space (state cohomology) \\
$H^0_\op(Q)$ & Space of physical operators (operator cohomology)
\end{tabular}

\section{The harmonic oscillator}
\label{sec:harmosc}

\subsection{Conventional canonical quantization}

For consistency with later developments, we denote the time
coordinate by $x$ (to be thought of as $x^0$) and the oscillator
degree of freedom by $\phi(x)$. The action
\be
S = \half \int dx\ ( \phi'(x)^2 - \w^2 \phi(x)^2),
\ee
where $\phi' = d\phi/dx$,
leads to the Euler-Lagrange (EL) equation
\be
\EE(x) \equiv -{\dlt S\/\dlt \phi(x)} = \phi''(x) + \w^2 \phi(x) = 0.
\label{ELharm}
\ee
In the Hamiltonian formalism, we introduce canonical momenta
\be
\pi(x) = \phi'(x),
\label{momharm}
\ee
subject to the equal-time commutation relations
\be
[\phi(x), \pi(x)] = i, \qquad
[\phi(x), \phi(x)] = [\pi(x), \pi(x)] = 0.
\ee
The phase space $\PP$ is spanned by pairs of variables $(\phi,\pi)$.
The time evolution of any function $F(\phi,\pi)$ over $\PP$ is governed
by Hamilton's equations of  motion $F'(x) = i[\H,F(x)]$, where
the time-independent Hamiltonian is
\be
\H &=& \half (\pi^2 + \w^2 \phi^2).
\label{Hamharm}
\ee
In particular,
\be
\phi'(x)= \pi(x), \qquad \pi'(x) = -\w^2 \phi(x),
\ee
which is equivalent to the EL equation (\ref{ELharm}).

Alternatively, we can choose a basis in $\PP$ of annihilation and
creation operators,
\be
a = \sw (\w \phi + i\pi), \qquad \adag = \sw (\w \phi - i\pi).
\label{aadag}
\ee
They satisfy canonical commutation relations (CCR)
\be
[a, \adag] = 1.
\ee
In terms of these variables, the Hamiltonian (\ref{Hamharm}) becomes
\be
\H = \w \adag a,
\label{Haa}
\ee
and time evolution is simply given by
\be
a'(x) = -i\w a(x), \qquad (\adag)'(x) = i\w \adag(x).
\ee
The creation and annihilation operators thus carry energy
$\w$ and $-\w$, respectively, i.e.
\be
[\H, \adag] = \w\adag, \qquad [\H, a] = -\w a.
\ee

To quantize the theory, we introduce a vacuum state $\ket 0$, which
is annihilated by all annihilation operators, i.e.
$a\ket 0 = 0$. A basis for the Hilbert space is given by the
$n$-quanta states
\be
\ket n = {1\/\sqrt{n!}} (\adag)^n\ket 0.
\label{ketn1}
\ee
$\ket n$ is an eigenvector of the Hamiltonian with energy $n\w$:
\be
\H\ket n = n\w\ket n.
\label{Hn1}
\ee
Dually, we define
\be
\bra n = {1\/\sqrt{n!}} \bra 0 a^n
\label{bran1}
\ee
If we choose normalization such that $\bracket 0 0 = 1$, the inner
product in the Hilbert space is
\be
\bracket m n = \dlt_{mn},
\ee
which obviously is positive-definite.

\subsection{History phase space}

Let us now explain how to recover this well-known story by treating
dynamics as a constraint in the history phase space.
Introduce virtual histories $\phi(x)$ with canonical momenta
$\pi(x) = -i\dlt/\dlt \phi(x)$, subject to the Heisenberg algebra
\be
[\phi(x), \pi(x')] = i \delta(x-x'), \qquad
[\phi(x), \phi(x')] = [\pi(x), \pi(x')] = 0.
\label{CCRhist}
\ee
Note that these brackets vanish whenever $x \neq x'$; (\ref{CCRhist})
are the CCR in history phase space $\HP$, which has the basis
$(\phi(x),\pi(x))_{x\in\RR}$. $\HP$ has an alternative basis of Fourier
modes $(\phi_k, \pi_k)$, in terms of which the CCR become
\be
[\phi_k, \pi_{k'}] = i \delta_{k+k'}, \qquad
[\phi_k, \phi_{k'}] = [\pi_k, \pi_{k'}] = 0.
\label{CCRk}
\ee
To avoid problems with the continuum limit, we assume for the time
being that the Fourier variable $k \in \ZZ$ is discrete, and that
$\pm\w \in \ZZ$; the continuum case is discussed in subsection
\ref{ssec:functional} below.
The EL equation (\ref{ELharm}) clearly has two solutions, $\phi_\w$
and $\phi_{-\w}$, which correspond to the solutions in real space
\bes
\phi(x) &=& \phi_\w \e^{i\w x} + \phi_{-\w}\e^{-i\w x}, \nle
\pi(x) &=& i\w\phi_\w \e^{i\w x} - i\w\phi_{-\w}\e^{-i\w x}.
\eens

It is clear that the history phase space $\HP$ is vastly larger than
the usual phase space $\PP$; the former has dimension $2\times\infty$
and the latter has dimension $2$. To properly describe the harmonic
oscillator in $\HP$, we must thus eliminate all but two of these
$2\times\infty$ variables. The strategy for doing this is to view
dynamics as a constraint in the history phase space.
The EL equations (\ref{ELharm}) become {\em the dynamics constraint},
\be
\EE_k \equiv \kw \phi_k \approx 0,
\label{Ekharm}
\ee
and the definition of momentum becomes {\em the momentum constraint}
\be
\MM_k \equiv \pi_k - ik \phi_k \approx 0.
\ee
As usual, wiggly equality signs indicate equality modulo constraints.
$\EE_k$ and $\MM_k$ are collected into the constraint vector
\be
\chi_k = \begin{pmatrix} \EE_k \\ \MM_k \end{pmatrix}
= \begin{pmatrix} \kw \phi_k \\ \pi_k - ik \phi_k \end{pmatrix}
\ee
with Poisson-bracket matrix
\be
\Delta_{kk'}
= [\chi_k, \chi_{k'}]
= \begin{pmatrix}
0 & i\kw \\
-i\kw & 2k
\end{pmatrix} \delta_{k+k'}.
\label{PBmatrix}
\ee
$\Delta_{kk'}$ clearly has a block diagonal form, so we may consider
each $(k, -k)$ pair separately. For $k \neq \pm\w$, the matrix is
non-singular and the constraint $\chi_k \approx 0$ is second class.
We can then define Dirac brackets
\be
[F, G]^* = [F,G] - \sum_k \sum_{k'}\ [F, \chi_k] \Delta^{kk'} [\chi_{k'}, 
G],
\ee
where $\Delta^{kk'}$ is the inverse of $\Delta_{kk'}$, and eliminate
the constraints. This leads to $\phi_k \approx \pi_k \approx 0$.

For $k = \pm \w$ the dynamics constraint $\EE_\w = \EE_{-\w}$ vanishes,
(\ref{PBmatrix}) is singular, and we can only impose the momentum
constraints $\MM_\w \approx \MM_{-\w} \approx 0$. The non-singular part
of the Poisson bracket matrix is now
\be
\Delta_{\w,-\w} = [\MM_\w, \MM_{-\w}] = 2\w.
\ee
The corresponding Dirac brackets commute with the momentum constraints,
and we may eliminate two of the four variables
$\phi_\w, \phi_{-\w}, \pi_\w, \pi_{-\w}$ in terms of the two others.
Denote the independent solutions by
\bes
a_\w &=& \sw (\pi_\w + i\w\phi_\w) \approx \sqrt{2\/\w}\pi_\w
\approx i\sqrt{2\w}\phi_\w,
\nlb{awaw}
a_{-\w} &=& \sw (\pi_{-\w} - i\w\phi_{-\w}) \approx \sqrt{2\/\w}\pi_{-\w}
\approx -i\sqrt{2\w}\phi_{-\w}.
\eens
Since $[a_\w, a_{-\w}] = -1$, we can identify $a_\w = \adag$,
$a_{-\w} = a$, where $a$ and $\adag$ are defined as in (\ref{aadag}).

The Hamiltonian in $\HP$ is simply the generator of rigid time
translations, i.e.
\be
\H = \int dx\ \phi'(x) \pi(x)
= \sumall{k}\ ik\phi_k \pi_{-k}.
\ee
On the constraint surface,
\be
\H \approx i\w (\phi_\w \pi_{-\w} - \phi_{-\w}\pi_\w)
= \w a_\w a_{-\w},
\label{Hawaw}
\ee
which of course is the same as (\ref{Haa}).

\subsection{Cohomological description I}
\label{ssec:cohom1}

In the previous section we described the harmonic oscillator as a
constrained system in the history phase space, and then recovered the
standard formulation by eliminating these second-class constraints.
By itself, this amounts to nothing. The advantage of this viewpoint
is that it suggests a different procedure: divide the constraints
into two sets, and regard one set as a first class constraint and the
other as a gauge fixation. This allows us to employ powerful
cohomological (BRST) methods.

The division of the constraints into first class and gauge conditions
is not unique, and will be done in different ways. In this
subsection we deal with a corrected version of the method presented
in \cite{Lar04}, and in subsections \ref{ssec:cohom2} and
\ref{ssec:cohom3} we introduce other cohomological descriptions.

To eliminate the unphysical variables in cohomology, we introduce
fermi\-onic antifields $\phi^*_k$ with canonical momenta $\pi^*_k$,
subject to the canonical anti-commutation relations (CAR)
\be
\{\phi^*_k, \pi^*_{k'}\} = \delta_{k+k'}, \qquad
\{\phi^*_k, \phi^*_{k'}\} = \{\pi^*_k, \pi^*_{k'}\} = 0.
\ee
Denote the the extended history phase space by $\HP^*$, and the space
of functions over it by $C(\HP^*) = C(\phi, \phi^*, \pi, \pi^*)$.
The dynamics constraint is implemented by the BRST charge
$Q = Q_D$,
\be
Q_D = \sumall{k} \EE_k \pi^*_{-k} = \sumall{k} \kw \phi_k \pi^*_{-k},
\label{Q1}
\ee
which acts as $\dlt_D F = [Q_D, F]$:
\bes
\dlt_D \phi_k = 0 & \qquad & \dlt_D \pi_k = i\kw \pi^*_k, \nle
\dlt_D \phi^*_k = \kw \phi_k, && \dlt_D \pi^*_k = 0.
\eens

Let us fix $k\neq \pm\w$, so $k^2 - \w^2 \neq 0$. It is clear that
$\ker \dlt_D$ consists of $\phi_k$ and $\pi^*_k$, because $\dlt_D
\pi_k$ and $\dlt_D \phi^*_k$ are nonzero. However, both $\phi_k =
\dlt_D(\phi^*_k/\kw)$ and $\pi^*_k = -i \dlt_D(\pi_k/\kw)$ are exact,
and hence all four variables $\phi_k$, $\phi^*_k$, $\pi_k$ and
$\pi^*_k$ vanish in cohomology.

In contrast, when $k^2 - \w^2 = 0$,
\bes
&&\dlt_D \phi_\w = \dlt_D \pi_\w = \dlt_D \phi^*_\w = \dlt_D \pi^*_\w = 0, 
\nle
&&\dlt_D \phi_{-\w} = \dlt_D \pi_{-\w} = \dlt_D \phi^*_{-\w} =
\dlt_D \pi^*_{-\w} = 0.
\eens
Hence the cohomology groups are generated by all eight variables with
$k = \pm\w$. This is clearly an overcounting, because we know that
the correct phase space is generated by only two variables $a$ and
$\adag$ (\ref{aadag}). It was argued in \cite{Lar04} that this
overcounting would disappear when interactions are introduced. However,
this argument was wrong. To eliminate overcounting, further terms must
be added to the BRST charge.

To eliminate $\phi^*_\w, \phi^*_{-\w}, \pi^*_\w, \pi^*_{-\w}$ we
introduce bosonic antifields $\theta_\w$ and $\theta_{-\w}$ with
canonical momenta $\chi_\w$ and $\chi_{-\w}$. They satisfy the CCR
\be
[\theta_k, \chi_{k'}] = i \delta_{k+k'}, \qquad
[\theta_k, \theta_{k'}] = [\chi_k, \chi_{k'}] = 0,
\ee
where $k, k' = \pm\w$.
The {\em antifield constraint} $\phi^*_\w \approx \phi^*_{-\w}
\approx \pi^*_\w \approx \pi^*_{-\w} \approx 0$ is implemented by the
BRST operator
\be
Q_A = \phi^*_\w \chi_{-\w} + \phi^*_{-\w} \chi_\w,
\label{Q2}
\ee
which acts in an extended phase space with basis
$\phi_k$, $\pi_k$, $\phi^*_k$, $\pi^*_k$, $\theta_{\pm\w}$,
$\chi_{\pm\w}$; this space is still denoted by $\HP^*$.
It is clear that $\{Q_D, Q_A\} = 0$, because $Q_D$ is independent of
all modes with $k = \pm\w$.
The antifield constraint acts as $\dlt_A F = [Q_A, F]$, where
\bes
\dlt_A \phi^*_\w = 0, &\qquad&
\dlt_A \phi^*_{-\w} = 0, \nl
\dlt_A \pi^*_\w = \chi_\w, &\qquad&
\dlt_A \pi^*_{-\w} = \chi_{-\w}, \nle
\dlt_A \theta_\w = -i\phi^*_\w, &\qquad&
\dlt_A \theta_{-\w} = -i\phi^*_{-\w}, \nl
\dlt_A \chi_\w = 0, &\qquad&
\dlt_A \chi_{-\w} = 0.
\eens
Hence $\ker \dlt_A = \im \dlt_A =
C(\phi^*_\w, \phi^*_{-\w},\chi_\w, \chi_{-\w})$.
The kernel and image are the same, and all variables vanish in
cohomology.
We are thus left with the four variables $\phi_{\pm\w}$ and
$\pi_{\pm\w}$, two of which need to be eliminated. To this end, we
consider the {\em momentum constraints}
\be
\MM_\w = \pi_\w - i\w \phi_\w, \qquad
\MM_{-\w} = \pi_{-\w} + i\w \phi_{-\w},
\label{Mw}
\ee
which satisfy
\be
[\MM_\w, \MM_{-\w}] = 2\w.
\label{MMMM}
\ee
Since these constraints have a nonzero Poisson bracket, they are second
class. We must therefore consider one of them, say $\MM_\w$, as a
first-class constraint, and the other as a gauge fixation.
To implement this in cohomology, we can introduce two canonically
conjugate fermi\-onic antifields $\bt_\w$ and $\bt_{-\w}$:
\be
\{\bt_\w, \bt_{-\w}\} = 1, \qquad
\{\bt_\w, \bt_\w\} = \{\bt_{-\w}, \bt_{-\w}\} = 0.
\label{bw}
\ee
One way to impose the momentum constraints is to choose the BRST charge
\be
\QM = \MM_\w \bt_{-\w}
\label{Q3a}
\ee
It is clear that $\{\QM, \QM\} = 0$, since $[\MM_\w, \MM_\w] = 0$
in view of (\ref{Mw}).
Moreover, $\{Q_D, \QM\} = \{Q_A, \QM\} = 0$, because $Q_D$ is
independent of all modes with $k = \pm\w$ and $Q_A$ is independent of
$\phi_{\pm\w}$ and $\pi_{\pm\w}$.
$\QM$ acts on functions over $\HP^*$ as $\dltM F = [\QM, F]$; the
relevant piece is on the modes with $k = \pm\w$:
\bes
\dltM \phi_\w = 0 &\qquad&
\dltM \phi_{-\w} = -i\bt_{-\w} \nl
\dltM \pi_\w = 0 &&
\dltM \pi_{-\w} = \w\bt_{-\w}
\label{dlt3} \\
\dltM \bt_\w = \MM_\w && \dltM \bt_{-\w} = 0.
\eens
In particular,
\be
\dltM \MM_\w = 0, \qquad \dltM \MM_{-\w} = 2\w\bt_{-\w}.
\ee
$\ker \QM$ is generated by $\phi_\w$, $\pi_\w$, $\bt_{-\w}$,
and $\pi_{-\w} - i\w\phi_{-\w}$. $\im \QM$ is generated by
$\MM_\w = \pi_\w - i\w\phi_\w$ and $\bt_{-\w}$. The cohomology
$H^\bullet_\cl(\QM)$ is thus generated by
\bes
a_\w &=& \sw(\pi_w + i\w\phi_\w) + x\MM_\w,
\nlb{aw}
a_{-\w} &=& \sw(\pi_{-\w} - i\w\phi_{-\w}),
\eens
where $x$ is an arbitrary constant.
They have the nonzero brackets
\be
[a_\w, a_{-\w}] = -1.
\ee
In particular, $a_{-\w}$ commutes with $\MM_\w$, so the bracket
is independent of the parameter $x$, as is necessary because it must
be well-defined in cohomology. If we make the particular choice $x=0$,
$a_\w$ also commutes with $\MM_{-\w}$, and we can identify
$a_\w = \adag$, $a_{-\w} = a$ in (\ref{aadag}) and (\ref{awaw}).

Let us summarize the construction in this subsection.
We started from the space of functions over the extended phase space,
\break $C(\phi_k, \pi_k, \phi^*_k, \pi^*_k, \theta_\w, \chi_{-\w},
\theta_{-\w}, \chi_\w, \bt_\w, \bt_{-\w})$. We then introduced the
BRST charge $Q = Q_D + Q_A + \QM$, (\ref{Q1}), (\ref{Q2}), (\ref{Q3a}),
which is nilpotent. It is clear from the discussion above that
cohomology groups are
\be
H^0_\cl(Q) = C(a_\w, a_{-\w}), \qquad
H^n_\cl(Q) = 0, \ \forall n \neq 0.
\label{classcohom}
\ee
In other words, we have obtained a resolution of the classical phase
space $C(a_\w, a_{-\w})$.

In the extended history phase space $\HP^*$, we can define a natural
Hamiltonian as the generator of rigid time translations:
\be
\H = \sumall{k} k (i\phi_k \pi_{-k} + \phi^*_k \pi^*_{-k})
+ i\w \theta_\w \chi_{-\w} - i\w \theta_{-\w} \chi_\w
+ \w \bt_\w \bt_{-\w}
\label{HamHP}
\ee
This Hamiltonian picks out the Fourier frequency $k$; for any
monomial $A_k B_{k'} ...$ on $\HP^*$, we have $[\H, A_k B_{k'} ...]
= (k+k'+...)A_k B_{k'} ...$. In particular, the BRST charge $Q$ carries
zero frequency, and hence it commutes with the Hamiltonian; $[\H, Q] = 0$.
Hence the Hamiltonian acts in a well-defined manner on the cohomology
groups. We verify that the Hamiltonian can be written as
\be
\H = \w a_\w a_{-\w} + \{Q, O\},
\ee
where
\bes
O &=& \sum_{k^2\neq\w^2} {ik\/\kw} \fs_k \pi_{-k} \nle
&&+\  i\w(\theta_\w\pi^*_{-\w} - \theta_{-\w}\pi^*_\w)
+ \half \bt_\w\MM_{-\w}.
\eens
In particular, on the zeroth cohomology
$H^0_\cl(Q) = C(a_\w, a_{-\w})$
it is equivalent to the operator $\H = \w a_\w a_{-\w}$, which is
(\ref{Hawaw}).

\subsection{Quantization}

So far, our considerations were completely classical.
The idea behind MCCQ is to quantize in the extended history phase
space first, and impose the constraints by passing to cohomology
afterwards. We quantize the theory by introducing a vacuum state $\ket0$
which is annihilated by all negative-frequency operators. Hence
\bes
\phi_{-k}\ket0 = \pi_{-k}\ket0 = \phi^*_{-k}\ket0 = \pi^*_{-k}\ket0 = \nle
\theta_{-\w}\ket0 = \chi_{-\w}\ket0 = \bt_{-\w}\ket0 = 0.
\label{killneg}
\eens
Recall from (\ref{classcohom}) that the physical phase space could
be identified with the classical cohomology
$H^0_\cl(Q) = C(a_\w, a_{-\w})$. Since all negative-frequency operators
annihilate the vacuum, quantization leaves us with the state space
$H^0_\st(Q) = C(a_\w)$, $H^n_\st(Q) = 0$, $n \neq 0$. A basis
for the state space $H^0_\st(Q)$ is thus given by
\be
\ket n = {1\/\sqrt{n!}} a_\w^n\ket 0
\label{ketn2}
\ee
An operator $A$ is physical if $[Q, A] = 0$, and two physical operators
$A$ and $A'$ are equivalent if $A' = A + [Q,B]$, $B$ arbitrary. This
is precisely the definition of the classical cohomology of $Q$, and
hence the operator cohomology is given by
$H^0_\op(Q) = H^0_\cl(Q) = C(a_\w, a_{-\w})$,
$H^n_\op(Q) = H^n_\cl(Q) = 0$, $n \neq 0$.

The Hamiltonian commutes with the BRST charge, $[\H,Q] = 0$,
and hence the energy of the $n$-quanta state (\ref{ketn2}) is given by
\be
\H\ket n = n\w \ket n.
\label{Hn2}
\ee
Note the complete isomorphism between (\ref{aadag}), (\ref{ketn1}),
(\ref{Hn1}) and (\ref{aw}), (\ref{ketn2}), (\ref{Hn2}), provided that
we identify
\be
a_\w = \adag, \qquad a_{-\w} = a.
\ee
This shows that the state space $H^0_\st(Q)$ is isomorphic, as a linear
space, to the Hilbert space of the harmonic oscillator, with the same
energy eigenvalues.

The first major flaw in \cite{Lar04}, the overcounting of states in the
harmonic oscillator, has thus been corrected. This was done by
introducing bosonic antifields
$\theta_\w$, $\theta_{-\w}$, $\chi_\w$, $\chi_{-\w}$ to eliminate
$\phi^*_\w$, $\phi^*_{-\w}$, $\pi^*_\w$, $\pi^*_{-\w}$, and by introducing
fermi\-onic antifields $\bt_\w$, $\bt_{-\w}$ to eliminate the constraints
$\MM_\w$, $\MM_{-\w}$, i.e. to identify momenta and velocities.

\subsection{Involution and cohomological description II}
\label{ssec:cohom2}

In the previous subsection we succeeded in constructing a resolution of
the Hilbert space of the harmonic oscillator, {\em regarded as a linear
space}. However, a Hilbert space is not just a linear space; it is
a metric space equipped with an positive-definite inner product. In
addition to the overcounting of states, which is a direct error,
\cite{Lar04} has a serious omission: no attempt was made to construct
the inner product.

Involution can be naturally defined in the history phase space as
negation of frequency:
\bes
\phi^\dagger_k = \phi_{-k}, &\qquad&
\pi^\dagger_k = \pi_{-k}, \nl
\phi^{*\dagger}_k = \phi^*_{-k}, &\qquad&
\pi^{*\dagger}_k = \pi^*_{-k},
\nlb{invharm}
\theta^\dagger_{\pm\w} = \theta_{\mp\w}, &\qquad&
\chi^\dagger_{\pm\w} = \chi_{\mp\w}, \nl
\bt^\dagger_{\pm\w} = \bt_{\mp\w}.
\eens
$\Qdag_D = Q_D$ and $\Qdag_A = Q_A$, but
\be
\Qdag_M = \MM_{-\w}\bt_\w \neq \QM = \MM_\w\bt_{-\w}.
\ee
The BRST operator is thus not selfadjoint, $Q^\dagger \neq Q$, which
ruins the cohomological setup.
The source of this problem is clearly that the $\MM_{\pm\w}$ constraints
are second class, cf. (\ref{MMMM}). To make them first class, we employ
a stardard trick. Introduce a new pair of canonically conjugate
bosonic variables $\al_{\pm\w}$, with nonzero brackets
\be
[\al_\w, \al_{-\w}] = -2\w.
\label{alal}
\ee
We replace (\ref{Mw}) by
\bes
\MM'_\w = \MM_\w + \al_\w
&=& \pi_\w - i\w \phi_\w + \al_\w ,
\nlb{Mw2}
\MM'_{-\w} = \MM_{-\w} + \al_{-\w}
&=& \pi_{-\w} + i\w \phi_{-\w} + \al_{-\w},
\eens
which are first class: $[\MM'_\w,\MM'_{-\w}] = 0$.
Consequently, the corresponding antifields must also be first class.
To this end, we let the antifields $\bt_{\pm\w}$ anticommute, and
introduce their canonical momenta $\gm_{\pm\w}$.
The brackets (\ref{bw}) are now replaced by
\bes
&&\{\bt_\w, \gm_{-\w}\} = \{\bt_{-\w}, \gm_\w\} = 1, \nle
&&\{\bt_{\pm\w}, \bt_{\pm\w}\} = \{\gm_{\pm\w}, \gm_{\pm\w}\} = 0,
\eens
and the BRST charge (\ref{Q3a}) by
\be
Q'_M = \MM'_\w \gm_{-\w} +  \MM'_{-\w} \gm_\w.
\label{Q'M}
\ee
The relevant part of the BRST charge is $Q = Q_A + Q'_M$, which acts
as follows on the modes with $k=\w$:
\bes
\dlt \phi_\w = -i\gm_\w, &\qquad&
\dlt \pi_\w = -\w\gm_\w, \nl
\dlt \phi^*_\w = 0, &\qquad&
\dlt \pi^*_\w = \chi_\w, \nl
\dlt \bt_\w = \MM'_\w, &\qquad&
\dlt \gm_\w = 0,
\label{dltw}\\
\dlt \theta_\w = -i\fs_\w, &\qquad&
\dlt \chi_\w = 0, \nl
\dlt \al_\w = 2\w\gm_\w, &\qquad&
\eens
We note that $\dlt\phi_\w$, $\dlt\pi_\w$ and $\dlt\al_\w$ are all
proportional to $\gm_\w$. Two linearly independent combinations of
these variables thus belong to the kernel, e.g.	$\pi_\w + i\w\phi_\w$
and $\MM'_\w$. Furthermore, $\fs_\w$, $\gm_\w$ and $\chi_\w$ all
belond to the kernel, so $\dim\ker Q = 5$. $\im Q$ is generated by
$\MM'_\w$, $\fs_\w$, $\gm_\w$ and $\chi_\w$, and $\dim\im Q = 4$.
Hence $\dim H^\bullet_\cl(Q) = 1$.
Since the action on modes with $k=-\w$ is completely analogous, and
$Q_D$ eliminates all modes with $k^2\neq\w^2$, we find that
$H^\bullet_\cl(Q)$ is generated by
\bes
a_\w &=& \sw(\pi_w + i\w\phi_\w) + x\MM'_\w,
\nlb{aw2}
a_{-\w} &=& \sw(\pi_{-\w} - i\w\phi_{-\w}) + y\MM'_{-\w},
\eens
where $x$ and $y$ are arbitrary constants.
Appending the involution rule (\ref{invharm}) by
\be
\bt^\dagger_{\pm\w} = \bt_{\mp\w},
\qquad \gm^\dagger_{\pm\w} = \gm_{\mp\w},
\qquad \al^\dagger_{\pm\w} = \al_{\mp\w},
\ee
we see that the improved BRST charge
$Q = Q_D + Q_A + Q'_M$ is indeed self-adjoint, $\Qdag = Q$.

To ensure that the Hamiltonian in $\HP^*$ still commutes with the
BRST charge, we must give energy to the new modes; (\ref{HamHP})
must therefore be replaced by
\bes
\H &=& \sumall{k} k (i\phi_k \pi_{-k} + \phi^*_k \pi^*_{-k})
+ i\w \theta_\w \chi_{-\w} - i\w \theta_{-\w} \chi_\w
\nlb{Ham2}
&&+\  \w \bt_\w \gm_{-\w} - \w \bt_{-\w} \gm_\w
+ \half \al_\w \al_{-\w}.
\eens
We have $[\H, Q]=0$ and $\H = \w a_\w a_{-\w} + \{Q, O\}$, where
\bes
O &=& \sum_{k^2\neq\w^2} {ik\/\kw} \fs_k \pi_{-k} \nl
&&+\  i\w(\theta_\w\pi^*_{-w} - \theta_{-\w}\pi^*_\w)
+ \half(\bt_\w\al_{-\w} + \bt_{-\w}\al_\w) \\
&&+\  x(\bt_\w\MM_{-\w} - \bt_{-\w}\MM_\w).
\eens
Equation (\ref{Ham2}) thus induces a well-defined Hamiltonian in
cohomology.

\subsection{Cohomological description III}
\label{ssec:cohom3}

The separation of the second-class constraints into first class
constraints plus gauge conditions is not unique. In this subsection we
illustrate this with a non-minimal cohomological formulation of the
harmonic oscillator.

Introduce new bosonic fields $\psi_k$ with
canonical conjugates $\rho_k$, defined for all $k\in\RR$.
For $k^2 \neq \w^2$ we define the dynamics and momentum constraints
\bes
\EE_k &=& (k^2 - \w^2)(\phi_k + \psi_k), \nle
\MM_k &=& \pi_k - ik \phi_k - \rho_k - ik\psi_k,
\eens
and for $k = \pm\w$,
\bes
\EE_{\pm\w} &=& 0, \nl
\MM'_{\pm\w}
&=& \pi_{\pm\w} \mp i\w \phi_{\pm\w} + \al_{\pm\w} \\
&=& \MM_{\pm\w} + \rho_{\pm\w} \pm i\w \psi_{\pm\w} + \al_{\pm\w}.
\eens
The bosonic fields $\psi_k$, $\rho_k$ satisfy the nonzero CCR
$[\psi_k, \rho_{k'}] = i\dlt_{k+k'}$, and $\al_{\pm\w}$ still satisfy
(\ref{alal}). Moreover, we have to introduce new fermionic antifields
$\ka_{\pm\w}$ and their momenta $\la_{\pm\w}$, with the nonzero brackets
$\{\ka_\w,\la_{-\w}\} = \{\ka_{-\w},\la_\w\} = 1$.

The BRST charge is
\bes
Q &=& \sum_k \big( \EE_k \pi^*_{-k} + \MM_k \gm_{-k} \big)
+ (\rho_\w + i\w \psi_\w + \al_\w) \gm_{-\w} \nle
&&+\ \phi^*_\w \chi_{-\w} + \psi_\w\la_{-\w} + \w \leftrightarrow -\w.
\eens
It acts on functions $F \in C(\HP^*)$ as $\dlt F = [Q, F]$.
For $k^2 \neq \w^2$, the action on the basis vectors reads
\bes
\dlt \phi_k = -i\gm_k, &\qquad&
\dlt \pi_k = i\kw\pi^*_k - k\gm_k, \nl
\dlt \psi_k = i\gm_k, &\qquad&
\dlt \rho_k = i\kw\pi^*_k - k\gm_k, \nle
\dlt \phi^*_k = \EE_k, &&
\dlt \pi^*_k = 0, \nl
\dlt \bt_k = \MM_k, &&
\dlt \gm_k = 0.
\eens
$\ker Q$ is generated by $\phi_k+\psi_k$,
$\pi_k-\rho_k$, $\pi^*_k$ and $\gm_k$.
$\im Q$ is generated by $\EE_k$, $\MM_k$, $\pi^*_k$ and $\gm_k$.
Since $\dim \ker Q = \dim \im Q = 4$, $\dim H^\bullet = 0$ and there
is no contribution to cohomology for such $k$.

For $k^2 = \w^2$, say $k=\w$, we have instead
\bes
\dlt \phi_\w = -i\gm_\w, &\qquad&
\dlt \pi_\w = -\w\gm_\w, \nl
\dlt \phi^*_\w = 0, &&
\dlt \pi^*_\w = \chi_\w, \nl
\dlt \psi_\w = 0, &&
\dlt \rho_\w = i\la_\w, \nl
\dlt \bt_\w = \MM'_\w, &&
\dlt \gm_\w = 0, \\
\dlt \ka_\w = \psi_\w, &&
\dlt \la_\w = 0, \nl
\dlt \theta_\w = -i\fs_\w, &&
\dlt \chi_\w = 0, \nl
\dlt \al_\w = 2\w\gm_\w, &&
\eens
This is the same as in (\ref{dltw}), except for the additional
quadruple $\psi_\w$, $\rho_\w$, $\ka_\w$, $\la_\w$, which cancels in
cohomology ($\psi_\w$ and $\la_\w$ are exact and $\rho_\w$ and $\ka_\w$
are not closed). The cohomology is thus the same as in the previous
section. To the Hamiltonian in (\ref{Ham2}) we must add terms which
give energy to the new fields, i.e.
\be
\Delta \H = i\w\psi_\w\rho_{-\w} + \w\ka_\w\la_{-\w}
- \w\leftrightarrow-\w.
\ee

It is impossible to construct the correct phase space algebra
$C(\PP) = C(a, \adag)$ in terms of fields and antifields which
depend on $k$ only; we must also introduce antifields which are
explicitly labelled by the solutions $\pm\w$. In the field theory
context, it means that the momentum constraint is bound to break manifest
covariance. To see this, it suffices to count the degrees of
freedom. Assume for simplicity that the history fields $\phi_k$ only have
finitely many component, i.e. $k$ belongs to a finite but large set with
$N$ elements. In the previous subsection, we introduced antifields and
momenta with the following number of components, where fermions count
negative:
\be
\barr{crc|cr}
\hbox{Field} & \hbox{\# comp} && \hbox{Momentum} & \hbox{\# comp} \\
\hline
\phi_k	& N && \pi_k & N \\
\phi^*_k & -N && \pi^*_k & -N \\
\bt_{\pm\w} & -2 && \gm_{\pm\w} & -2 \\
\theta_{\pm\w} & 2 && \chi_{\pm\w} & 2 \\
\alpha_{\pm\w} & 2 && & \\
\earr
\label{harmcount}
\ee
Hence $\dim H^0_\cl(Q) = 2N - 2N - 4 + 4 + 2 = 2$, which is the correct
dimension for the physical phase space spanned by $a$ and $\adag$.
We can introduce additional antifields, like $\bt_k$ and $\gm_k$ in
this subsection, but then they must be cancelled by two other antifields
$\psi_k$ and $\rho_k$ with $2N$ components. At the end, the balance
is always maintained by the two components of $\alpha_{\pm\w}$.

A similar conclusion holds generally, both for interacting theories
and field theories, as discussed in subsection \ref{ssec:mancov}.

\subsection{Continuous Fourier variable}
\label{ssec:functional}

In the previous subsections the Fourier variable $k$ was assumed to
be discrete. The precise domain of $k$ was unimportant, as long as
$\pm\w$ belongs to this domain. In this subsection we will take the
formal continuum limit of the cohomological formulation in
subsection \ref{ssec:cohom2}. This mainly amounts to replacing
sums by integrals, but there is a slight complication: the
infinite constant $\dlt(0)$ enters in some formulas. This problem is
purely formal, and we will ignore it henceforth.

The dynamics and momentum constraints read
\bes
\EE(k) &=& \kw\phi(k), \nle
\MM(k) &=& \pi(k) - ik\phi(k) + \al(k),
\eens
where
\be
[\al(k), \al(k')] =  -2k\dlt(k+k').
\ee
However, $\MM(k)$ is only defined for $k^2 = \w^2$, and so is
$\al(k)$.
We note that $[\MM(k), \MM(k')] = 0$, and $[\MM(k), \EE(k')] = 0$,
provided that $k^2 = \w^2$, which is the case.
Introduce antifields $\phi^*(k)$ for all $k$, and additional
antifields $\theta(k)$ and $\bt(k)$ for $k^2 = \w^2$, and their
canonical momenta. The complete list of nonzero CCRs is
\bes
[\phi(k), \pi(k')] =  i\dlt(k+k'), &&
\{\phi^*(k), \pi^*(k')\} = \dlt(k+k'), \nl
{[}\theta(k), \chi(k')] = i\dlt(k+k'), &&
\{\bt(k), \gm(k')\} = \dlt(k+k'), \\
{[}\al(k), \al(k')] = -2k\dlt(k+k'). &&
\eens
The BRST charge $Q = Q_D + Q_A + \QM$, where
\bes
Q_D &=& \int dk\ \EE(k)\pi^*(-k), \nl
Q_A &=& \int dk\ \phi^*(k)\chi(-k)\dlt\kw \nl
&=& {1\/2\w} \big(\phi^*(\w)\chi(-\w) + \phi^*(-\w)\chi(\w)),
\label{QBRST3}\\
\QM &=&	\int dk\ \MM(k)\gm(-k)\dlt\kw \nl
&=& {1\/2\w} \bigg( \MM(\w)\gm(-\w) + \MM(-\w)\gm(\w) \bigg),
\eens
acts on the variables defined for all $k$ as
\bes
\dlt \phi(k) &=& -i\gm(k)\dlt\kw, \nl
\dlt \pi(k) &=& i\kw\pi^*(k) - k\gm(k)\dlt\kw, \nle
\dlt \phi^*(k) &=& \kw\phi(k), \nl
\dlt \pi^*(k) &=& \chi(k)\dlt\kw,
\eens
and on the variables defined for $k^2 = \w^2$ as
\bes
\dlt \bt(k) &=& \MM(k)\dlt\kw, \nl
\dlt \gm(k) &=& 0,  \nl
\dlt \theta(k) &=& -i\fs(k)\dlt\kw, \\
\dlt \chi(k) &=& 0, \nl
\dlt \al(k) &=& 2k\gm(k)\dlt\kw.
\eens
The cohomology is computed as in subsection \ref{ssec:cohom2}.
For $k \neq \pm\w$, $\phi(k)$ and $\pi^*(k)$ belong to both the
kernel and the image, and $\pi(k)$ and $\phi^*(k)$ to neither; hence
$H^\bullet_\cl(Q) = 0$.
For $k^2 = \w^2$, $k = \w$ say, the BRST charge acts like in
(\ref{dltw}), except that the RHS is multiplied by the infinite
constant $\dlt\kw \propto \dlt(0)$, e.g.
\bes
\dlt \phi(\w) &=& -i\gm(\w)\dlt(0), \nle
\dlt \al(\w) &=& 2\w\gm(\w)\dlt(0).
\eens
We will consider $\dlt\kw$ simply as a non-zero constant, and ignore
that it is infinite when $k = \pm\w$. The cohomology is then generated
by $a_\w$ and $a_{-\w}$ as in (\ref{aw2}).
The Hamiltonian in $\HP^*$ is
\bes
\H &=& \int dk\ \bigg( ik\phi(k) \pi(-k) + k\phi^*(k) \pi^*(-k) \\
&&+\  \big(ik \theta(k) \chi(-k) + k \bt(k) \gm(-k)
+ \half \al(k) \al(-k)\big)\dlt\kw \bigg),
\eens
We have $[\H, Q]=0$ and $\H = \w a_\w a_{-\w} + \{Q, O\}$, where
and $O$ is some operator.

\subsection{Time-plane formulation}

It is instructive to write down the formalism in the time-plane.
The dynamics constraint becomes
\be
\EE(x) = \phi''(x) + \w^2 \phi(x) \approx 0.
\ee
The following identities hold:
\be
\int dx\ \e^{i\w x} \EE(x) = \int dx\ \e^{-i\w x} \EE(x) = 0.
\ee
We introduce the conjugate momentum $\pi(x)$, the continuous antifield
$\phi^*(x)$, and its canonical momentum $\pi^*(x)$; the nonzero brackets
are
\be
[\phi(x), \pi(x')] = i\dlt(x-x'), \qquad
\{\phi^*(x), \pi^*(x')\} = \dlt(x-x').
\ee
As in section
\ref{ssec:cohom2}, we also introduce the discrete antifields
$\theta_{\pm\w}$, $\bt_{\pm\w}$ and $\al_\w$ with momenta
$\chi_{\mp\w}$, $\gm_{\mp\w}$ and $\al_{-\w}$, respectively.
The BRST charge is $Q = Q_D + Q_A + \QM$, where
\bes
Q_D &=& \int dx\ (\phi''(x) + \w^2 \phi(x)) \pi^*(x), \nl
Q_A &=& \chi_{-\w} \int dx\ \e^{i\w x} \phi^*(x) +
\chi_{\w} \int dx\ \e^{-i\w x} \phi^*(x), \nle
\QM &=& \MM_\w\gm_{-\w} + \MM_{-\w}\gm_\w, \nl
\MM_{\pm\w} &=& \al_{\pm\w} + \int dx\ \e^{\pm i\w x} (\pi(x) - \phi'(x)).
\eens
The cohomology $H^\bullet_\cl(Q)$ is generated by
\bes
a_\w &=& x_\w\MM_\w + \int dx\ \e^{i\w x} (\pi(x) + \phi'(x)), \nle
a_{-\w} &=& x_{-\w}\MM_{-\w} + \int dx\ \e^{-i\w x} (\pi(x) + \phi'(x)),
\eens
where $x_\w$ and $x_{-\w}$ are arbitrary constants.
The Hamiltonian in $\HP^*$,
\bes
\H &=& \int dx\ (i\phi'(x)\pi(x) + {d\/dx}\phi^*(x)\pi^*(x)) \nle
&&+\  \w(\bt_\w\gm_{-\w} - \bt_{-\w}\gm_\w) + \half\al_\w\al_{-\w},
\eens
can be written as $\H = \w a_\w a_{-\w} + \{Q, O\}$,
for some operator $O$.
Involution sends
\bes
\phi^\dagger(x) = \phi(x), &\qquad& \pi^\dagger(x) = \pi(x), \nle
\phi^{*\dagger}(x) = \phi^*(x), && \pi^{*\dagger}(x) = \pi^*(x),
\eens
whereas the involution of the discrete variables is as in (\ref{invharm}).
$Q$ is clearly invariant under this transformation.

\section{ The free particle }

\subsection{ Cohomological construction }
\label{ssec:cohomFree}

The free particle moving in $d$-dimensional space is described by $d$
harmonic oscillators with zero frequency.
It will become of considerable interest below, as
it describes the observer's trajectory in QJT. Anticipating this
application, we denote the independent time parameter by $t$, the
histories by $q^\mu(t)$, and the history momentum $p_\mu(t)$,
subject to non-zero CCR
\be
[q^\mu(t), p_\nu(t')] = i \dlt^\mu_\nu \dlt(t-t').
\label{qpbracket}
\ee
We work in Minkowski space with a flat, background metric $\eta_\mn$
with inverse $\eta^\mn$, which we freely use to raise and lower indices,
e.g. $q_\mu(t) = \eta_\mn q^\nu(t)$, $p^\mu(t) = \eta^\mn p_\nu(t)$.

The equations of motion read
\be
\EE^\mu(t) = \ddot q^\mu(t) \approx 0.
\label{Epart}
\ee
To implement these in cohomology, we introduce fermionic antifields
$q^\mu_*(t)$ with momenta $p^*_\mu(t)$, satisfying non-zero brackets
\be
\{q^\mu_*(t), p^*_\nu(t')\} = \dlt^\mu_\nu \dlt(t-t').
\ee
The equations (\ref{Epart}) are subject to the redundancies
\be
\int dt\ \EE^\mu(t) = \int dt\ t\, \EE^\mu(t) \equiv 0,
\label{redpart}
\ee
which require that we introduce two new, bosonic antifields
$\theta^\mu_1$ and $\theta^\mu_0$ with momenta $\chi^\mu_1$
and $\chi^\mu_0$. We assume that the integral in (\ref{redpart})
vanishes if the integrand is a total derivative.

Moreover, we need to identify momenta and velocities, which is
accomplished by the momentum constraint
\be
\MM_\mu(t) = p_\mu(t) - \dot q_\mu(t).
\ee
Most of the momentum constraints do not commute with dynamics, because
\be
[\MM_\mu(t), \EE^\nu(t')] = -i\dlt^\nu_\mu\ddot\dlt(t-t').
\ee
However, integrating by parts we immediately see that the expressions
\bes
\MM^0_\mu &=& \int dt\ \MM_\mu(t) = \int dt\ p_\mu(t), \nle
\MM^1_\mu &=& \int dt\ t\, \MM_\mu(t) \approx \int dt\ t\, p_\mu(t).
\eens
do commute with $\EE^\nu(t')$. Here we used that
\be
\int dt\ t\, \dot q^\mu(t) = \dlt(-\half\int dt\ t^2\, q_*^\mu(t))
\ee
is BRST exact in view of the dynamics constraint (\ref{Epart}), and
thus it can be ignored. A novel feature compared to the harmonic
oscillator is that the momentum constraints are first class, because
\be
[\MM^0_\mu, \MM^1_\nu] = 0.
\ee
After the dynamics and antifield constraints have been implemented,
four degrees of freedom remain: the two solutions to $\ddot q^\mu(t) = 0$
and their associated momenta.  The task of the momentum constraints is
to identify velocities and momenta, and thus to cut down the number of
degrees of freedom to two. This would be accomplished by two second
class constraints. However, $\MM^0_\mu$ and $\MM^1_\mu$ are first
class and count twice. Hence only of them can be implemented in
cohomology.

Collecting everything, the extended history phase space $\HP^*$ is
spanned by the following fields, antifields, and canonical momenta:
\bes
\barr{c|c|c|l}
\hbox{Field} & \hbox{Momentum} & \hbox{Parity} & \hbox{Constraint} \\
\hline
q^\mu(t) & p_\mu(t) & B & \\
q^\mu_*(t) & p^*_\mu(t) & F & \EE^\mu(t) = \ddot q^\mu(t)\\
\theta^\mu_1 & \chi_\mu^1 & B & \int dt\ q^\mu_*(t) \\
\theta^\mu_0 & \chi_\mu^0 & B & \int dt\ t\, q^\mu_*(t) \\
\bt_\mu^0 & \gm^\mu_0 & F & \MM^0_\mu \\
\bt_\mu^1 & \gm^\mu_1 & F & \MM^1_\mu \\
\earr
\eens
where only one of the antifields $\bt_\mu^0$ and $\bt_\mu^1$ are used.
The total BRST operator is $Q_q = Q_d + Q_a + Q_m$, where
\bes
Q_d &=& \int dt\  \ddot q^\mu(t) p^*_\mu(t), \nl
Q_a &=& \chi^1_\mu \int dt\ q^\mu_*(t) +
\chi^0_\mu \int dt\ t\, q^\mu_*(t),
\label{Qpart}\\
Q_m &=& \begin{cases}
\MM^0_\mu \gm^\mu_0, \qquad \hbox{or}\\
\MM^1_\mu \gm^\mu_1.
\end{cases}
\eens
The analysis becomes simpler if we expand all fields in a Taylor series,
to which we turn next.

\subsection{Mode expansion }
\label{ssec:modes}

We assume that all fields can be expanded in a Laurent series in $t$,
and define the modes $q^\mu_m$ and $p^m_\mu$ by
\bes
q^\mu(t) &=& \sumall{m} q^\mu_m t^m,
\nlb{qpmodes}
p_\mu(t) &=& \sumall{m} p^m_\mu t^{-m-1}.
\eens
The Laurent modes satisfy the non-zero brackets
\be
[q^\mu_m, p_\nu^n] = -i\dlt^\mu_\nu \dlt^n_m.
\label{qpmbracket}
\ee
Using the expansion of the delta function around $s=1$,
\be
\dlt(1-s) = \sumall{m} s^m
\ee
and that
\be
\dlt(t-t') = -{1\/t'}\dlt(1-{t\/t'}),
\ee
we find that the definition (\ref{qpmbracket}) leads to the correct CCR
(\ref{qpbracket}) for the fields (\ref{qpmodes}).
The modes can be recovered from the fields by taking moments, e.g.
\be
q^\mu_m = \int dt\ t^{-1-m} q^\mu(t).
\ee
In particular, $q^\mu_{-1}$ is the residue.

The dynamics constraint takes the form
\be
\EE^\mu_m = (m+2)(m+1)q^\mu_{m+2}.
\ee
Hence we introduce an antifield $q^{*\mu}_m$ with momentum $p^m_{*\mu}$
and nonzero brackets $\{q^{*\mu}_m, p^n_{*\nu}\} = \dlt^\mu_\nu \dlt^n_m$,
and define the dynamics constraint
\be
Q_d = \sumall{m} \EE^\mu_m p^m_{*\mu}
= \sumall{m} m(m-1) q^\mu_m p^{m-2}_{*\mu}.
\label{Qdmode}
\ee
$Q_d$ acts like
\bes
\dlt_d q^\mu_m = 0, &&
\dlt_d p_\mu^m = -im(m-1)p^{m-2}_{*\mu}, \nle
\dlt_d q^{*\mu}_m = (m+2)(m+1)q^\mu_{m+2}, &&
\dlt_d p^m_{*\mu} = 0.
\eens
$\ker Q_d$ is generated by all $q^\mu_m$, $q^{*\mu}_{-2}$,
$q^{*\mu}_{-1}$, $p^0_\mu$, $p^1_\mu$ and all $p^m_{*\mu}$.
$\im Q_d$ is generated by all $q^\mu_m$ except $m=0$ and $m=1$,
and by all $p^m_{*\mu}$ except $m = -2$ and $m = -1$.
$H^\bullet_\cl(Q_d)$ is hence generated by $q^\mu_0$, $q^\mu_1$,
$q^{*\mu}_{-2}$, $q^{*\mu}_{-1}$, $p^0_\mu$, $p^1_\mu$,
$p^{-2}_{*\mu}$ and $p^{-1}_{*\mu}$.

The antifields constraints correspond to
\be
q^{*\mu}_{-1} = \int dt\ q^\mu_*(t), \qquad
q^{*\mu}_{-2} = \int dt\ t\, q^\mu_*(t).
\label{residues}
\ee
To kill the antifields in cohomology, we introduce new bosonic antifields
$\theta^\mu_1$, $\theta^\mu_0$ and their momenta
$\chi_\mu^1$, $\chi_\mu^0$. The antifield BRST charge,
\be
Q_a = q^{*\mu}_{-1} \chi_\mu^1 +  q^{*\mu}_{-2} \chi_\mu^0,
\label{Qamode}
\ee
acts on the relevant fields as
\bes
\dlt_a q^{*\mu}_{-1} = \dlt_a q^{*\mu}_{-2} = 0, &
\dlt_a p^{-1}_{*\mu} = \chi^1_\mu, &
\dlt_a p^{-2}_{*\mu} = \chi^0_\mu, \\
\dlt_a \theta^\mu_1 = q^{*\mu}_{-1}, &
\dlt_a \theta^\mu_0 = q^{*\mu}_{-2}, &
\dlt_a \chi^1_\mu = \dlt_a \chi^0_\mu = 0.
\eens
It is clear that the antifields cancel in quadruplets,
and $H^\bullet_\cl(Q_d+Q_a)$ is thus generated by $q^\mu_m$ and $p^m_\mu$,
$m = 0,1$.

The modes of the momentum constraints read
\be
\MM^m_\mu = p^m_\mu + m\eta_\mn q^\nu_{m+1},
\ee
for $m = 0,1$. They commute with $\EE^\mu_m$:
\be
[\MM^m_\mu, \EE^\nu_n] = i m(m-1) \dlt^\nu_\mu \dlt^m_{n+2} = 0.
\ee
Moreover, since $q^\mu_{-1} = \dlt(1/2 q^{*\mu}_{-3}) \approx 0$, we
can take the momentum constraints to be
\be
\MM^0_\mu = p^0_\mu, \qquad\hbox{or}\qquad
\MM^1_\mu = p^1_\mu.
\label{MM01}
\ee
$\MM^0_\mu$ and $\MM^1_\mu$ commute, and hence they are first
class constraints. We can therefore only hope to implement one of them
in cohomology.

To implement $\MM^0_\mu \approx 0$, we introduce a pair of antifields
$\bt_\mu^0$ and $\gm_0^\mu$, and the momentum part of the
BRST operator becomes
\be
Q_m \equiv Q^0_m = p^0_\mu\gm^\mu_0.
\label{Qmmode0}
\ee
It acts on the relevant fields as
\bes
\dlt_m q^\mu_0 = i\gm^\mu_0, &&
\dlt_m p^0_\mu =
\dlt_m q^\mu_1 =
\dlt_m p^1_\mu = 0, \nl
\dlt_m \bt^\mu_0 = p^0_\mu &&
\dlt_m \gm^0_\mu = 0.
\ees
The quadruplet $q^\mu_0$, $p^0_\mu$, $\bt^\mu_0$ and $\gm_\mu^0$ cancels,
and the cohomology is generated by
\bes
u_\mu &=& q^1_\mu + x p^0_\mu, \nle
s_\mu &=& p^1_\mu,
\eens
for some constant $x$. The brackets
\be
[s_\mu, u_\nu] = i\eta_\mn,
\ee
are well defined in $H^\bullet_\cl(Q^0_m)$, i.e. independent of $x$.

Alternatively, we could implement $\MM^1_\mu \approx 0$ in cohomology.
We then introduce a pair of antifields
$\bt_\mu^1$ and $\gm_1^\mu$, and the momentum part of the
BRST operator becomes
\be
Q_m \equiv Q^1_m = p^1_\mu\gm^\mu_1.
\label{Qmmode1}
\ee
It acts on the relevant fields as
\bes
\dlt_m q^\mu_1 = i\gm^\mu_1, &&
\dlt_m p^1_\mu =
\dlt_m q^\mu_0 =
\dlt_m p^0_\mu = 0, \nl
\dlt_m \bt^\mu_1 = p^1_\mu &&
\dlt_m \gm^1_\mu = 0.
\ees
The quadruplet $q^\mu_1$, $p^1_\mu$, $\bt^\mu_1$ and $\gm_\mu^1$ cancels,
and the cohomology is generated by
\bes
s_\mu &=& q^0_\mu + x p^1_\mu, \nle
u_\mu &=& p^0_\mu,
\eens
for some constant $x$. The brackets
\be
[s_\mu, u_\nu] = -i\eta_\mn,
\ee
are well defined in $H^\bullet_\cl(Q^1_m)$, i.e. independent of $x$.

\subsection{Involution and quantization}
\label{ssec:qpart}

Three different notions of involution are conceivable.

1. $m \to 1-m$, e.g. $q^\mu_m \to q^\mu_{1-m}$ and
$p_\mu^m \to p_\mu^{1-m}$. Then
\be
Q^0_m \to Q^1_m \to Q^0_m, \qquad
q^\mu(t) \to t q^\mu({1\/t}), \qquad
p_\mu(t) \to {1\/t}p_\mu({1\/t}).
\ee
Since the momentum part of the BRST operator is not invariant, this
notion of involution leads to difficulties for quantization.

2. $m \to -m$, e.g. $q^\mu_m \to q^\mu_{-m}$ and
$p_\mu^m \to p_\mu^{-m}$. Then
\be
Q^0_m \to Q^0_m, \qquad
q^\mu(t) \to q^\mu({1\/t}), \qquad
p_\mu(t) \to p_\mu({1\/t}).
\ee
Since $Q^0_m$ is invariant under involution, we choose it as the
momentum part of the BRST operator, and the corresponding vacuum $\ket0$
in $\HP^*$ annihilates all modes with $m < 0$, e.g.
$q^\mu_{-m}\ket 0 = p_\mu^{-m}\ket0 = 0$ for all $-m < 0$. We must also
decide whether $q^\mu_0$ or $p_\mu^0$ should annihilate the vacuum.

3. $m \to 2-m$, e.g. $q^\mu_m \to q^\mu_{2-m}$ and
$p_\mu^m \to p_\mu^{2-m}$. Then
\be
Q^1_m \to Q^1_m, \qquad
q^\mu(t) \to t^2 q^\mu({1\/t}), \qquad
p_\mu(t) \to {1\/t^2}p_\mu({1\/t}).
\ee
Since $Q^1_m$ is invariant under involution, we choose it as the
momentum part of the BRST operator, and the corresponding vacuum $\ket0$
in $\HP^*$ annihilates all modes with $m < 1$, e.g.
$q^\mu_{-m}\ket 0 = p_\mu^{-m}\ket0 = 0$ for all $-m \leq 0$. We must also
decide whether $q^\mu_1$ or $p_\mu^1$ should annihilate the vacuum.

\subsection{ Scale symmetry }

Scale transformations are generated by the dilatations $D = t d/dt$,
which satisfy the one-dimensional abelian algebra $[D,D] = 0$. To fix
notation, let us recall some basic facts about representations of
the scaling group.
A constant operator $A$ has weight $\la$ if it transforms under
dilatations as
\be
[D, A] = - \la A.
\label{wtA}
\ee
We write $\wt A = \la$. The weight is preserved by brackets; if
$\wt A = \la_A$ and $\wt B = \la_B$, $\wt [A,B] = \la_A + \la_B$.
A function $\phi(t)$ has weight $\la$ if it transforms as
\be
[D, \phi(t)] = - t \dot\phi(t) - \la \phi(t)
= \sumall{m} -(m+\la) \phi_m t^m.
\label{wtphi}
\ee
Again we write $\wt \phi(t) = \la$. We see that the mode $\phi_m$
has weight $m+\la$ in the sense (\ref{wtA}). If $\wt \phi(t) = \la$, then
$\wt \dot \phi(t) = \la + 1$, and more generally
$\wt d^m\phi/dt^m = \la+m$. Moreover, if $\wt \phi(t) = \la$, then its
$m$:th moment is a constant operator and
$\wt (\int dt\ t^m \phi(t)) = \la-m-1$
in the sense (\ref{wtA}). In particular, the residue
$\int dt\ t^{-1+\la} \phi(t)$ is invariantly defined.

If $\pi(t)$ is canonical conjugate
to $\phi(t)$, i.e. $[\phi(t), \pi(t')] = i\dlt(t-t')$, then
$\wt \pi(t) = 1-\la$. For the corresponding modes, $\wt \pi^m = -m-\la$.
This is compatible with the brackets, because if
\be
[D, \phi_m] = -(m+\la) \phi_m, \qquad
[D, \pi^m] = (m+\la) \pi^m,
\ee
then $[D, [\phi_m, \pi^n]] = 0 = [D, -i\dlt^n_m]$.

Dilatations act on the fields and momenta according to the following
table:
\bes
\barr{cr|cr|cr}
\hbox{Field} & \la &\hbox{Momentum} & \la & \hbox{Constraint} & \la\\
\hline
q^\mu(t) & 0 & p_\mu(t) & 1 && \\
q^\mu_m & m & p^m_\mu & -m && \\
q^\mu_*(t) & 2 & p^*_\mu(t) & -1 & \ddot q^\mu(t) & 2\\
q^{*\mu}_m & m+2 & p^m_{*\mu} & -m-2 & (m+2)(m+1)q^\mu_{m+2} & m+2\\
\theta^\mu_0 & 0 &\chi_\mu^0 & 0 &
q^{*\mu}_{-2} = \int dt\ t\, q^\mu_*(t) & 0 \\
\theta^\mu_1 & 1 &\chi_\mu^1 & -1 &
q^{*\mu}_{-1} = \int dt\ q^\mu_*(t) & 1 \\
\bt_\mu^0 & 0 & \gm^\mu_0 & 0 & \MM^0_\mu = p^0_\mu  & 0 \\
\bt_\mu^1 & -1 & \gm^\mu_1 & 1 & \MM^1_\mu = p^1_\mu& -1 \\
\earr
\eens
Thus, the indices in (\ref{Qdmode}), (\ref{Qamode}), (\ref{Qmmode0})
and (\ref{Qmmode1}) equal the weight of the corresponding mode, except
for $q^{*\mu}_m$ which is shifted by two units. Hence $\wt Q = 0$,
and the weight is preserved by the BRST charge.

\subsection{ Energy-momentum tensor }

The energy-momentum tensor
\be
T(t) = \eta_\mn \dot q^\mu(t) \dot q^\nu(t)
\ee
is preserved by time evolution; $\dot T(t) = 0$ when the equations of
motion $\ddot q^\mu(t) = 0$ are taken into account.
In the present framework, this is
expressed by noting that the time derivative $\dot T(t)$ is BRST
exact, which is straightforward to verify:
\bes
\dot T(t) &=& 2\eta_\mn \ddot q^\mu(t) \dot q^\nu(t) = \dlt U(t),
\nle
U(t) &=& 2\eta_\mn q_*^\mu(t) \dot q^\nu(t).
\eens

Using that $\dot q^\mu_m = (m+1) q^\mu_{m+1}$, the modes of the
energy-momentum tensor become
\be
T_m = \eta_\mn \sumall{n} (n+1)(m-n+1) q^\mu_{n+1} q^\mu_{m-n+1}.
\ee
One verifies that $T_m$ is BRST closed, and that
$(m+1) T_{m+1} = \dlt U_m$, where
\be
U_m = 2\eta_\mn \sumall{n} (m-n) q^{*\mu}_{n+1} q^\nu_{m-n}.
\ee
Hence $T_m$ is BRST exact, except for
\be
T_0 = \eta_\mn \sumall{n} n(2-n) q^\mu_n q^\nu_{2-n},
\ee
which thus belongs to cohomology.

We check that the weights match.
\be
\barr{c|cccc|cccc}
\hbox{Field} & q^\mu(t) & T(t) & q_*^\mu(t) & U(t)
& q^\mu_m & T_m & q^{*\mu}_m & U_m \\
\hline
\hbox{Weight} & 0 & 2 & 2 & 3 & m & m+2 & m+2 & m+3
\earr
\ee
Since $\wt T(t) = 2$, the physical operator $T_0$ equals the residue
\be
T_0 = \int dt\ t^{-1}T(t).
\ee
Thus it acts in a well-defined manner on the physical Hilbert space.
For reasons that will become apparent below, we refer to the $T_0$
eigenvalue as {\em the observer's mass} squared and denote it by
$\Mobs^2$.

\subsection{ Hamiltonian}

Since the Hamiltonian is the generator of rigid time translations, we
are tempted to define the Hamiltonian in $\HP^*$ as
\be
\H_0 = \int dt\ \dot q^\mu(t) p_\mu(t)
-i \int dt\ \dot q^\mu_*(t)p^*_\mu(t).
\label{Hpart}
\ee
This definition would lead to the expected time evolution
\bes
[\H_0, q^\mu(t)] = -i\dot q^\mu(t), &&
[\H_0, p_\mu(t)] = -i\dot p_\mu(t), \nle
{[}\H_0, q_*^\mu(t)] = -i\dot q_*^\mu(t), &&
[\H_0, p^*_\mu(t)] = -i\dot p^*_\mu(t).
\eens
However, $\H_0$ does not commute with the BRST operator. Namely,
because
\bes
[\H_0, \int dt\ q_*^\mu(t)] = 0, &&
[\H_0, \int dt\ t\,q_*^\mu(t)] = i\int dt\ q_*^\mu(t), \nle
{[}\H_0, \MM^0_\mu] = 0, &&
[\H_0, \MM^1_\mu] = i\MM^0_\mu,
\eens
we find
\bes
[\H_0, Q_d] &=& 0, \nl
{[}\H_0, Q_a] &=& i\chi^1_\mu \int dt\ q_*^\mu(t), \\
{[}\H_0, Q_m] &=& i\MM^0\gm^\mu_1.
\eens
To cancel these contributions, we add an extra term to the Hamiltonian:
\be
\H_1 = \chi^1_\mu \theta^\mu_1 + i\gm^\mu_1\bt^0_\mu.
\ee
One verifies that the total Hamiltonian $\H = \H_0 + \H_1$ does
commute with the BRST charge $Q_q = Q_d + Q_a + Q_m$. It is therefore
well defined on the cohomology $H^\bullet_\cl(Q_q)$.

The Hamiltonian $\H = \H_0 + \H_1$ translates into
\bes
\H_0 &=& - i \sumall{m} (m+1)q^\mu_{m+1} p^m_\mu
+ \sumall{m} (m+1)q^{*\mu}_{m+1} p^m_{*\mu} \nl
&&+\ i \theta^\mu_0 \chi^{-2}_\mu
- \bt^0_\mu \gm^\mu_1.
\ees
$\H_0$ acts on the modes as
\bes
[\H_0, q^\mu_m] = (m+1)q^\mu_{m+1} &&
[\H_0, p^m_\mu] = -mp^{m-1}_\mu, \nl
{[}\H_0, q^{*\mu}_m] = (m+1)q^{*\mu}_{m+1} &&
[\H_0, p^m_{*\mu}] = -mp^{m-1}_{*\mu}, \nl
{[}\H_0, \theta^\mu_1] = -\theta^\mu_0 &&
[\H_0, \chi^{-2}_\mu] = 0, \nl
{[}\H_0, \theta^\mu_0] = 0 &&
[\H_0, \chi^{-1}_\mu] = \chi^{-2}_\mu, \\
{[}\H_0, \gm^\mu_0] = \gm^\mu_1 &&
[\H_0, \bt^0_\mu] = 0, \nl
{[}\H_0, \gm^\mu_1] = 0 &&
[\H_0, \bt^1_\mu] = -\bt^0_\mu.
\eens
We verify that $[\H_0, Q_d+Q_a] = 0$, but
\be
[\H_0, Q_m] = -2\eta_\mn q^\mu_2 \gm^\nu_0 - \eta_\mn q^\mu_1 \gm^\nu_1.
\label{H0Qm}
\ee
The first term equals $[\H_1, Q_d]$, where
\be
\H_1 = -\eta_\mn q^{*\mu}_0\gm^\nu_0,
\ee
and it can hence be cancelled by taking the Hamiltonian to be $\H_0+\H_1$.
It does not seem possible to cancel the second term in (\ref{H0Qm}),
though.

\subsection{ Reparametrizations }
\label{ssec:repar}

We denote the repara\-metrization generators by $L(t)$, or equivalently
use the smeared generators
\be
L_f = \int dt\ f(t) L(t).
\label{L_f}
\ee
They generate the algebra $\vect(1)$:
\bes
[L(t), L(t')] &=& (L(t) + L(t')) \dot\dlt(t-t'),
\nlb{vect1}
[L_f, L_g] &=& L_{[f,g]},
\eens
where $[f,g] = f\dot g - g \dot f$. The representations of $\vect(1)$
are densities (primary fields) $\phi(t)$ of weight $\la$, transforming as
\be
[L_f, \phi(t)] &=& -f(t)\dot\phi(t) - \la \dot f(t)\phi(t).
\label{Lff}
\ee
We write $\wt \phi(t) = \la$; in particular, $\wt L(t) = 2$.
If $\wt\phi(t) = 0$, then $\dot \phi(t)$
is also primary and $\wt \dot\phi(t) = 1$. On the other hand, if the
weight of $\phi(t)$ is some $\la\neq 0$, then $\dot\phi(t)$ is not
primary, but there is an additional term proportional to $\phi(t)$ in
the RHS of (\ref{Lff}). The integral $\int dt\ \phi(t)$ can only be
defined if $\wt\phi(t) = 1$; in that case, the integral transforms
trivially.

To give a repara\-metrization invariant formulation of the free particle,
we add a new independent variable, the einbein $e(t)$, with momentum
$p_e(t)$, satisfying $[e(t), p_e(t')] = i\dlt(t-t')$. We introduce the
suggestive notation
\be
\dot\tau(t) = e^{-1}(t), \quad
{df\/d\tau}(t) = e(t)\dot f(t), \quad
\int d\tau\ f(t) = \int dt\ \dot\tau(t) f(t).
\ee
$\tau(t)$ can be thought of as proper time. $\wt \tau(t) = 0$,
and hence $\wt \dot\tau(t) = 1$ and $\wt e(t) = -1$.

To obtain a repara\-metrization invariant formulation of the free
particle, we simply replace $t \to \tau(t)$ and $df/dt \to df/d\tau$
everywhere in the formulas above.
The dynamics constraint (\ref{Epart}) becomes
\be
\EE^\mu(t) &=& {d^2 q^\mu\/d\tau^2}(t) \equiv
e(t) \ddt(e(t)\dot q^\mu(t))
\label{Epart2}
\ee
and the redundancies (\ref{redpart})
\bes
\int d\tau\ \EE^\mu(t) = \int dt\ \dot\tau(t)\, \EE^\mu(t) \equiv 0, \nle
\int d\tau\ \tau\,\EE^\mu(t) = \int dt\ \dot\tau(t)\, \tau(t)\EE^\mu(t)
\equiv 0.
\label{redtau}
\eens
The mode expansion (\ref{qpmodes}) is replaced by
\bes
q^\mu(t) &=& \sumall{m} q^\mu_m \tau^m,
\nlb{taumodes}
p_\mu(t) &=& \sumall{m} p^m_\mu \tau^{-m-1},
\eens
and the Taylor modes still satisfy the brackets (\ref{qpmbracket}).
When formulated in terms of these Taylor modes, all formulas in
subsection \ref{ssec:modes} remain unchanged.

The repara\-metri\-za\-tions $L(t)$ generate a gauge symmetry, which we want
to eliminate in cohomology. To this end,
we introduce a fermionic ghost $c(t)$ with canonical momentum $b(t)$,
satisfying the CAR $\{c(t), b(t')\} = \dlt(t-t')$.
The total list of antifields is thus given by
\be
\barr{ccr|cr|r|cr}
\hbox{Field} & \hbox{Parity} & \la &\hbox{Momentum} & \la &
c & \hbox{Constraint} & \la \\
\hline
q^\mu(t) & B & 0 & p_\mu(t) & 1 & 2d &&\\
e(t) & B & -1 & p_e(t) & 2 & 24 &&\\
q^\mu_*(t) & F & 0 & p^*_\mu(t) & 1 & -2d & \EE^\mu(t) & 0\\
c(t) & F & 2 & b(t) & -1 & -24 & L(t) & 2\\
\theta^\mu_1 & B & - &\chi_\mu^1 & - & 0 & \int d\tau\ q^\mu_* & 0 \\
\theta^\mu_0 & B &- &\chi_\mu^0 & - & 0 & \int d\tau\ \tau\, q^\mu_* & 0 \\
\bt_\mu^0 & F & - & \gm^\mu_0 & - & 0 & p^0_\mu  & 0 \\
\bt_\mu^1 & F & - & \gm^\mu_1 & - & 0 & p^1_\mu & 0 \\
\earr
\label{cpart}
\ee
Note that $\wt \EE^\mu(t) = 0$, and hence $\wt q^\mu_*(t) = 0$. This
difference compared to subsection \ref{ssec:modes} arises because the
Euler-Lagrange equation (\ref{Epart2}) is the second derivative
w.r.t. $\tau$ rather than w.r.t. $t$.

The ghost contribution to the BRST operator reads
\be
Q_c = \int dt\ L(t)b(t) + \half \int dt\ L^\gh(t)b(t)
= L_b + \half L^\gh_b,
\label{Q_c}
\ee
where
\bes
L(t) &=& -i\dot q^\mu(t)p_\mu(t) -i \dot e(t)p_e(t)
+ 2i \ddt(e(t)p_e(t))
- \dot q_*^\mu(t)p^*_\mu(t), \nl
L^\gh(t) &=& \dot b(t) c(t),
\label{Lpart}
\ees
and $L_b$ and $L^\gh_b$ denote the generators smeared
with $b(t)$, in analogy with (\ref{L_f}).
$Q_c$ commutes with $Q_q = Q_d+Q_a+Q_m$ because $[L(t), Q_q] = 0$ and
$[Q_q, e(t)] = 0$, and $Q_c^2 = 0$ by construction. Hence the total
BRST operator $Q' = Q_d + Q_a + Q_m + Q_c$ is also nilpotent, and its
cohomology defines the physical phase space by $C(\PP) = H^0_\cl(Q')$.

We quantize as usual by introducing a vacuum state $\ket0$. As we
discussed in subsection \ref{ssec:qpart}, there is some freedom in doing
so, but the details are not very important. We may define modes as in
(\ref{qpmodes}) or as in (\ref{taumodes}), and demand that all negative
modes annihilate the vacuum. What matters is that we divide the
generators of the history phase space $\HP^* = \HP^*_+ \oplus \HP^*_-$
into two subspaces of equal size, and that the involution maps $\HP^*_+$
and $\HP^*_-$ into each other. We can then define the vacuum $\ket0$
by $\HP^*_-\ket0 = 0$.

There is one complication. When acting on the vacuum, the constant part
of the total repara\-metrization generators
$L^\TOT(t) = L(t) + L^\gh(t)$ gives infinity,
$\int dt\ L^{TOT}(t) \ket0 = \infty\ket0$. To remedy this, we must
normal order, i.e. move all negative modes in (\ref{Lpart}) to the
right and the positive modes to the left. As is well known, this leads
to a central extension, and $\vect(1)$ must be replaced by the
Virasoro algebra with central charge $c$:
\be
[L_f, L_g] = L_{[f,g]} + {c\/24\pi i} \int dt\
\big( \ddot f(t) \dot g(t) - \dot f(t) g(t) \big).
\label{Vir}
\ee
This is a potential problem, because if the central charge of the total
repara\-metrization operator $L^\TOT_f = L_f + L^\gh_f$ is $c_\TOT \neq 0$,
the total BRST operator $Q_\TOT = Q_q + Q_c$ is no longer nilpotent after
quantization. This does not have to be a disaster; it would just mean
that after quantization repara\-metri\-za\-tions become a global symmetry,
which act on the Hilbert space rather than reducing it. This point is
further emphasized in subsection \ref{ssec:needanom}.

However, anomalies are not a problem in this case. The contribution to
the central charge from the various fields and antifields is listed in
the sixth column in the table (\ref{cpart}). Adding the $c$'s together,
we see that the total central charge is
\be
c_\TOT = 2d + 24 - 2d - 24 = 0.
\ee
The total Virasoro algebra is thus anomaly free, and $Q_\TOT^2 = 0$.
There is no obstruction to the elimination of repara\-metri\-za\-tions for
the free particle.

\section{Free scalar field}
\label{sec:scalar}

\subsection{Conventional canonical quantization}

The action
\be
S = \half \int d^4x\ ( \eta^\mn\d_\mu\phi(x)\d_\nu\phi(x)
- \w^2 \phi(x)^2),
\ee
leads to the EL equation
\be
\EE(x) \equiv \eta^\mn\d_\mu\d_\nu\phi(x) + \w^2 \phi(x) = 0.
\label{Exfield}
\ee
From the corresponding expression in Fourier space,
\be
\EE(k) \equiv (k_0^2 - \kk^2 - \w^2) \phi(k) = 0,
\label{Efield}
\ee
we see that the solution is given by Fourier modes
$\phi(k) = \phi(k_0, \kk)$ with	$k_0 = \pm \wk$, where
\be
\wk \equiv \sqrt{\kk^2 + \w^2}.
\ee
The canonical momenta are defined by $\pi(x) = \d_0\phi(x)$, i.e.
\be
\MM(k) = \pi(k) - ik_0\phi(k) \approx 0.
\label{Mfield}
\ee
They are subject to the equal-time CCR
\bes
[\phi(\kk), \pi(\kk')] &=& i\dlt(\kk+\kk'), \nle
{[}\phi(\kk), \phi(\kk')] &=& [\pi(\kk), \pi(\kk')] = 0,
\eens
where $\phi(\kk) = \phi(k_0,\kk)$ for fixed $k_0$, etc.
The time-independent Hamiltonian is
\bes
\H &=& \half \int d^3\kk\ (\pi(\kk)\pi(-\kk)
+ \wk^2 \phi(\kk)\phi(-\kk))
\nlb{Hfield}
&=& \int d^3\kk\ \wk \adag(\kk) a(-\kk),
\eens
where the annihilation and creation operators are
\bes
a(\kk) &=& \swk (\wk \phi(\kk) + i\pi(\kk)),
\nlb{aafield}
\adag(\kk) &=& \swk (\wk \phi(\kk) - i\pi(\kk)).
\eens
They satisfy the non-zero CCR,
\be
[a(\kk), \adag(\kk')] = \dlt(\kk+\kk'),
\ee
and carry energy $-\wk$ and $\wk$, respectively, i.e.
\be
[\H, \adag(\kk)] = \wk\adag(\kk), \qquad
[\H, a(\kk)] = -\wk a(\kk).
\ee

\subsection{Non-covariant cohomological formulation}

As for a single harmonic oscillator, there are several alternative
cohomological descriptions. The description in
subsection \ref{ssec:cohom2} is most suitable for generalization
to field theory, and we proceed in analogy with that.
The extended history phase space $\HP^*$ is spanned by
$\phi(k)$, $\pi(k)$, $\phi^*(k)$ and $\pi^*(k)$, for all
$k = (k_0, \kk) \in \RR^4$, and by $\theta(k)$, $\chi(k)$,
$\bt(k)$, $\gm(k)$, and $\al(k)$, for
$k$ on shell, i.e. $k^2 = \w^2$.
As usual, we take the nonzero brackets to be, for bosonic fields
\be
[\phi(k), \pi(k')] = [\theta(k), \chi(k')] = i\dlt^4(k+k'),
\ee
for fermionic fields
\be
\{\phi^*(k), \pi^*(k')\} = \{\bt(k), \gm(k')\} = \dlt^4(k+k'),
\ee
and for the second-class bosonic field $\al(k)$
\be
[\al(k), \al(k')] = -2k_0\dlt^4(k+k').
\label{alfield}
\ee
$\al(k)$ is needed to cancel the RHS of
\be
[\MM(k), \MM(k')] = 2k_0\dlt^4(k+k').
\ee
Note that e.g.
\be
[\theta(k_0, \kk), \chi(k_0', \kk')] = i\dlt(k_0+k_0') \dlt^3(\kk+\kk'),
\ee
and thus
\be
[\theta(\w_\kk, \kk), \chi(-\w_{\kk'}, \kk')] =
i\dlt(\w_\kk-\w_{\kk'}) \dlt^3(\kk+\kk')
= i\dlt(0) \dlt^3(\kk+\kk')
\ee
As in subsection \ref{ssec:functional}, on-shell constraints will thus
be accompanied by a harmless factor $\dlt(0)$, which could be eliminated
by putting the system in a finite box.

The field theory generalization of BRST charge (\ref{QBRST3}) is
$Q = Q_D + Q_A + \QM$, where
\bes
Q_D &=& \intdk\EE(k) \pi^*(-k) \nl
&=& \intdk\kw \phi(k) \pi^*(-k), \nl
Q_A &=& \intdk\phi^*(k) \chi(-k) \dlt\kw,
\label{Qfield1}\\
\QM &=& \intdk(\MM(k) + \al(k)) \gm(-k) \dlt\kw \nl
&=& \intdk(\pi(k) - ik_0\phi(k) + \al(k)) \gm(-k)\dlt\kw ;
\eens
the dynamics constraint $\EE(k)$ was defined in (\ref{Efield})
and momentum constraint $\MM(k)$ in (\ref{Mfield}).
This BRST charge acts on $C(\HP^*)$ as $\dlt F = [Q,F]$:
\bes
\dlt \phi(k) &=& -i\gm(k)\dlt\kw, \nl
\dlt \pi(k) &=& i\kw\pi^*(k) - k_0\gm(k)\dlt\kw, \nl
\dlt \phi^*(k) &=& \kw\phi(k), \nl
\dlt \pi^*(k) &=& \chi(k)\dlt\kw, \nl
\dlt \bt(k) &=& (\MM(k) + \al(k))\dlt\kw, \\
\dlt \gm(k) &=& 0,  \nl
\dlt \theta(k) &=& -i\fs(k)\dlt\kw, \nl
\dlt \chi(k) &=& 0, \nl
\dlt \al(k) &=& -2k_0\gm(k)\dlt\kw.
\eens
The cohomology is computed exactly as for the harmonic oscillator.
For $k$ off shell, $\phi(k)$ and $\pi^*(k)$ are both closed and exakt,
whereas $\phi^*(k)$ and $\pi(k)$ are neither, and the cohomology
vanishes. For $k$ on shell, say $k = (\w_\kk, \kk)$,
\bes
\dlt \phi(\kk) &=& -i\gm(\kk)\dlt(0), \nl
\dlt \pi(\kk) &=& - \w_\kk\gm(\kk)\dlt(0), \nl
\dlt \phi^*(\kk) &=& 0, \nl
\dlt \pi^*(\kk) &=& \chi(\kk)\dlt(0), \nl
\dlt \bt(\kk) &=& (\MM(\kk) + \al(\kk))\dlt(0), \\
\dlt \gm(\kk) &=& 0,  \nl
\dlt \theta(\kk) &=& -i\fs(\kk)\dlt(0), \nl
\dlt \chi(\kk) &=& 0, \nl
\dlt \al(\kk) &=& 2\w_\kk\gm(\kk)\dlt(0).
\eens
The cohomology $H^\bullet_\cl(Q)$ is generated by
\be
a(\w_\kk,\kk) = \swk(\pi(\kk) + i\wk\phi(\kk)) + x_\kk\MM(\kk),
\ee
where $x_\kk$ is an arbitrary constant. Analogously, the solution for
the opposite on-shell value $k = (-\wk, \kk)$ is
\be
a(-\w_\kk,\kk) = \swk(\pi(\kk) - i\wk\phi(\kk)) + y_\kk\MM(\kk),
\ee
where $y_\kk$ is another constant.

The Hamiltonian in $\HP^*$,
\bes
\H &=& \intdk\bigg( ik_0 \phi(k) \pi(-k)
+ k_0 \phi^*(k) \pi^*(-k)
+ \big( ik_0 \theta(k) \chi(-k) \nl
&&+\ k_0 \bt(k) \gm(-k)
+ \half \al(k) \al(-k) \big)\dlt\kw \bigg),
\ees
reduces to
\be
\H = \int d^3\kk\ \wk a(\w_\kk,\kk) a(-\w_\kk,-\kk) + \{Q,O\},
\ee
$O$ some operator, which is identical to (\ref{Hfield}) in cohomology.
We note that $Q$ and $\H$ are self-adjoint under involution, which
sends $k = (k_0, \kk)$ to $-k = (-k_0, -\kk)$.

To quantize the theory, we introduce a vacuum $\ket0$ which is
anniliated by all oscillators in $\HP^*$ with negative $k_0$. The
dual vacuum $\bra0$ is annihilated by all modes with $k_0>0$.

\subsection{Covariance and the observer's velocity}

It is clear that the cohomological formulation in the previous
subsection violates manifest covariance, and thus it violates the
fundamental guiding principle behind MCCQ. The source of
non-covariance is as usual the 3+1 decomposition $k = (k_0, \kk)$.
In particular, the momentum constraint (\ref{Mfield}) and the related
definition (\ref{alfield}) depend explicitly on $k_0$ and are thus
non-covariant.

We can restore apparent covariance by introducing a vector
$u=(1,0,0,0)$, which is naturally identified as {\em the observer's
4-velocity}. The time component of the momentum can then be written as
$k_0 = k_\mu u^\mu$, i.e. as the projection of $k$ onto $u$. However,
this is of course a purely cosmetic improvement. We have only made
explicit that the definition of energy depends on a hidden background
structure, namely the observer's velocity. The key lesson from general
relativity is that physics should not depend on background structures.
Hence we need to eliminate the observer's velocity from the background
by bringing it to the foreground, i.e. by giving it quantum dynamics.

The simplest way to do so is to assume that that the observer moves
along a straight line, i.e. the the observer's trajectory has the
form
\be
q^\mu(t) = u^\mu t + s^\mu.
\ee
From the natural definition of the observer's momentum,
$p^\mu = \Mobs u^\mu$, we find that $u^\mu$ and $s^\mu$ are canonically
conjugate, \viz
\be
[s^\mu, u^\nu] = {i\/\Mobs} \eta^\mn.
\label{obsccr}
\ee
On the classical level, this modification is enough to ensure manifest
covariance. We simply replace $k_0 \to k_\mu u^\mu$ everywhere in
the previous subsection.  Since $\dlt u^\mu = 0$, the observer's
velocity belongs to $H^\bullet_\cl(Q)$ and is observable. However,
$\dlt s^\mu\neq 0$, and hence we can not observe the observer's position.
This conclusion is of course absurd.

The problem arises from the momentum part of the BRST charge
(\ref{Qfield1}),
\be
\QM = \intdk(\pi(k)-ik_\mu u^\mu\phi(k)+\al(k)) \gm(-k)\dlt\kw.
\ee
The second term acts as follows on $s^\mu$:
\be
\dlt s^\mu = {-1\/\Mobs} \intdk k^\mu \phi(k)\gm(-k)\dlt\kw.
\ee
The third term above is more subtle but also acts nontrivially on
$s^\mu$, since the brackets for
$\al(k)$ involve $k_0$ and thus $u^\mu$, cf. (\ref{alfield}).
The problem becomes even more acute on the quantum level. Here we
need to represent the Heisenberg algebra on a Hilbert space with
energy bounded from below. The positive-energy condition $k_0 > 0$
is thus replaced by
\be
k_\mu u^\mu > 0,
\label{kugt0}
\ee
which makes no sense because the observer's velocity is now a
quantum operator.

The physical origin of these problems is the assumption that the
observers mass $\Mobs$ is finite. If we instead assume the
$\Mobs = \infty$, we see from (\ref{obsccr}) that $u^\mu$ and $s^\mu$
commute, and hence $\dltM s^\mu = 0$. Since $u^\mu$ and $s^\mu$
commute with all other operators as well, they are central and can
be represented by c-numbers. The condition (\ref{kugt0}) then makes
sense, and we are back to the situation in the previous subsection.

This discussion leads to an important conclusion: underlying the
usual formulation of canonical quantization is a hidden assumption
about a macroscopic, and thus infinitely massive, observer. This is
of course an excellent approximation to most experimental situations,
where the measurement apparatus is much heavier than the phenomenon
being observed. However, this approximation is fundamentally flawed,
since all physical objects including the observer are fundamentally
quantum. We argue in subsection \ref{ssec:obsQG} that this problem
becomes serious when gravity is taken into account.

\subsection{The need for QJT}

We saw in the previous subsection that the explicit introduction of
the observer's position and velocity is necessary but not sufficient
for a manifestly covariant canonical formulation of quantum physics.
The key problem resides in the condition (\ref{kugt0}). This is an
operator equation for the indeterminate $k$. However, only operators
like $\phi(k)$ can depend on operator equations, not c-numbers like
$k$ itself.

To circumvent this problem, and finally obtain a formulation of the
free field which is manifestly covariant, we need to introduce Quantum
Jet Theory (QJT). The idea behind QJT is to formulate physics in terms
of jet data rather than field data. A $p$-jet $J^p\phi$ is normally
defined as an equivalence class of functions; two functions are
equivalent if their derivatives up to order $p$, evaluated at the
point $q$, agree. Locally, a jet can be canonically identified with a
truncated Taylor series around the point $q$, and this is sufficient for
our purposes. We thus expand every field
$\phi(x)$ in a multi-dimensional Taylor series around an
operator-valued curve $q^\mu(t)$ in spacetime - {\em the observer's
trajectory}:
\be
\phi(x) = \sum_M {1\/M!} \phi_M(t) (x-q(t))^M,
\label{FieldTaylor}
\ee
where $M$ is a multi-index; in one dimension, $M$ is just an integer
and the sum runs over positive $M$.

It is now rather straightforward to translate the formalism above to
jet space. E.g., the free-field equation (\ref{Exfield}) translates
into the hierarchy
\be
\EE_M(t) = \sum_{\mu,\nu=0}^3 \eta^\mn \phi_{M+\mu+\nu}(t) + \w^2\phi_M(t)
\approx 0.
\ee
The crucial advantage of the passage to jet space is that it allows
us to replace the vacuum definition (\ref{kugt0}) by a c-number
equation; the quantization step does not involve an illegal operator
equation. Introduce the frequency $m$ by making a Fourier
transformation w.r.t. $t$,
\be
\phi_M(t) = \int dm\ \exp(imt) \phi_M(m).
\ee
We now posit that the vacuum $\ket0$ be annihilated by all
oscillators of negative frequency, e.g.
\be
\phi_M(-m)\ket0 = \phi^*_M(-m)\ket0 = q^\mu(-m)\ket0 = ... = 0,
\ee
for all $-m < 0$. The important property is that this frequncy is a
c-number, and we do not need any operators to distinguish between
positive and negative frequency. The price is that we must
explicitly introduce the observer's trajectory into the formalism.
This amounts to a substantial modification of physics, e.g. because
new diff and gauge anomalies arise.

In the next section, we implement this idea for the harmonic
oscillator.

\section{Harmonic oscillator in Quantum Jet Theory}

\subsection{Definition of constraints}

We introduce an operator-valued curve $q(t)$, called {\em the observer's
trajectory}, expand the field $\phi(x)$ in a Taylor series around it,
and truncate at some fixed but arbitrary finite order $p$:
\be
\phi(x) = \sum_{M=0}^p {1\/M!} \phi_M(t) (x-q(t))^M.
\label{HarmTaylor}
\ee
We identify $\phi_M(t) = d^M\phi/dx^M |_{x=q(t)}$. The relevant
history phase space is spanned by the observer's
trajectory $q(t)$ and its momentum $p(t)$, the Taylor
coefficient functions $\phi_M(t)$ for $0 \leq M \leq p$, as well as the
corresponding momenta $\pi^M(t)$. The nonzero Poisson brackets are
\be
[q(t), p(t')] = i\dlt(t-t'), \qquad
[\phi_M(t), \pi^N(t')] = i\dlt^N_M \dlt(t-t').
\ee
This space thus consists of histories of $p$-jets and their momenta.
It will be referred to as {\em the $p$-jet history phase space} and
denoted by $J^p\HP$. Apart from the truncation order $p$, the definition
of $p$-jets also depend on the expansion point $q(t)$ as well, although
the notation does not make this explicit.

As a matter of nomenclature, QJT refers to Quantum Jet Theory in general,
whereas QJT($p$) specifically refers to the quantum theory of $p$-jets.

Since we are dealing with the harmonic oscillator, we assume that there
is a one-dimensional metric $\dlt_\MN$ (simply the Kronecker delta),
which can be used to raise and lower indices. We will use this freedom
to put all Taylor indices $M,N,...$ downstairs. E.g., the momentum
$\pi_M(t) = \dlt_\MN \pi^N(t)$ satisfies
\be
[\phi_M(t), \pi_N(t')] = i\dlt_\MN \dlt(t-t').
\ee

The dynamics constraint,
\be
\EE(x) = \phi''(x) + \w^2\phi(x),
\ee
turns into the following hierarchy:
\be
\EE_M(t) \equiv \phi_{M+2}(t) + \w^2\phi_M(t).
\label{EMt}
\ee
Since $\phi_M(t)$ is only defined for $M\leq p$, $\EE_M(t)$ is
defined for $M\leq p-2$.

We must also impose a new constraint, which expresses that
(\ref{HarmTaylor}) is independent of the trajectory parameter $t$. The
RHS really defines a field $\phi(x,t)$; to ensure that this is
independent of $t$, we impose the constraint $\d_t \phi(x,t) = 0$,
which leads to {\em the time constraint}
\be
\TT_M(t) = \dot\phi_M(t) - \dot q(t)\phi_{M+1}(t).
\label{TMt}
\ee
$\TT_M(t)$ is defined for $M \leq p-1$.
If we make the assumption that $\dot q(t) = 1$, the time constaint is
simply replaced by
\be
\TT_M(t) = \dot\phi_M(t) - \phi_{M+1}(t).
\label{TTm}
\ee
This assumption amounts to a special choice of parametrization, namely
$q(t)=t+s$, and the velocity $\dot q(t)$ is a c-number.
In subsection \ref{ssec:repar} we discuss the
modifications necessary when the original repara\-metri\-zation invariance
is not gauge fixed.

The dynamics and time constraints are not independent, since
\be
\phi_{M+2}(t) \approx -\w^2\phi_M(t) \approx \dot\phi_{M+1}(t)
\approx \ddot\phi_M(t).
\label{red}
\ee
In the cohomological formulation below, this redundancy gives
rise to extra terms in the BRST operator.

Finally we turn to the momentum constraint. We would like to impose
\be
\MM(x) = \pi(x) - \phi'(x),
\ee
but expressing a momentum $\pi(x)$ in terms of the jet momenta $\pi_M(t)$
is unnatural. Instead, we define the following momentum
constraint, for $M \leq p$:
\be
\MM_M(t) \equiv \pi_M(t) - \dot\phi_M(t) = 0.
\label{MMt}
\ee
The momentum constraint is second class in view of
\be
[\MM_M(t), \MM_N(t')] = -2 \dlt_\MN\dot\dlt(t-t').
\ee
Moreover, it does not commute with the dynamics and time constraints;
rather
\bes
[\MM_M(t), \EE_N(t')] &=& -i(\dlt_{M,N+2} + \w^2\dlt_\MN)\dlt(t-t'),
\nle
{[}\MM_M(t), \TT_N(t')] &=&
i(\dlt_\MN\dot\dlt(t-t') + \dlt_{M,N+1}\dlt(t-t')).
\eens

It is convenient to perform a Fourier transform w.r.t. the trajectory
parameter $t$. To avoid complications analogous to those in subsection
\ref{ssec:functional}, we should assume that the Fourier variable $m$
is discrete, but $\pm\w$ belongs to the domain of $m$; this amounts to
putting the system in a large box. However, we ignore this technicality
here, and define
\be
\phi_M(t) = \intm \e^{imt} \phi_M(m).
\label{Fourierm}
\ee
The Fourier transformed fields satisfy the nonzero brackets
\be
[\phi_M(m), \pi_N(n)] = i\dlt_\MN\dlt(m+n).
\ee
The constraints now become
\bes
\EE_M(m) &=& \phi_{M+2}(m) + \w^2\phi_M(m), \nl
\TT_M(m) &=& im\phi_M(m) - \phi_{M+1}(m),
\label{constrjet} \\
\MM_M(m) &=& \pi_M(m) - im\phi_M(m).
\eens
Since the constraints for different values of $m$ decouple, the
Fourier description is much simpler, and we use it for the rest of
this section.
The following linear combination of momentum constraints,
\be
\MM_k(m) = \sum_{M=0}^p (-ik)^M \MM_M(m),
\label{MMk}
\ee
satisfy
\be
[\MM_k(m), \MM_{k'}(n)] = 2m \dlt_p(k,k')\dlt(m+n),
\ee
where
\be
\dlt_p(k,k') \equiv {1-(-kk')^{p+1} \/ 1+kk'} = \dlt_p(k',k).
\ee
Most of the  constraints (\ref{constrjet}) and (\ref{MMk}) are second
class since
$\MM_k(m)$ has nonzero brackets with $\EE_N(n)$ and $\TT_N(n)$;
the exceptions are
\be
\MM_\w \equiv \MM_\w(\w), \qquad
\MM_{-\w} \equiv \MM_{-\w}(-\w).
\label{MMw}
\ee
It is readily verified that
\bes
[\MM_k(m), \EE_N(n)] &=& -i(-ik)^N (-k^2 + \w^2) \dlt_{m+n}, \\
{[}\MM_k(m), \TT_N(n)] &=& (-ik)^N (-m+k) \dlt_{m+n},
\eens
which vanish when $m = k$ and $k = \pm\w$. In particular,
$[\MM_{\pm\w}, \EE_N(n)] = [\MM_{\pm\w}, \TT_N(n)] = 0$.

\subsection{Solution of constraints}

{F}rom the time constraint, we find $\phi_{M+1}(m) \approx im\phi_M(m)$,
i.e.
\be
\phi_M(m) \approx (im)^M\phi_0(m).
\ee
On the other hand, the dynamics constraint leads to
\be
\phi_M(m) \approx
\begin{cases}
(-\w^2)^{M/2}\phi_0(m), \qquad &\hbox{if $m$ even}, \\
(-\w^2)^{(M-1)/2}\phi_1(m), \qquad &\hbox{if $m$ odd}.
\end{cases}
\ee
Consistency then requires that
\be
(im)^{M+2}\phi_0(m) + \w^2 (im)^M \phi_0(m) \approx 0,
\ee
which holds if $m = \pm \w$. The momentum constraint then gives the
expression for the canonical momentum,
$\pi_M(m) \approx im\phi_M(m) \approx (im)^{M+1}\phi_0(m)$.
There are many equivalent expressions in $J^p\HP$ which correspond to
the same elements in $\PP$ after the constraints have been taken into
account. The independent solutions can be taken to be
\bes
a_\w &\approx& (i\w)^{-M} \phi_M(\w)
\approx (i\w)^{-M-1}\pi_M(\w) \nl
&\approx& \half (i\w)^{-M-1}(\pi_M(\w)+i\w\phi_M(\w)), 
\nlb{aaDirac}
a_{-\w} &\approx& (-i\w)^{-M} \phi_M(-\w)
\approx (-i\w)^{-M-1}\pi_M(-\w) \nl
&\approx& \half (-i\w)^{-M-1}(\pi_M(-\w)-i\w\phi_M(-\w)).
\eens
Thus
\bes
\phi_M(\w) \approx (i\w)^M a_\w, &&
\phi_M(-\w) \approx (-i\w)^M a_{-\w}, \nl
\pi_M(\w) \approx (i\w)^{M+1}a_\w, &&
\pi_M(-\w) \approx (-i\w)^{M+1}a_{-\w}, \\
\phi_M(k) \approx 0, &&
\pi_M(k) \approx 0, \qquad \hbox{if $k \neq \pm\w$.}
\eens
The nonzero Dirac bracket between the independent solutions is as
expected:
\be
[a_\w, a_{-\w}] &=& -1,
\ee
which shows that they behave as standard harmonic oscillators. The
evaluation of the brackets is simplest if we choose the last 
representatives for $a_\w$ and $a_{-\w}$ in (\ref{aaDirac}), because
the these commute with all constraints and hence there is no difference
between Dirac and Poisson brackets.
The natural Hamiltonian in $J^p\HP$ reads
\be
\H = \sum_{M=0}^p \intm im\phi_M(m) \pi_M(-m).
\ee
It performs a rigid time translation of all histories, which in Fourier
space amounts to a multiplication:
\be
[\H, \phi_M(m)] = m\phi_M(m) \qquad
[\H, \pi_M(m)] = m\pi_M(m).
\ee
In particular, it acts as follows on the constraint surface,
\be
[\H, a_\w] = \w a_\w, \qquad
[\H, a_{-\w}] = -\w a_{-\w},
\ee
and thus acts as the correct harmonic oscillator Hamiltonian.

\subsection{ Cohomological formulation }
\label{ssec:cohomjet}

To employ the BRST formalism, we need to make all constraints
(\ref{constrjet}) commute. To this end, we first discard
all momentum constraints except $\MM_\w$ and $\MM_{-\w}$,
as defined in (\ref{MMk}) and (\ref{MMw}). These constraints commute with
$\EE_M(m)$ and $\TT_M(m)$, but not with each other; instead
\be
[\MM_\w, \MM_{-\w}] = \si_p(\w),
\label{MMwwjet}
\ee
where
\be
\si_p(\w) = -\si_p(-\w) = \si_p(\w,-\w) = 2\dlt_p(\w,-\w)
={1-\w^{2p+2} \/ 1-\w^2}.
\ee
To rectify this, we introduce new variables $\al_\w$ and $\al_{-\w}$,
subject to the nonzero CCR
\be
[\al_\w, \al_{-\w}] = -\si_p(\w).
\ee
The modified momentum constraints
\be
\MM_\w' = \MM_\w + \al_\w, \qquad
\MM_{-\w}' = \MM_{-\w} + \al_{-\w},
\ee
do commute with each other, and our full set of constraints is thus
first class.

To eliminate the dynamics and time constraints, we introduce fermionic
antijets $\fs_M(m)$ and $\bfi_M(m)$, respectively, with canonical
momenta $\ps_M(m)$ and $\bpi_M(m)$, and brackets
\be
\{\fs_M(m), \ps_N(n)\} = \dlt_\MN\dlt_{m+n}, \quad
\{\bfi_M(m), \bpi_N(n)\} = \dlt_\MN\dlt_{m+n}.
\label{fsbfs}
\ee
We note that $\phi_M(m)$ only belongs to $J^p\HP$
if $M \leq p$. Hence $\EE_M(m)$ is defined for $M \leq p-2$ and
$\TT_M(m)$ for $M \leq p-1$, and so are the corresponding antijets.

These constraints are not independent in view of (\ref{red}). This
leads to the secondary constraint,
\be
\R_M(m) = \bfi_{M+2}(m) + \w^2\bfi_M(m) - im\fs_M(m) + \fs_{M+1}(m).
\label{RMm}
\ee
To deal with this, we introduce the bosonic, second-order antijet
$\bfs_M(m)$, with momentum $\bps_M(m)$.
Like $\R_M(m)$, it is defined for $M \leq p-3$.
Finally, it turns out that yet another expression will be BRST closed,
namely $\B_M(\w)$ and $\B_M(-\w)$, where
\be
\B_M(m) = \fs_M(m) + \bfi_{M+1}(m) + im\bfi_M(m),
\ee
which is defined for $M \leq p-2$.
To eliminate the corresponding unwanted cohomology, we introduce yet
another bosonic antifields $\theta(\w)$ and $\theta(-\w)$ with
momenta $\chi(-\w)$ and $\chi(\w)$.

To summarize, we have introduced antijets and
canonical momenta according to the following table:
\be
\barr{c|c|c|c|l}
\hbox{Field} & \hbox{Momentum} & \hbox{Parity} & \hbox{Order} &
\hbox{Constraint} \\
\hline
\phi_M(m) & \pi_M(-m) & B & p & -\\
\fs_M(m) & \ps_M(-m) & F & p-2 & \EE_M(m) \\
\bfi_M(m) & \bpi_M(-m) & F & p-1 & \TT_M(m) \\
\bfs_M(m) & \bps_M(-m) & B & p-3 & \R_M(m) \\
\theta_M(\w) & \chi_M(-\w) & B & p-2 & \B_M(\w) \\
\theta_M(-\w) & \chi_M(\w) & B & p-2 & \B_M(-\w) \\
\bt_\w & \gm_{-\w} & F & - & \MM_\w + \al_\w \\
\bt_{-\w} & \gm_\w & F & - & \MM_{-\w} + \al_{-\w} \\
\al_\w & \al_{-\w} & B & - & - \\
\earr
\ee
The third column indicates the Grassmann parity (bose/fermi), and the
fourth the order where the constraints have to be truncated. Note that
there are only two momentum constraints irrespective of $p$.

The BRST charge is $Q = Q_D + Q_T + Q_R + Q_B + \QM$, where
\bes
Q_D &=& \sum_{M=0}^{p-2} \intm \EE_M(m)\ps_M(-m), \nl
Q_T &=&  \sum_{M=0}^{p-1} \intm \TT_M(m)\bpi_M(-m), \nl
Q_R &=& \sum_{M=0}^{p-3} \intm i\R_M(m)\bps_M(-m),
\label{Qhjet}\\
Q_B &=& \sum_{M=0}^{p-2}
\big( \B_M(\w)\chi_M(-\w) + \B_M(-\w)\chi_M(\w) ), \nl
\QM &=& (\MM_\w + \al_\w)\gm_{-\w} + (\MM_{-\w}+ \al_{-\w})\gm_\w,
\eens
$Q$ acts one the jets as $\dlt F = [Q, F]$.
For $m \neq \pm\w$, we have
\bes
\dlt \phi_M(m) &=& 0, \nl
\dlt \fs_M(m) &=& \EE_M(m) = \phi_{M+2}(m) + \w^2\phi_M(m), \nl
\dlt \bfi_M(m) &=& \TT_M(m) = im\phi_M(m) - \phi_{M+1}(m), \nl
\dlt \bfs_M(m) &=& \R_M(m) 
\label{dhjet}\\
&=& \bfi_{M+2}(m) + \w^2\bfi_M(m) - im\fs_M(m) + \fs_{M+1}(m), \nl
\dlt \pi_M(m) &=& i\big( \ps_{M-2}(m) + \w^2\ps_M(m) - im\bpi_M(m)
- \bpi_{M-1}(m) \big), \nl
\dlt \ps_M(m) &=& im\bps_M(m) + \bps_{M-1}(m), \nl
\dlt \bpi_M(m) &=& \bps_{M-2}(m) + \w^2\bps_M(m), \nl
\dlt \bps_M(m) &=& 0.
\eens
The action on the momenta ($\pi_M(m)$, $\pi^*_M(m)$, etc.) is only
written down for generic values of $M$. For $M$ close to $0$, some
terms in the RHS are not defined; if so, these terms are discarded.
For $m^2 = \w^2$, say $m = \w$, we have instead
\bes
\dlt \phi_M(\w) &=& -i(i\w)^M\gm_\w, \nl
\dlt \fs_M(\w) &=& \phi_{M+2}(\w) + \w^2\phi_M(\w), \nl
\dlt \bfi_M(\w) &=&im\phi_M(\w) - \phi_{M+1}(\w), \nl
\dlt \bfs_M(\w) &=&
\bfi_{M+2}(\w) + \w^2\bfi_M(\w) - im\fs_M(\w) + \fs_{M+1}(\w), \nl
\dlt \theta_M(\w) &=& \fs_M(\w) + \bfi_{M+1}(\w) + i\w\bfi_M(\w), \nl
\dlt \bt_\w &=& \MM_\w + \al_\w, \\
\dlt \al_\w &=&	\si_p(\w)\gm_\w, \nl
\dlt \pi_M(\w) &=& i\big( \ps_{M-2}(\w) + \w^2\ps_M(\w) - im\bpi_M(\w)
- \bpi_{M-1}(\w) - (i\w)^{M+1}\gm_\w \big), \nl
\dlt \ps_M(\w) &=& m\bps_M(\w) + i\bps_{M-1}(\w) + \chi_M(\w), \nl
\dlt \bpi_M(\w) &=& i\bps_{M-2}(\w) + i\w^2\bps_M(\w)
+ \chi_{M-1}(\w) + im\chi_M(\w), \nl
\dlt \bps_M(\w) &=& 0, \nl
\dlt \gm_\w &=& 0.
\eens
To compute the cohomology, we first focus on the fields and ignore the
momenta.
Start with the time constraint. The kernel is generated by the functions
$\phi_M(m)$, $\dlt_T\phi_M(m) = 0$, and the image by the functions
$\bfi_M(m)$, $\dlt_T \bfi_M(m) = \TT_M(m)$. Since there are $p+1$
functions $\phi_M(m)$ but only $p$ functions $\bfi_M(m)$, one $\phi_M(m)$
must remain in $H^\bullet_\cl(Q_T)$. Solving $\TT_M(m)$, we see that it
has many representatives in $J^p\HP$, namely
\be
\phi_0(m) \approx (im)^{-M} \phi_M(m).
\label{cohomT}
\ee
We next consider the dynamics constraint. The kernel is generated by the
functions $\phi_M(m)$, $\dlt_D\phi_M(m) = 0$, and the image by the
functions $\fs_M(m)$, $\dlt_D \fs_M(m) = \EE_M(m)$. Since there are $p+1$
functions $\phi_M(m)$ but only $p-1$ functions $\fs_M(m)$, two $\phi_M(m)$
must remain in $H^\bullet_\cl(Q_D)$. Solving $\EE_M(m)$, their
representatives in $J^p\HP$ are $\phi_0(m)$ and $\phi_1(m)$, and
\be
\phi_M(m) \approx
\begin{cases}
(-\w^2)^{M/2} \phi_0(m), \qquad &\hbox{if $M$ even}, \\
(-\w^2)^{(M-1)/2} \phi_1(m), \qquad &\hbox{if $M$ odd}.
\end{cases}
\label{cohomD}
\ee
Now consider the combined cohomology $H^\bullet_\cl(Q_T+Q_D)$. The
expressions (\ref{cohomT}) and (\ref{cohomD}) are clearly only
compatible if $m^2 = \w^2$, i.e. $m = \pm \w$, and the cohomology is
generated by $\phi_w$ and $\phi_{-\w}$, which have the representatives
in $J^p\HP$:
\bes
\phi_\w = \phi_0(\w) &\approx& (i\w)^{-M} \phi_M(\w),
\nlb{phiw}
\phi_{-\w} = \phi_0(-\w) &\approx& (-i\w)^{-M} \phi_M(-\w).
\eens
All antijets vanish in cohomology. The only combinations which belong
to the kernel are $\R_M(m)$ and $\B_M(\pm\w)$, but they are also exact.

Now we turn to the corresponding momenta. The time constraint gives
\bes
\dlt_T \pi_p(m) &=& - i \bpi_{p-1}(m), \nl
\dlt_T \pi_M(m) &=& m\bpi_M(m) - i \bpi_{M-1}(m),
\qquad \hbox{if $1\leq M\leq p-1$.} \nl
\dlt_T \pi_0(m) &=& m\bpi_0(m).
\ees
There are $p+1$ functions $\pi_M(m)$ but only $p$ functions 
$\dlt_T\bpi_M(m)$.
All of the latter are exact, but only $p$ of the former can be non-closed,
by linear independence. The linear combination which is closed is
\be
\dlt_T( \sum_{M=0}^p (-im)^M \pi_M(m) ) = 0.
\label{piT}
\ee
Analogously, the dynamics constraint gives
\bes
\dlt_D \pi_p(m) &=& -i\ps_{p-2}(m), \nl
\dlt_D \pi_{p-1}(m) &=& -i\ps_{p-3}(m), \nl
\dlt_D \pi_M(m) &=& -i(\ps_{M-2}(m) + \w^2\ps_M(m)),
\qquad \hbox{if $2\leq M\leq p-2$,} \nl
\dlt_D \pi_1(m) &=& -i\w^2\ps_1(m) \\
\dlt_D \pi_0(m) &=& -i\w^2\ps_0(m).
\eens
There are $p+1$ functions $\pi_M(m)$ but only $p-1$ functions 
$\dlt_D\ps_M(m)$.
All of the latter are exact, but only $p-1$ of the former can be non-closed,
by linear independence. The linear combinations which are closed are
\be
\dlt_D( \sum_{M=0}^p (i\w)^M \pi_M(m) ) =
\dlt_D( \sum_{M=0}^p (-i\w)^M \pi_M(m) ) = 0.
\label{piD}
\ee
Combining (\ref{piT}) and (\ref{piD}), we see that the only combinations
which belong to the kernal of $Q_T + Q_D$ are
\bes
\pi_\w &=& \sum_{M=0}^p (-i\w)^M \pi_M(\w),
\nle
\pi_{-\w} &=& \sum_{M=0}^p (i\w)^M \pi_M(-\w).
\eens
These linear combinations satisfy the following brackets with the elements
in (\ref{phiw}):
\bes
[\phi_\w, \pi_{-\w}] = [\phi_{-\w}, \pi_\w] &=& i,
\nlb{ppbracket}
{[}\phi_\w, \phi_{-\w}] = [\phi_\w, \phi_{-\w}] &=& 0.
\eens
It is straightforward to verify that these brackets are independent of
which representatives for $\phi_\w$ and $\phi_{-\w}$ we choose from
(\ref{phiw}), and hence they induce well-defined relations in cohomology.

The antijet momenta all vanish in cohomology, and hence
$H^\bullet_\cl(Q_D + Q_T + Q_R + Q_B) =
C(\phi_\w, \phi_{-\w},\pi_\w, \pi_{-\w})$ with the Poisson bracket
(\ref{ppbracket}). Remains the momentum constraints, defined in
(\ref{MMk}) and (\ref{MMw}). It is clear that they can be written as
\bes
\MM_\w &=& \sum_{M=0}^p (-i\w)^M (\pi_M(t) -i\w\phi_M(t)) \nl
&\approx& \pi_\w - i\w\sum_{M=0}^p (-i\w)^M (i\w)^M \phi_\w
\label{Mwdef}\\
&=& \pi_\w - i\w\dlt_p(\w,-\w) \phi_\w,
\eens
and similar for $\MM_{-\w}$.
The relations (\ref{MMwwjet}) follow immediately from (\ref{Mwdef})
The remaining part of the BRST operator reads
\be
\QM = (\MM_\w + \al_\w)\gm_{-\w} + (\MM_{-\w} + \al_{-\w})\gm_\w.
\ee
It acts as
\bes
\dltM \phi_\w = -i\gm_\w, &&
\dltM \pi_\w = \w\dlt_p(\w,-\w)\gm_\w, \nl
\dltM \bt_\w = \MM_\w + \al_\w, &&
\dltM \gm_\w = 0, \\
\dltM \al_\w = \si_p(\w)\gm_\w = \w\dlt_p(\w,-\w)\gm_\w,
\eens
and analogously on the modes with index $-\w$. The kernel is generated by
\bes
a_\w &=& {1\/\sqrt{2\w\dlt_p(\w)}}(\pi_\w + i\w\dlt_p(\w,-\w)\phi_\w)
+ x (\MM_\w + \al_\w),
\nlb{awjet}
a_{-\w} &=& {1\/\sqrt{2\w\dlt_p(\w)}}(\pi_{-\w} - i\w\dlt_p(\w,-\w)\phi_{-\w})
+ y (\MM_{-\w} + \al_{-\w}),
\eens
where $x$ and $y$ are arbitrary constants, multiplying terms belonging
to $\im \QM$. The operators $a_\w$ and
$a_{-\w}$ satisfy the brackets $[a_\w, a_{-\w}] = -1$ and are thus
well-defined in cohomology. Hence the BRST operator (\ref{Qhjet})
selects the correct phase space $\PP = C(a_\w, a_{-w}) = H^0_\cl(Q)$.

\subsection{ Antijet constraints}

One may wonder why there is no antifield constraint in the jet
formulation. The reason is that we have chosen to truncate the jet
$\phi_M(m)$ at order $p$. This forces us to truncate the time constraint
$\TT_M(m)$ at order $p-1$ and the dynamics constraint $\EE_M(m)$ at
order $p-2$, and hence the corresponding antijets are truncated at the
same orders. That there are fewer constraints than fields is clearly
necessary in order to have solutions, and non-vanishing cohomology. However,
if we want we can add extra antijets, so that $\fs_M(m)$ and $\bfi_M(m)$
are defined for all $m$, $0\leq m\leq p$. Since the dynamics and
time constraints are not a priori defined when they involve a jet
$\phi_M(m)$ with $M>p$, we must choose new constraints which are
compatible with the solutions. The introduction of new antijets then
leads to linear relations between the constraints, which require further
antijets to be cancelled. At the end, the cohomology stays the same.

Let us see in detail how this happens.
The constraints $\EE_M(m)$ for $M\leq p-2$ and $\TT_M(m)$ for
$M \leq p-1$ were defined in (\ref{constrjet}). Now define
\bes
\TT_p(m) &=& im (\phi_p(m) - (im)^p \phi_0(m)), \nl
\EE_p(m) &=& \w^2 (\phi_p(m) - (i\w)^p\phi_0(m)), \\
\EE_{p-1}(m) &=& \w^2 (\phi_{p-1}(m) - (i\w)^{p-2}\phi_1(m)).
\eens
For simplicity we assume that $p$ is even, so that the relations
$\phi_p(m) \approx (i\w)^p\phi_0(m)$ and
$\phi_{p-1}(m) \approx (i\w)^{p-2}\phi_1(m)$
should be implemented in cohomology. We introduce new antijets
$\bfi_p(m)$, $\fs_p(m)$ and $\fs_{p-1}(m)$, with momenta
$\bpi_p(m)$, $\ps_p(m)$ and $\ps_{p-1}(m)$, to eliminate the new
constraints.

The constraints are no longer independent; rather, they are subject to
the relations
\bes
\sum_{M=0}^p (im)^{-M} \TT_M(m) &=& 0, \nle
\sum_{M=0}^p (i\w)^{-M} \EE_M(m) &=&
\sum_{M=0}^p (-i\w)^{-M} \EE_M(m) = 0.
\eens
Hence there are new expressions which belong to the kernel, namely
\bes
\dlt(\sum_{M=0}^p (im)^{-M} \bfi_M(m)) &=& 0, \nle
\dlt(\sum_{M=0}^p (i\w)^{-M} \fs_M(m)) &=&
\dlt(\sum_{M=0}^p (-i\w)^{-M} \fs_M(m)) = 0.
\eens
To eliminate these expressions in cohomology, we introduce new antijets
$\bar\theta(m)$, $\theta^*_+(m)$ and $\theta^*_-(m)$, with momenta
$\bar\chi(m)$, $\chi^*_+(m)$ and $\chi^*_-(m)$, respectively, and add
the following term to the BRST operator:
\bes
\Delta Q &=&
\intm \bigg( \bar\chi(-m)\,\sum_{M=0}^p (im)^{-M} \bfi_M(m) \\
&&+\ \chi^*_+(-m)\,\sum_{M=0}^p (i\w)^{-M} \fs_M(m) +
\chi^*_-(-m)\,\sum_{M=0}^p (-i\w)^{-M} \fs_M(m) \bigg).
\eens
The new antijets cancel in quadruplets, and we return to the situation
in the previous subsection.

\subsection{Hamiltonian, involution, quantization}

The natural Hamiltonian in $J^p\HP^*$ reads
\bes
\H &=& \intm \bigg(\sum_{M=0}^p im\phi_M(m) \pi_M(-m)
+\sum_{M=0}^{p-2} m\phi^*_M(m) \pi^*_M(-m) \nl
&&+\ \sum_{M=0}^{p-1} m\bfi_M(m) \bpi_M(-m)
+\sum_{M=0}^{p-3} im\bfs_M(m) \bps_M(-m) \bigg).
\label{HJp}
\ees
It picks out the Fourier variable,
\bes
[\H, \phi_M(m)] = m\phi_M(m), &&
[\H, \pi_M(m)] = m\pi_M(m), \nle
{[}\H, \phi^*_M(m)] = m\phi^*_M(m), &&
[\H, \pi^*_M(m)] = m\pi^*_M(m),
\eens
etc. Since the cohomology generators are linear combinations of
terms with $m = \pm\w$, the Hamiltonian (\ref{HJp}) commutes with the
BRST charge and acts in a well-defined manner on $H^\bullet_\cl(Q)$:
\be
[\H, a_\w] = \w a_\w, \qquad
[\H, a_{-\w}] = -\w a_{-\w}.
\ee
To quantize, we introduce a vacuum $\ket0$ which is annihilated by all
negative frequency modes in $J^p\HP^*$, e.g.
\be
\phi_M(-m)\ket0 = \pi_M(-m)\ket0 =
\phi^*_M(-m)\ket0 = \pi^*_M(-m)\ket0 = 0,
\ee
for all $-m < 0$. We also have to divide the set of zero modes into creation
and annihilation operators, but this choice is not important since
only modes with $m = \pm\w$ will survive in cohomology. The physical
Hilbert space is identified with the quantum state cohomology $H^0_\st(Q)$,
which is spanned by the vectors $\ket n = (1/\sqrt{n!})a_\w^n\ket0$; clearly,
the Hamiltonian acts in the well-defined manner $\H\ket n = n\w\ket n$,
as appropriate for the harmonic oscillator.

We define involution in $J^p\HP^*$ by $m \to -m$, i.e.
\be
\phi_M(m) \to \phi_M(-m), \qquad
\pi_M(m) \to \pi_M(-m),
\ee
etc. Involution commutes with the BRST charge and sends $\H \to -\H$;
hence it acts in the correct way on the physical phase space,
$a_\w \to a_{-\w}$ and $a_{-\w} \to a_\w$. If we denote the dual
vacuum by $\bra0$, we obtain the correct inner product
$\bracket m n = \dlt_{mn}$.

We thus conclude that the quantum cohomology of the BRST operator
(\ref{Qhjet}) is a resolution of the correct Hilbert space of the
harmonic oscillator: $H^0_\st(Q) = \HH$ and $H^n_\st(Q) = 0$ for all 
$n\neq0$.
This result is independent of the truncation order $p$.

\subsection{ Keeping the observer's trajectory }

We now want to modify the analysis where we do not make the assumption
$\dot q(t) = 1$, but keep $q(t)$ as a dynamical variable, subject to
the Euler-Lagrange equation $\ddot q(t) = 0$.
The total BRST charge is $Q = Q_\phi + Q_q$, where $Q_\phi$ is the
field contribution and $Q_q$ is the observer contribution.
Here $Q_\phi = Q_D + Q_T + Q_R + Q_B + \QM$ is given by (\ref{Qhjet}),
which written in $t$-space reads
\bes
Q_D &=& \sum_{M=0}^{p-2} \int dt\ \EE_M(t)\ps_M(t), \nl
Q_T &=&  \sum_{M=0}^{p-1} \int dt\ \TT_M(t)\bpi_M(t), \nl
Q_R &=& \sum_{M=0}^{p-3} \int dt\ i\R_M(t)\bps_M(t), \\
Q_B &=& \sum_{M=0}^{p-2}
\big( \B_M(\w)\chi_M(-\w) + \B_M(-\w)\chi_M(\w) ), \nl
\QM &=& (\MM_\w+\al_\w)\gm_{-\w} + (\MM_{-\w}+\al_{-\w})\gm_\w,
\eens
where
\bes
\EE_M(t) &=& \phi_{M+2}(t) + \w^2\phi_M(t), \nl
\TT_M(t) &=& \dot\phi_M(t) - \dot q(t)\phi_{M+1}(t), \\
\R_M(t) &=& \bfi_{M+2}(t) + \w^2\bfi_M(t)
- \dot\phi^*_M(t) + \dot q(t)\fs_{M+1}(t),
\eens
and $\MM_{\pm\w}$ and $\B_M(\pm\w)$ are unchanged.
$Q_q = Q_d + Q_a + Q_m$ was written down in (\ref{Qpart}).

The Fourier transform (\ref{Fourierm}) no longer makes sense,
because it is $q(t)$ rather than $t$ which is the physical time variable.
Instead, we make the ansatz
\be
\phi_M(t) = \intm \e^{imq(t)} \phi_M(m).
\label{ansatz}
\ee
If follows that
\bes
\EE_M(t) &=& \intm \e^{imq(t)} \EE_M(m), \nl
\TT_M(t) &=& \intm \dot q(t) \e^{imq(t)} \TT_M(m), \nle
\R_M(t) &=& \intm \dot q(t) \e^{imq(t)} \R_M(m), \nl
\MM_M(t) &=& \intm \dot q(t) \e^{imq(t)} \MM_M(m),
\eens
where $\EE_M(m)$, $\TT_M(m)$ and $\MM_M(m)$ where defined in
(\ref{constrjet}) and $\R_M(m)$ was defined in (\ref{RMm}).
$q(t)$ and hence $\exp(imq(t))$ have zero
weight under repara\-metri\-za\-tions, and $\dot q(t)$ transforms as a 
density
of weight 1; hence $\wt \EE_M(t) = 0$ and $\wt \TT_M(t) = \wt \R_M(t) = 1$.
We introduce antijets with the same weights as the corresponding
constraints, i.e. $\wt \fs_M(t) = 0$, $\wt \bfi_M(t) = 1$, and
$\wt \bfs_M(t) = 1$, e.g.
\be
\bfi_M(t) &=& \intm \dot q(t) \e^{imq(t)} \bfi_M(m).
\ee
The weights of the canonical momenta are one minus the weight of the
corresponding field, i.e. $\wt \pi_M(t) = 1$, $\wt \ps_M(t) = 1$,
$\wt \bpi_M(t) = 0$, and $\wt \bps_M(t) = 0$.
In particular,
\be
\pi_M(t) &=& \intm \dot q(t) \e^{imq(t)} \pi_M(m).
\ee
Using that
\be
\dlt(q(t) - q(t')) = \dot q^{-1}(t)\dlt(t-t')
\ee
and the properties of the Fourier transform, we show that the CCR
\newline $[\phi_M(m), \pi_N(n)] = i\dlt_{MN}\dlt(m+n)$
lead to the expected brackets
\newline$[\phi_M(t), \pi_N(t)] = i\dlt_{MN}\dlt(t-t')$.

When expressed in terms of the Fourier variables $\phi_M(m)$, $\fs_M(m)$,
$\bfi_M(m)$ and $\bfs_M(m)$ and their momenta, the BRST operator
$Q_\phi$ exactly reproduces the BRST charge (\ref{Qhjet}) in subsection
\ref{ssec:cohomjet}. The analysis in that subsection thus applies, and
we obtain a phase space $\PP_\phi$ with
$C(\PP_\phi) = H^0_\cl(Q_\phi) = C(a_\w, a_{-\w})$, as in (\ref{awjet}).
However, we also have to take the observer's trajectory into account.
According to the analysis in subsection \ref{ssec:qpart}, the space of
functions over the observer phase space $\PP_q$ is
$C(\PP_q) = H^0_\cl(Q_q) = C(u,s)$, where the observer's velocity and
position satisfy the CCR $[u,s] = i$. Hence the total phase space
$\PP_\TOT = \PP_\phi + \PP_q$ has the function algebra
\be
C(\PP_\TOT) = H^0_\cl(Q_\phi+Q_q) = C(a_\w, a_{-\w}, u, s)
= C(\PP_\phi) \otimes C(\PP_q).
\ee

\subsection{Hamiltonian and the observer}
\label{ssec:HamObs}

In the previous subsection we computed the cohomology of the total
BRST charge $Q = Q_\TOT = Q_\phi + Q_q$, but we did not specify the
Hamiltonian. A first guess would be to take the
{\em total Hamiltonian} $\H_\TOT = \H^0_\phi + \H^0_q$, where
\bes
\H^0_\phi &=& \sum_{M=0}^p \int dt\ \dot \phi_M(t) \pi_M(t)
- i \sum_{M=0}^{p-2} \int dt\ \dot \phi^*_M(t) \pi^*_M(t) \nl
&&-\ i\sum_{M=0}^{p-1} \int dt\ \ddt\bfi_M(t) \bpi_M(t)
+ \sum_{M=0}^{p-3} \int dt\ \ddt\bfi^*_M(t) \bpi^*_M(t). \nl
\H^0_q &=& \int dt\ \dot q(t) p(t)-i \int dt\ \dot q_*(t) p^*(t).
\ees
$\H_\TOT$ commutes with the total BRST operator $Q_\TOT$:
$[Q_\TOT, \H_\TOT] = 0$,
and would therefore seem as a good candidate Hamiltonian. However, it is
not want we want, because it commutes with the fields,
\bes
[\H_\TOT, \phi(x)] &=& -i\sum_{M=0}^p {1\/M!} \big( \dot\phi_M(t)
- \dot q(t)\phi_{M+1}(t)\big)(x-q(t))^M \nle
&\equiv& -i\ddt\phi(x) \approx 0,
\eens
in view of the time constraint $\TT_M(t)$. Instead, the natural
generalization of the physical Hamiltonian would translate the field
$\phi(x)$ in the time direction, i.e. in the direction of $\dot q(t)$.
Natural candidates would be $\H^0_\phi$ are $-\H^0_q$, since
\be
[\H^0_\phi, \phi(x)] = [-\H^0_q, \phi(x)] = -i\dot q(t)\phi'(x).
\ee
The physical interpretation of this relations is that the physical
Hamiltonian translates the fields relative to the observer, or vice
versa.

Unfortunately, the operators $\H^0_\phi$ are $-\H^0_q$ are not separately
well defined in cohomology. Since
\bes
[\H^0_\phi, \EE_M(t)] &=& -i\dot \EE_M(t), \nl
{[}\H^0_\phi, \TT_M(t)] &=&
-i\dot \TT_M(t)- i \ddot q(t)\phi_{M+1}(t), \\
{[}\H^0_\phi, \R_M(t)] &=& -i\dot \R_M(t)+ i \ddot q(t)\phi^*_{M+1}(t),
\eens
we see that
\bes
{[}\H^0_\phi, Q_T] &=&
- i \sum_{M=0}^{p-1} \int dt\ \ddot q(t)\phi_{M+1}(t)\bpi_M(t), \nle
{[}\H^0_\phi, Q_R] &=&
i \sum_{M=0}^{p-3} \int dt\ \ddot q(t)\phi^*_{M+1}(t)\bps_M(t).
\eens
Hence $[\H^0_\phi, Q_\phi] = [\H^0_\phi, Q_T + Q_R] \neq 0$, and
$\H^0_\phi$ is not BRST closed.
To fix this problem, we add two more terms to the Hamiltonian:
\bes
\H^1_\phi &=&
- i \sum_{M=0}^{p-1} \int dt\ q_*(t)\phi_{M+1}(t) \bpi_M(t), \nle
\H^2_\phi &=&
i \sum_{M=0}^{p-3} \int dt\ q_*(t)\phi^*_{M+1}(t) \bpi_M(t).
\eens
Since $[q_*(t), Q_d] = -\ddot q(t)$, we verify that
\bes
{[}\H^1_\phi, Q_q] &=& -{[}\H^0_\phi, Q_T], \nl
{[}\H^2_\phi, Q_q] &=& -{[}\H^0_\phi, Q_R], \\
{[}\H^1_\phi, Q_R] &=& -{[}\H^2_\phi, Q_D].
\eens
Hence $\H_\phi = \H^0_\phi + \H^1_\phi + \H^2_\phi$ is BRST closed:
\be
[\H_\phi, Q_\TOT] = [\H^0_\phi, Q_T + Q_R]
+[\H^1_\phi, Q_R] + [\H^2_\phi, Q_D] = 0.
\ee
The physical Hamiltonian acts on the modes (\ref{ansatz}) as
\be
[\H_\phi, \phi_M(m)] = m \phi_M(m),
\ee
i.e. the mode label $m$ is nothing but the physical energy. In
particular, the physical oscillators $a_\w$ and $a_{-\w}$ carry energy
$\w$ and $-\w$, respectively:
\be
[\H_\phi, a_\w] = \w a_\w, \qquad
[\H_\phi, a_{-\w}] = -\w a_{-\w}.
\ee

\subsection{ Quantization }

With the phase space and Hamiltonian at hand, we can now proceed to
quantization. We first divide the extended history phase space
$J^p\HP^* = J^p\HP^*_+ \oplus J^p\HP^*_-$ into positive and negative
parts, where $J^p\HP^*_+$ is generated by modes (\ref{ansatz}) with
$m > 0$ and $J^p\HP^*_-$ by modes with $m < 0$. As usual, there is some
ambiguity with zero modes, but this is not important because it does
not affect the physical modes with $m = \pm\w$. Involution replaces
$m$ and $-m$ and thus exchanges $J^p\HP^*_+$ and
$J^p\HP^*_-$. The vacuum $\ket 0$ is mapped onto the dual vacuum $\bra 0$,
which is annihilated by positive modes.

We must also introduce some
quantization prescription for the observer's trajectory and associated
antijets. After this, the constraints are implemented by passing to
the quantum BRST state cohomology, which becomes our Hilbert space.

If we choose a quantization for the observer's trajectory such that
$s = q_0$ annihilates the vacuum, the total Hilbert space becomes
\be
\HH_\TOT = H^0_\st(Q_\phi+Q_q) = C(a_\w, u) = C(a_\w) \otimes C(u)
= \HH_\phi \otimes \HH_q.
\ee
We thus recover the correct Hilbert space for the harmonic oscillator,
$\HH_\phi = C(a_\w)$, as a factor in the total Hilbert space. There is
another factor $C(u)$ which is the Hilbert space for the observer's
trajectory, but it does not influence calculations in $C(a_\w)$. QJT
is hence equivalent to the standard quantization of the harmonic oscillator.

\subsection{ Reparametrization}
\label{ssec:reparjet}

The parameter $t$, which labels the points on the observer's
trajectory, only becomes physical once the observer's equations of
motion $\ddot q(t) = 0$ are taken into account. Instead, we can
follow the route in subsection \ref{ssec:repar} and  introduce
an einbein $e(t) \approx q^{-1}(t)$. The observer's dynamics constraint
(\ref{Epart}) is replaced by (\ref{Epart2}),
$d^2 q/d\tau^2 \approx 0$. The repara\-metrization algebra $\vect(1)$
acts on the fields $\phi_M(t)$, $\pi_M(t)$, etc. E.g.,
\be
[L_f, \phi_M(t)] = -f(t) \dot \phi_M(t), \quad
{[}L_f, \pi_M(t)] = -{d\/dt}\big( f(t)\pi_M(t) \big).
\ee
The repara\-metrization weights $\la$ of the $p$-jets and their antijets and
momenta is given by the following table; in addition, $\vect(1)$ also
acts on the observer's trajectory as in (\ref{cpart}).
\be
\barr{ccr|cr|cc}
\hbox{Field} & \hbox{Parity} & \la &\hbox{Momentum} & \la & n & c \\
\hline
\phi_M(t) & B & 0 & \pi_M(t) & 1 & p+1 & 2(p+1) \\
\fs_M(t) & F & 0 & \ps_M(t) & 1 & p-1 & -2(p-1) \\
\bfi_M(t) & F & 1 & \bpi_M(t) & 0 & p & -2p \\
\bfs_M(t) & B & 1 & \bps_M(t) & 0 & p-2 & 2(p-2) \\
\bt_k & F & - & \gm_k & - & 0 & 0 \\
\al_k & B & - & \al_{-k} & - & 0 & 0
\earr
\ee
In the last two columns, the number of components and the corresponding
value of the central charge $c$ are listed; note that the truncation order
$p=n-1$. The normal-ordered repara\-metri\-zation operator is
$L(t) = L^q(t) + L^\phi(t)$, where $L^q(t)$ is the contribution
(\ref{Lpart}) from the observer's trajectory and its antifields, and
\bes
L^\phi(t) &=& \sum_{M=0}^p i\no{\dot\phi_M(t)\pi_M(t)}
+ \sum_{M=0}^{p-2} \no{\dot\phi^*_M(t)\ps_M(t)}
\nlb{Lphit}
&&+\  \sum_{M=0}^{p-1} \no{\ddt\bfi_M(t)\bpi_M(t)}
+ \sum_{M=0}^{p-3} i\no{\ddt\bfs_M(t)\bps_M(t)}.
\eens
We noted in (\ref{cpart}) that $L^q(t)$ generates a centerless Virasoro
algebra when ghosts are taken into account. The central charge of
$L^\phi(t)$ also vanishes, because
\be
c_\phi = 2(p+1) - 2(p-1) - 2p + 2(p-2) = 0,
\ee
and hence the total repara\-metrization generators $L(t)$ satisfy a
Virasoro algebra with $c=0$. There is no obstruction to the
elimination of repara\-metri\-za\-tions.

\section{Free scalar field in Quantum Jet Theory}

\subsection{Definition of constraints}

We now turn to QJT($p$) for the free field in $d$ spacetime dimensions, 
which
differs from the harmonic
oscillator only in that the field $\phi(x)$ is defined for $x\in\RR^d$
rather than $x\in\RR$. The observer's trajectory $q^\mu(t)$ is now
an operator-valued curve in $\RR^d$, and the Taylor series
(\ref{HarmTaylor}) is replaced by
\be
\phi(x) = \sum_M^p {1\/M!} \phi_M(t) (x-q(t))^M.
\label{FreeTaylor}
\ee
We employ standard multi-index notation.
$M = (M_0,M_1,...,M_{d-1})$ is a multi-index of length
$|M| = \sum_{\mu=0}^{d-1} M_\mu$, $M! = M_0! M_1! ... M_{d-1}!$,
and $(x-q)^M = (x^0-q^0)^{M_0} (x^1-q^1)^{M_1} ...
(x^{d-1}-q^{d-1})^{M_{d-1}}$.
Moreover, the notation $\sum_M^p$ is an abbreviation, meaning that the
sum extends over all $M$ such that $|M|\leq p$.

The history phase space $J^p\HP$ is now spanned by the observer's
trajectory $q^\mu(t)$ and its momentum $p_\mu(t)$, and the Taylor
coefficient functions $\phi_M(t)$ and $\pi^M(t)$ for $0 \leq |M| \leq p$.
The nonzero Poisson brackets are
\be
[q^\mu(t), p_\nu(t')] = i\dlt^\mu_\nu\dlt(t-t'), \qquad
[\phi_M(t), \pi^N(t')] = i\dlt^N_M \dlt(t-t').
\ee
We assume a flat, Minkowski metric $\eta_\mn$ with inverse $\eta^\mn$.
By symmetrization, it induces a metric $\eta_\MN$ for the jets, which
we use to lower all jet indices:
\be
[q_\mu(t), p_\nu(t')] = i\eta_\mn\dlt(t-t'), \qquad
[\phi_M(t), \pi_N(t')] = i\eta_\MN \dlt(t-t').
\ee

The dynamics constraint,
\be
\EE(x) = \d^\mu \d_\mu \phi(x) + \w^2\phi(x),
\ee
turns into the following hierarchy:
\bes
\EE_M(t) &=& \sum_{\mu\nu} \eta^\mn \phi_{M+\mu+\nu}(t) + \w^2\phi_M(t) \nle
&\equiv& \sum_\mu \hm \phi_{M+2\mu}(t) + \w^2\phi_M(t),
\eens
where $\mu$ denotes a unit vector in the $\mu$:th direction, and
$\hm = +1$ for $\mu = 0$ and $\hm = -1$ for $\mu = 1,2,...,d-1$,
i.e. $\hm = \eta_{\mu\mu}$ (no sum on $\mu$).
Since $\phi_M(t)$ is only defined for $|M|\leq p$, $\EE_M(t)$ is
defined for $|M|\leq p-2$.

The time constraint reads
\be
\TT_M(t) = \dot\phi_M(t) - \sum_\mu \dot q^\mu(t)\phi_{M+\mu}(t).
\ee
$\TT_M(t)$ is defined for $|M| \leq p-1$.
For the harmonic oscillator, we could make the assumption
$\dot q(t) = 1$. We could try to impose a similar condition in higher
dimensions, i.e.
$\dot q^\mu(t) \dot q_\mu(t) = 1$, but it is preferable to keep manifest
repara\-metri\-zation invariance as in subsection \ref{ssec:repar}.
Again, we have a redundancy between the dynamics and time constraints,
corresponding to (\ref{red}). This redundancy is most easily expressed
if we assume that $\dot q^0(t) = 1$, $\dot q^i(t)=0$ for all $i>0$:
\be
\phi_{M+2\hat0}(t) \approx \ddot \phi_M(t)
\approx \sum_{i=1}^{d-1} \phi_{M+2\hat\imath}(t) + \w^2\phi_M(t).
\ee

We choose the momentum constraint, for all $|M| \leq p$, to be
\be
\MM_M(t) \equiv \pi_M(t) - {1\/M!}\dot\phi_M(t) = 0.
\label{MMMt}
\ee
Compared to the analogous expression (\ref{MMt}) for the harmonic 
oscillator,
we have included a factor $1/M!$ in order to make $\MM_M(t)$ Lorentz
covariant, cf. subsection \ref{ssec:Poincare} below.
$\MM_M(t)$ neither commutes with itself nor with the other constraints; 
rather
\bes
[\MM_M(t), \EE_N(t')] &=&
-i( \sum_\mu\hm\eta_{M,N+2\mu} + \w^2\eta_\MN)\dlt(t-t'),
\nl
{[}\MM_M(t), \TT_N(t')] &=& i( \eta_\MN\dot\dlt(t-t')
+ \sum_\mu \dot q^\mu(t') \eta_{M,N+\mu}\dlt(t-t') ), \nl
{[}\MM_M(t), \MM_N(t')] &=& -{2i\/M!}\eta_\MN\dot\dlt(t-t').
\ee
The linear combinations
\be
\MM_k(t) = \sum_M^p (-ik)^M \MM_M(t),
\label{MMkt}
\ee
commute with the dynamics constraint provided that $k^2 = \w^2$.
Finally, the expression
\bes
\MM_k &=& \int dt\ \e^{ik\cdot q(t)} \MM_k(t)
\nlb{MMk2}
&=& \sum_M^p (-ik)^M \int dt\ \e^{ik\cdot q(t)} \MM_M(t)
\eens
commutes with the time constraint as well.
However, $\MM_k$ for different values of $k$ do not commute among
each other; instead
\be
[\MM_k, \MM_{k'}] = \si_p(k,k') =  -\si_p(k',k),
\label{MMkk}
\ee
where
\bes
\si_p(k,k') &=& (k_\mu'-k_\mu)\dlt_p(k,k')
\int dt\ \dot q^\mu(t)\e^{i(k+k')\cdot q(t)},
\label{sikk} \\
\dlt_p(k,k') &=& \dlt_p(k',k) =\sum_M^p {(-1)^M\/M!} k^M (k')^M.
\label{dltp1}
\ees
The integral $\MM_k$ commutes with all
dynamics and time constraints for all on-shell values of $k$, but not
all such operators are linearly independent. This is clear because
$\MM_k$ is linear in the jet momenta
$\pi_M(t)$ with $|M|\leq p$, and there are
only ${d+p \choose d}$ such values of $M$. In fact, there are fewer.
As we will see below, the dynamics constraint leaves $\phi_M(t)$ with
$p-1 \leq |M| \leq p$ undetermined. Since there are
\be
r_{p,d} = {d+p \choose d} - {d+p-2 \choose d}
\ee
such values, we pick $r_{p,d}$ values $k_i$ and define an index set
\be
\KK = \{ k_1, k_2, ..., k_{r_{p,d}}\}.
\label{indexset}
\ee
The precise values of the $k_i$ are immaterial, as long as $k_i^2 = \w^2$
and the $k_i$ are linearly independent. We will assume that all components
$k^\mu_i \neq 0$, so that we can define
$k_i^{-M} = \prod_\mu (k^\mu_i)^{-M_\mu}$ without problems.
Moreover, we assume that if  $k\in\KK$, then $-k\in\KK$, so that the
Kronecker delta $\dlt_{k+k'}$ is not identically zero.
If we specialize to one dimension, $r_{p,1} = 2$, and the index set
$\KK = \{\w, -\w\}$ corresponds to the momentum constraints $\MM_\w$ and
$\MM_{-\w}$ for the harmonic oscillator.

Finally, we must also introduce constraints for the observer's trajectory.
We assume that the observer behaves as a free particle, i.e. that it
is subject the dynamics constraints $\EE^\mu(t) = \ddot q^\mu(t)$.

\subsection{ Cohomological formulation }

To employ the BRST formalism, we need to make all constraints
commute. To this end, we first discard all momentum constraints except
$\MM_k$, $k\in\KK$. $k$ is restricted to belong to the
index set (\ref{indexset}). These constraints commute with
$\EE_M(m)$ and $\TT_M(m)$, but not with each other.
To rectify this, we introduce new variables $\al(k)$ for all
$k \in\KK$, subject to the nonzero CCR
\be
[\al_k, \al_{k'}] = -\si_p(k,k').
\ee
The modified momentum constraints
\be
\MM_k' = \MM_k + \al_k,
\ee
do commute with each other, and our full set of constraints is thus
first class.

To eliminate the dynamics and time constraints, we introduce fermionic
antijets as in (\ref{fsbfs}). The reducibility leads to the extra
constraint
\bes
\R_M(t) &=& \sum_\mu \hm\bfi_{M+2\mu}(t) + \w^2\bfi_M(t) \nle
&&-\  \dot\phi^*_M(t) + \sum_\mu \dot q^\mu(t)\fs_{M+\mu}(t).
\eens
To deal with this, we introduce the bosonic, second-order antijet
$\bfs_M(t)$. Like $\R_M(t)$, it is defined for $|M| \leq p-3$.
Finally, we introduce a pair of fermionic antifield $\bt_k$, $\gm_k$
for the momentum constraint $\MM_k$.

To summarize, we have introduced antijets and
canonical momenta according to the following table:
\be
\barr{c|c|c|c|l}
\hbox{Field} & \hbox{Momentum} & \hbox{Parity} & \hbox{Order}
& \hbox{Constraint} \\
\hline
\phi_M(t) & \pi_M(t) & B & p & \\
\fs_M(t) & \ps_M(t) & F & p-2 & \EE_M(t) \\
\bfi_M(t) & \bpi_M(t) & F & p-1 & \TT_M(t) \\
\bfs_M(t) & \bps_M(t) & B & p-3 & \R_M(t) \\
\bt_k & \gm_{-k} & F & - & \MM_{k} \\
\al_k & \al_{-k} & B & - & - \\
\earr
\label{table1}
\ee
The third column indicates the Grassmann parity (bose/fermi), and the
fourth the order where the constraints have to be truncated. The number of
momentum constraints is finite and equals $r_{p,d} = \dim\KK$, whereas the
other constraints are linearly infinite (there is one constraint for each
$t\in\RR$).

The BRST charge for the field $\phi(x)$ is
$Q_\phi = Q_D + Q_T + Q_R + \QM$, where
\bes
Q_D &=& \sum_M^{p-2} \int dt\ \EE_M(t)\ps_M(t), \nl
Q_T &=&  \sum_M^{p-1} \int dt\ \TT_M(t)\bpi_M(t),
\nlb{Qphi}
Q_R &=& \sum_M^{p-3} \int dt\ i\R_M(t)\bps_M(t), \nl
\QM &=& \sum_{k\in\KK} (\MM_k + \al_k)\gm_{-k},
\eens
$Q_\phi$ acts as $\dlt F = [Q_\phi, F]$.
\bes
\dlt \phi_M(t) &=&
-i\sum_{k\in\KK} (-ik)^M \e^{ik\cdot q(t)} \gm_{-k}, \nl
\dlt \fs_M(t) &=&
\sum_\mu\hm \phi_{M+2\mu}(t) + \w^2\phi_M(t), \nl
\dlt \bfi_M(t) &=& \dot\phi_M(t)
- \sum_\mu \dot q^\mu(t)\phi_{M+\mu}(t), \nl
\dlt \bfs_M(t) &=& \sum_\mu \hm\bfi_{M+2\mu}(t) + \w^2\bfi_M(t) \nl
&&-\ \dot\phi^*_M(t) + \sum_\mu \dot q^\mu(t)\fs_{M+\mu}(t), \nl
\dlt \bt_k &=& \MM_k + \al_k, \\
\dlt \pi_M(t) &=& i\big( \sum_\mu\hm\ps_{M-2\mu}(t) + \w^2\ps_M(t)
- \ddt\bpi_M(t) \nl
&&-\ \sum_\mu \dot q^\mu(t)\bpi_{M-\mu}(t)
+ \sum_{k\in\KK} {(-ik)^M\/M!} \e^{ik\cdot q(t)} \gm_{-k} \big), \nl
\dlt \ps_M(t) &=& \ddt\bps_M(t)
+ \sum_\mu \dot q^\mu(t)\bps_{M-\mu}(t), \nl
\dlt \bpi_M(t) &=& \sum_\mu\hm\bps_{M-2\mu}(t) + \w^2\bps_M(t), \nl
\dlt \bps_M(t) &=& \dlt \gm_k = 0, \nl
\dlt \al_k &=& \sum_{k'\in\KK} \si_p(k,k')\gm_{-k'}.
\eens
As in (\ref{dhjet}), we only write down the RHS for generic values of $M$.
If $|M|$ is sufficiently small, we replace
$\ps_{M-2\mu}(t)$ and $\bps_{M-2\mu}(t)$ by zero (if $M_\mu=0,1$), and
$\bpi_{M-\mu}(t)$ and $\bps_{M-\mu}(t)$ by zero (if $M_\mu=0$).
Conversely, if $M$ is sufficiently large, we replace
$\ps_M(t) \to 0$ (if $|M|=p-1,p$),
$\bpi_M(t) \to 0$ (if $|M|=p$), and
$\bps_M(t) \to 0$ (if $p-2\leq|M|\leq p$).
Moreover, the BRST charge also receives contributions from the observer's
trajectory. As explained in subsection \ref{ssec:cohomFree}, we must
then introduce further antifields $q^\mu_*(t)$, $\theta_0^\mu$,
$\theta_1^\mu$, $\bt^0_\mu$, $\bt^1_\mu$, $\al^0_\mu$,
and their conjugate momenta. The observer BRST charge is
$Q_q = Q_d + Q_a + Q_m$, as defined in (\ref{Qpart}). Hence the total
BRST charge is $Q = Q_\TOT = Q_\phi+ Q_q$, and it acts on the
extended history phase space $J^p\HP$ which contains all relevant
antifields and antifield momenta.

\subsection{Solution of constraints}
\label{ssec:cohomqjt}

To compute the cohomology, we make the simplifying assumption
$q^\mu(t) = 0$ if $\mu \neq 0$. We can then write
\be
\phi_M(t) = \intm \e^{imq^0(t)} \phi_M(m).
\label{assumption}
\ee
This assumption clearly breaks manifest covariance, since it singles out
the component $q^0(t)$ among the $q^\mu(t)$. Moreover, we will argue
below in subsection \ref{ssec:discjet} that this assumption is
unphysical, although it implicitly underlies conventional quantization of
fields. Nevertheless, equation (\ref{assumption}) leads to great
simplifications, and therefore we investigate it here.

Since $q^\mu(t)$ and hence $\exp(imq^0(t))$ has zero weight under
repara\-metri\-za\-tions, $\wt \phi_M(t) = 0$. As an example of an object
which transforms as a density, we quote
\be
\TT_M(t) &=& \intm \dot q^0(t) \e^{imq^0(t)} \TT_M(m).
\ee
The weights $\la$ of the different fields and momenta is listed in the
table below:
\be
\barr{cr|cr|cr}
\hbox{Field} & \la & \hbox{Momentum} & \la & \hbox{Constraint} & \la \\
\hline
\phi_M(t) & 0 & \pi_M(t) & 1 && \\
\fs_M(t) & 0 & \ps_M(t) & 1 & \EE_M(t) & 0\\
\bfi_M(t) & 1 & \bpi_M(t) & 0 & \TT_M(t) & 1 \\
\bfs_M(t) & 1 & \bps_M(t) & 0 & \R_M(t) & 1\\
\bt_k & - &\gm_{-k} & - & \MM_{k} & - \\
\al_k & - &\al_{-k} & - &  \\
\earr
\ee
The antifields associated to the momentum constraint
transform trivially under repara\-metri\-za\-tions.

The $\phi$ part of the BRST charge now becomes
$Q_\phi = Q_D + Q_T + Q_R + \QM$, where
\bes
Q_D &=& \sum_M^{p-2} \intm \EE_M(m)\ps_M(m), \nl
Q_T &=&  \sum_M^{p-1} \intm \TT_M(m)\bpi_M(m), \nl
Q_R &=& \sum_M^{p-3} \intm i\R_M(m)\bps_M(m), \nle
\EE_M(m) &=& \sum_\mu \hm \phi_{M+2\mu}(m) + \w^2\phi_M(m), \nl
\TT_M(m) &=& im\phi_M(m) - \phi_{M+\hat0}(m), \nl
\R_M(t) &=& \sum_\mu \hm\bfi_{M+2\mu}(m) + \w^2\bfi_M(m)
- im\fs_M(m) + \fs_{M+\hat0}(m).
\eens
We first focus on the time constraint, which acts on the relevant
fields as
\bes
\dlt_T \phi_M(m) = 0, &\quad& |M|\leq p, \nle
\dlt_T \bfi_M(m) = \TT_M(m), && |M|\leq p-1.
\eens
$\phi_M(m)$ belongs to the kernel and $\dlt_T\bfi_M(m)$ to the image.
Counting the number of components, we find that
$\dim\ker Q_T = {d+p \choose d}$ and
$\dim\im Q_T = {d+p-1\choose d}$, and hence
\be
\dim H^0_\cl(Q_T) = {d+p \choose d} - {d+p-1\choose d} = {d+p-1 \choose 
d-1}.
\ee
The solution to the constraint $\TT_M(m) \approx 0$ is
\be
\phi_M(m) \approx (im)^{M_0} \phi_{(0,\M)}(m) \equiv (im)^{M_0} \phi_\M(m),
\label{phiMT}
\ee
where multi-indices have been decomposed into temporal and spatial
components, i.e. $M = (M_0, \M)$. The set of
$\phi_\M(m) = \phi_{(0,\M)}(m)$ thus constitutes a basis for
$H^0_\cl(Q_T)$.

$Q_T$ acts on the conjugate momenta $\pi_M(m)$ as
\be
\dlt_T \pi_M(m) =
\begin{cases}
- i\bpi_{M-\hat0}(m) & \hbox{if $M_0 \geq 1$, $|M| = p$}, \\
m\bpi_M(m) - i\bpi_{M-\hat0}(m), & \hbox{if $M_0 \geq 1$, $|M|\leq p-1$}, \\
0, & \hbox{if $M_0 = 0$, $|M|=p$}, \\
m\bpi_M(m), & \hbox{if $M_0 = 0$, $|M|\leq p-1$}.
\end{cases}
\ee
The kernel is spanned by
\be
\pi_\M(m) \equiv \sum_{M_0=0}^{p-|\M|} (-im)^{M_0} \pi_{(M_0,\M)}(m),
\label{piMT}
\ee
$\dlt_T\pi_M(m) = 0$. These vectors also span $H^0_\cl(Q_T)$, because
the image is empty.
We verify that the bracket is well-defined in cohomology:
\be
[\phi_\M(m), \pi_\N(n)] = i\eta_{\M\N}\dlt(m+n),
\ee
irrespective of the choice of representative for $\phi_\M(m)$ in
(\ref{phiMT}).
The corresponding antijet $\bfi_M(m) \not\in \ker Q_T$, and its
momentum $\bpi_M(m) \in \im Q_T$, and hence both vanish in cohomology.

We next turn to the dynamics constraint, which acts on the relevant
fields as
\bes
\dlt_D \phi_M(m) = 0, &\quad& |M|\leq p, \nle
\dlt_D \fs_M(m) = \EE_M(m), && |M|\leq p-2.
\eens
$\phi_M(m)$ belongs to the kernel and $\dlt_D\fs_M(m)$ to the image.
Counting the number of components, we find that
$\dim\ker Q_D = {d+p \choose d}$ and
$\dim\im Q_D = {d+p-2\choose d}$, and hence
\be
\dim H^0_\cl(Q_D) = {d+p \choose d} - {d+p-2\choose d} = r_{p,d}.
\label{dimH0QD}
\ee
The solution to the constraint $\EE_M(m) \approx 0$ is
\be
\phi_M(m) \approx -\w^{-2}\sum_\mu \hm\phi_{M+2\mu}(m).
\label{phiMD}
\ee
Repeated use of this relation expresses $\phi_M(m)$ in terms of
$\phi_N(m)$ with $|N| = p-1, p$, which we can choose as a basis for
$H^0_\cl(Q_D)$. Clearly, the number of such fields equals $\dim 
H^0_\cl(Q_D)$
computed above.

$Q_D$ acts on the conjugate momenta $\pi_M(m)$ as
\bes
\dlt_D \pi_M(m) &=&
i(\sum_{\scriptstyle\mu \atop\scriptstyle M_\mu \geq 2}
\hm \ps_{M-2\mu}(m) + \w^2 \ps_M(m)), \quad
\hbox{if $|M|\leq p-2$}, \nl
&=& i\sum_{\scriptstyle \mu\atop \scriptstyle M_\mu \geq 2}
\hm \ps_{M-2\mu}(m), \qquad
\hbox{if $p-1\leq|M|\leq p$}.
\ees
Any vector of the form
\be
\pi_k(m) = \sum_M^p (-ik)^M \pi_M(m), \qquad k^2 = \w^2,
\label{piMD}
\ee
belongs to the kernel, $\dlt_D\pi_k(m) = 0$. However, not all $\pi_k(m)$
can be linearly independent, since there are infinitely many such vectors,
and only finitely many functions $\phi_M(m)$. Since $\pi_M(m)\not\in\im 
Q_D$,
a linearly independent subset with $r_{p,d}$ elements of the vectors
$\pi_k(m)$ in fact form a basis for the momentum part of $H^0_\cl(Q_D)$
as well. The natural
choice is to take $r_{p,d}$ vectors $k_i \in \KK$, where the index set
$\KK$ is the same as in (\ref{indexset}). A basis for $H^0_\cl(Q_D)$ is
thus given by $\phi_M(m)$, $p-1 \leq |M| \leq p$, and $\pi_{k_i}(m)$,
$k_i \in \KK$.
We verify that the bracket is well-defined in cohomology:
\bes
-\w^2[\phi_M(m), \pi_k(n)] &=& -i\w^2(-ik)^M\dlt(m+n) = \nle
{[}\sum_\mu \hm\phi_{M+2\mu}(m), \pi_k(n)]
&=& -i\sum_\mu\hm k^{2\mu}(-ik)^M\dlt(m+n),
\eens
because $\sum_\mu\hm k^{2\mu} \equiv k^2 = \w^2$ for all $k\in\KK$.

Combining (\ref{phiMT}) and (\ref{phiMD}), we obtain
\be
(\w^2 - m^2)\phi_\M(m) \approx \sum_i \phi_{\M + 2\ihat}(m).
\ee
A suitable representative for the $\phi_M(m)$ are
\be
\phi_k = \sum_\M^p {1\/\M!}(-i\kk)^\M \phi_\M(k_0),
\label{phikjet}
\ee
where $k = (k_0,\kk) \in \KK$, and $k_0^2 = \kk^2 + \w^2$.
Combining (\ref{piMT}) and (\ref{piMD}), we find that
\be
\pi_k = \sum_\M^p (-i\kk)^\M \sum_{M_0=0}^{p-|\M|} (-ik_0)^{M_0}
\pi_{(M_0,\M)}(k_0)
= \sum_\M^p (-i\kk)^\M \pi_\M(k_0)
\label{pikjet}
\ee
belongs to $\ker Q_T+Q_D$. They satisfy the bracket
\be
[\phi_k, \pi_{k'}] = i \dlt_p(k,k'),
\ee
where $\dlt_p(k,k')$ was defined in (\ref{dltp1}).

The purpose of the BRST charge $Q_R$ is to eliminate the unwanted
cohomology $\R_M(m)$. Finally, we turn to the momentum part. Note
that in view of (\ref{phikjet}) and (\ref{pikjet}), the momentum
constraint can be rewritten as
\be
\MM_k = \pi_k - ik_0 \phi_k,
\ee
which leads to the CCR
\be
[\MM_k, \MM_{k'}] = k_0\dlt_p(k,k')-k_0'\dlt_p(k',k) = \si_p(k,k').
\ee
This is the same as (\ref{MMkk}), since we made the assumption
$\dot q^\mu(t) = \dlt^\mu_0$ in (\ref{assumption}). The analogous
change is made in the CCR for $\al(k)$:
\be
[\al_k, \al_{k'}] = -\si_p(k,k') = \si_p(k',k).
\ee
$\QM$ acts on the remaining fields as
\bes
\dltM \phi_k = -i\gm_k, &&
\dltM \pi_k = -k_0\gm_k, \nl
\dltM \bt_k = \pi_k - ik_0 \phi_k + \al_k, &&
\dltM \gm_k = 0, \\
\dltM \al_k = \si_p(k,-k).
\eens
The kernel is generated by
\be
a_k = {1\/\sqrt{\si_p(k,-k)}}(\pi_k + ik_0\phi_k) + x_k \MM_k,
\label{akjet}
\ee
where the $x_k$ are arbitrary constants. These oscillators satisfy
well-defined brackets in $H^0_\cl(\QM)$:
\be
[a_k, a_{k'}] = \dlt_{k+k'},
\ee
which of course is the standard oscillators for the free field.
Note that the $a_k$ are not defined for all $k$, but only for
$k \in \KK$. Hence we have not obtained a resolution for the whole
physical phase space $\PP$ (or rather its $\phi$ part $\PP_\phi$),
but only for its $p$-jet approximation, which
we denote by $J^p\PP_\phi$. The corresponding function algebras are
$C(\PP_\phi) = C(\{a_k\}_{k^2 = \w^2})$ and
$C(J^p\PP_\phi) = C(\{a_k\}_{k\in\KK})$, respectively. This is not
surprising, since passing to $p$-jets amounts to a regularization.
The classical cohomology gives a $p$-jet approximation
to the correct phase space for the free field. In the limit $p\to\infty$,
the index set $\KK$ densely fills the surface $k^2=\w^2$, and
the correct classical theory is recovered.

The history phase space $J^p\HP^*$, from which we started, also depends
on the observer's trajectory, and hence the total phase space has the
form $\PP = \PP_\TOT = \PP_\phi \oplus \PP_q$. The corresponding
function algebra should be $C(\PP) = C(\PP_\phi) \otimes C(\PP_q)$.
However, there is a catch. Underlying the ansatz (\ref{assumption})
is the assumption that the spatial components of the observer's
trajectory vanish, $\dot q^i(t) = 0$ for $i = 1,2,3$. As we will discuss
in subsection \ref{ssec:discjet} below, this assumption becomes very
dubious after quantization. However, we only used it
to explicitly compute the cohomology. The history phase space
$J^p\HP^*$ and the BRST operator $Q_\TOT = Q_\phi + Q_q$ are defined
without reference to this assumption.

\subsection{Hamiltonian and quantization}
\label{ssec:QFTlim}

As in for the harmonic oscillator, subsection \ref{ssec:HamObs}, we
define the Hamiltonians
\bes
\H^0_\phi &=& \sum_M^p \int dt\ \dot \phi_M(t) \pi_M(t)
- i \sum_M^{p-2} \int dt\ \dot \phi^*_M(t) \pi^*_M(t) \nl
&&-\ i\sum_M^{p-1} \int dt\ \ddt\bfi_M(t) \bpi_M(t)
+ \sum_M^{p-3} \int dt\ \ddt\bfi^*_M(t) \bpi^*_M(t), \nl
\H^1_\phi &=&
- i \sum_M^{p-1} \int dt\ \sum_\mu q_*^\mu(t)\phi_{M+\mu}(t)\bpi_M(t), \\
\H^2_\phi &=&
i \sum_M^{p-3} \int dt\ \sum_\mu q_*^\mu(t)\phi^*_{M+\mu}(t)\bps_M(t), \nl
\H^0_q &=& \int dt\ \dot q^\mu(t) p_\mu(t)
-i \int dt\ \dot q_*^\mu(t) p^*_\mu(t).
\eens
There are two Hamiltonians which commute with the total BRST operator
$Q_\TOT = Q_\phi + Q_q$:
the total Hamiltonian $\H_\TOT = \H^0_\phi + \H^0_q$ and the
physical Hamiltonian $\H_\phi = \H^0_\phi + \H^1_\phi + \H^2_\phi$.
$\H_\TOT$ translates both the jets and the observer, and hence the
fields remain unchanged: $[\H_\TOT, \phi(x)] = 0$. In contrast, $\H_\phi$
translates the fields relative to the observer,
\be
[\H_\phi, \phi(x)] = -i\sum_\mu\dot q^\mu(t)\d_\mu\phi(x).
\ee
The proof is exactly as for the harmonic oscillator, except for
the appearence of some indices. E.g.
\be
{[}\H^0_\phi, \TT_M(t)] &=& -i\dot \TT_M(t)
- i \sum_\mu \ddot q^\mu(t)\phi_{M+\mu}(t).
\ee
The second term leads to ${[}\H^0_\phi, Q_T] \neq 0$, but since
$[q_*^\mu(t), Q_d] = -\ddot q^\mu(t)$, this contribution is
cancelled by $[\H^1_\phi, Q_d]$. As for the harmonic oscillator,
\bes
{[}\H^1_\phi, Q_q] &=& -{[}\H^0_\phi, Q_T], \nl
{[}\H^2_\phi, Q_q] &=& -{[}\H^0_\phi, Q_R], \\
{[}\H^1_\phi, Q_R] &=& -{[}\H^2_\phi, Q_D].
\eens
and the physical Hamiltonian $\H_\phi = H^0_\phi + H^1_\phi + H^2_\phi$
is a BRST closed observable.

We now quantize in $J^p\HP^*$ by declaring that all modes with
negative $m$ annihilate the vacuum $\ket0$. We must also choose some
representation for $q^\mu(t)$ and $p_\mu(t)$.

Let us now discuss some critical issues about the cohomological
construction of the physical phase space in subsection
\ref{ssec:cohomqjt}. To compute the cohomology,
we made the ansatz (\ref{assumption}), which amounts to the assumption
\be
\dot q^0(t) = 1, \qquad
\dot q^i(t) = 0, \qquad \forall i = 1, 2, ..., d-1.
\label{qfix}
\ee
However, the observer's trajectory $q^\mu(t) \approx u^\mu t + s^\mu$
in $H^0_\cl(Q_q)$. After quantization, $u^\mu$ and $s^\mu$ become quantum
operators, subject to the CCR
\be
[s^\mu, u^\nu] = {i\/\Mobs}\eta^\mn,
\label{suMobs}
\ee
where we have reinstated the observer's mass. Equation (\ref{qfix})
means that we assume that the velocity $u^\mu$ has a well-defined value,
which in view of (\ref{suMobs}) means that the position $s^\mu$ is
completely undetermined. This is not a disaster for the free field,
but an interacting we generally expect $q^\mu(t)$ to couple to the
field $\phi(x)$. Only in the limit $\Mobs\to\infty$ does the RHS of
(\ref{suMobs}) vanish, and $s^\mu$ and $u^\mu$ become c-numbers.

Note that the dubious ansatz (\ref{assumption}) is necessary to
recover the $p$-jet approximation to the correct QFT Hilbert space $\HH$.
This shows, or at least strongly suggests, that QFT implicitly
contains the assumption $\Mobs = \infty$.

\subsection{Reparametrizations}

We proceed in analogy with the treatment of the harmonic oscillator
in subsection \ref{ssec:reparjet}. The jets have repara\-metrization
weights $\la$ given by the following table:
\be
\barr{ccr|cr|ccc}
\hbox{Field} & \hbox{Parity} & \la &\hbox{Momentum} & \la &
\hbox{Order} & n & c \\
\hline
\phi_M(t) & B & 0 & \pi_M(t) & 1 & p & {d+p \choose d} & 2{d+p \choose d} \\
\fs_M(t) & F & 0 & \ps_M(t) & 1 & p-2 & {d+p-2 \choose d} & -2{d+p-2 \choose 
d} \\
\bfi_M(t) & F & 1 & \bpi_M(t) & 0 & p-1 & {d+p-1 \choose d} & -2{d+p-1 
\choose d} \\
\bfs_M(t) & B & 1 & \bps_M(t) & 0 & p-3 & {d+p-3 \choose d} & 2{d+p-3 
\choose d} \\
\bt_k & F & - & \gm_k & - & - & 0 & 0 \\
\al_k & B & - & \al_{-k} & - & - & 0 & 0
\earr
\ee
We have also listed the truncation order, the number of jet components $n$,
and the corresponding value of the central charge $c$.
After normal ordering, the repara\-metri\-za\-tions generators satisfy a
Virasoro algebra with central charge
\bes
c_\phi &=&  2{d+p \choose d} -2{d+p-2 \choose d}
-2{d+p-1 \choose d}+2{d+p-3 \choose d} \nl
&=& 2{d+p-1 \choose d-1} -2{d+p-3 \choose d-1}.
\label{cjet}
\ees
There is a significant difference between the harmonic oscillator, i.e.
$d=1$, and the free field in $d>1$ dimensions. When $d=1$, $c_\phi=0$,
and the repara\-metrization algebra is anomaly free. It is therefore
possible to eliminate repara\-metri\-za\-tions by adding ghosts as in
subsection \ref{ssec:repar}. In contrast, $c_\phi > 1$ when $d>1$. This
is not by itself devastating, because the Virasoro algebra has unitary
lowest-weight representations for all $c > 1$, so unitarity is not
necessarily lost. However, it means that it is not possible to eliminate
repara\-metri\-za\-tions.

What one could hope for is to eliminate repara\-metri\-za\-tions
for all jets with $|M|\leq p-3$. If we replace (\ref{Lphit}) by
\bes
L^\phi_0(t) &=& \sum_M^{p-3} \bigg( i\no{\dot\phi_M(t)\pi_M(t)}
+ \no{\dot\phi^*_M(t)\ps_M(t)} \nle
&&+\ \no{\ddt\bfi_M(t)\bpi_M(t)} + i\no{\ddt\bfs_M(t)\bps_M(t)} \bigg),
\eens
the counting now becomes
\[
c_\phi = 2{d+p-3 \choose d} -2{d+p-3 \choose d}
-2{d+p-3 \choose d}-2{d+p-3 \choose d} = 0,
\]
and the	repara\-metri\-za\-tions $L^0(t) = L^q(t) + L^\phi_0(t)$ can be
cancelled by adding a term $Q_c$ of the form (\ref{Q_c}) to the BRST charge.
This would leave the piece
\bes
L^\phi_1(t) &=& \sum_{|M|=p-2}^p i\no{\dot\phi_M(t)\pi_M(t)}
+ \sum_{|M|=p-2} \no{\dot\phi^*_M(t)\ps_M(t)}
\nle
&&+\  \sum_{|M|=p-2}^{p-1} \no{\ddt\bfi_M(t)\bpi_M(t)},
\eens
which also generates a Virasoro algebra, with central charge given by
(\ref{cjet}). Because of the anomaly, the operators $L^\phi_1(t)$ do
not generate a gauge symmetry after quantization. Instead, they act
as global symmetry generators on the physical Hilbert space.

On closer scrutiny, however, the idea to eliminate only jets with
$|M|\leq p-3$ does not seem to work. To see what goes wrong, consider the
dynamics constraint alone. It is then enough to remove jets with
$p-1 \leq |M| \leq p$ from the repara\-metrization generators, which now
acts as
\be
[L^0_f, \phi_M(t)] =
\begin{cases}
- f(t) \dot\phi_M(t), \quad& |M|\leq p-2\\
0 & p-1 \leq |M| \leq p
\end{cases}
\ee
Hence
\[
[L^0_f, \EE_M(t)] =
\begin{cases}
-f(t)\dot\EE_M(t) + f(t)\sum_\mu \hm \dot\phi_{M+2\mu}(t),
\quad & \hbox{if $p-3 \leq |M| \leq p-2$}, \\
-f(t)\dot\EE_M(t)
\quad & \hbox{if $|M| \leq p-4$}.
\end{cases}
\]
When calculating $[L^0_f, Q_D]$, terms with $|M|\leq p-4$ are not
affected by the modification and vanish as before, and hence
\be
[L^0_f, Q_D] = \sum_{|M|=p-3}^{p-2} \int dt\ f(t)
\sum_\mu \hm\dot\phi_{M+2\mu}(t) \pi^*_M(t).
\ee
To eliminate this term, we would need to introduce a term
\bes
L^1_f &=& \sum_{|M|=p-3}^{p-2} \int dt\ f(t)\dot\phi^*_M(t) \pi^*_M(t)
\ees
to $L_f = L^0_f + L^1_f$. But this brings back antijets with $|M|\geq p-3$,
which give rise to infinite anomalies in the $p\to\infty$ limit. Hence it
appears that it is impossible to eliminate part of the
repara\-metri\-za\-tions.

A different, and even more disturbing, problem is that the central charge
diverges in the field limit $p\to\infty$. The origin of
this difficulty is that in a true $p$-jet approximation to the physical
Hilbert space, all functions of $t$ should be eliminated in cohomology,
leaving only finitely many degrees of freedom. However, the time and
dynamics constraints are unable to do this, because they are not defined
for $|M| = p$ and $|M| = p-1, p$, respectively. There is thus a net surplus
of bosonic functions of $t$ ($c_\phi/2$ of them).

Similar anomalies which diverge when $p\to\infty$ also arise for
diffeomorphisms and gauge symmetries, as we discuss in subsection
\ref{ssec:anomaly} below. The proper treatment of these infinites remains
obscure to me.

\subsection{ Poincar\'e algebra}
\label{ssec:Poincare}

The Poincar\'e algebra is generated by Lorentz transformations $J_\La$
and translations $P_\vareps$, subject to the brackets
\be
[J_\La, J_{\La'}] = J_{[\La, \La']}, \qquad
{[}J_\La, P_\vareps] = P_{\La \vareps}, \qquad
{[}P_\vareps, P_{\vareps'}] = 0.
\ee
Here $\La_{\mu\nu} = -\La_{\nu\mu}$ and $\vareps^\mu$ are constants, and
\bes
[\La, \La']_\mn &=& \La_\mu{}^\rho \La_{\rho\nu}'
- \La_\nu{}^\rho \La_{\rho\mu}', \nle
(\La \vareps)_\mu &=& \La_\mu{}^\nu \vareps_\nu.
\eens
We consider Lorentz rotations around the observer's trajectory. Hence
the Poincar\'e algebra is realized by
\bes
J_\La &=& \La_\mn (x^\mu-q^\mu(t)) {\d\/\d x_\nu}
= \La_\mn k^\mu {\d\/\d k_\nu},
\nlb{JPreal}
P_\vareps &=& \vareps^\mu {\d\/\d x^\mu} = \vareps^\mu k_\mu.
\eens
{F}rom the action on the field $\phi(x)$,
\bes
[J_\La, \phi(x)] &=& -\La_\mn (x^\mu-q^\mu(t))\d^\nu \phi(x),
\nlb{JPphi}
{[}P_\vareps, \phi(x)] &=& -\vareps^\mu \d_\mu \phi(x),
\eens
its follows that the Poincar\'e algebra must act on $p$-jets as
\bes
[J_\La, \phi_M(t)] &=&
\sum_{\scriptstyle \mu \atop \scriptstyle M_\mu\leq p-1} 
\sum_{\scriptstyle \nu \atop \scriptstyle M_\nu\geq 1}
\hm \La_\mn M_\nu \phi_{M+\mu-\nu}(t),
\nlb{JPjet}
{[}P_\vareps, \phi_M(t)] &=&
- \sum_{\scriptstyle \mu \atop \scriptstyle M_\mu\leq p-1} 
\vareps^\mu \phi_{M+\mu}(t).
\eens
In particular, we note that the Lorentz transformations preserve the
jet order in the sense that $[J_\La, \phi_M(t)]$ is a linear combinations
of terms $\phi_N(t)$ with $|N| = |M|$. Note that the sums in (\ref{JPjet})
only extend over $\mu$ and $\nu$ such that the terms in the RHS exist
as $p$-jets.
{F}rom the CCR $[\phi_M(t), \pi_N(t')] = const.$ it follows
that the jet momentum must transform as
\bes
[J_\La, \pi_M(t)] &=&
\sum_{\scriptstyle \mu \atop \scriptstyle M_\mu\leq p-1} 
\sum_{\nu \atop \scriptstyle M_\nu\geq 1}
\hm \La_\mn (M_\mu+1) \pi_{M+\mu-\nu}(t),
\nlb{JPpi}
{[}P_\vareps, \pi_M(t)] &=&
\sum_{\scriptstyle \mu \atop \scriptstyle M_\mu\geq 1} 
\vareps_\mu \pi_{M-\mu}(t).
\eens
The behaviour under Lorentz transformations is compatible with the
identification $\pi_M(t) \approx (1/M!)\dot\phi_M(t)$, which motivates the
factor $1/M!$. However, the action of $P_\vareps$ is not compatible with 
this
identification; demanding the the momentum constraint is preserved by
translations instead leads to the transformation law
\be
{[}P_\vareps, \pi_M(t)] =
- \sum_{\scriptstyle \mu \atop \scriptstyle M_\mu\leq p-1} 
\vareps_\mu (M_\mu+1) \pi_{M+\mu}(t),
\label{Ppi}
\ee
which is clearly different from (\ref{JPpi}).

It also follows from (\ref{JPphi}) that
the observer's trajectory and its momentum transform as
\bes
[J_\La, q^\mu(t)] =  \La^\mu{}_\nu q^\nu(t), &\quad&
{[}P_\vareps, q^\mu(t)] = 0, \nle
{[}J_\La, p_\mu(t)] = -\La_\mu{}^\nu p_\nu(t), &&
{[}P_\vareps, p_\mu(t)] = 0.
\eens
Since the dynamics constraint $\EE_M(t)$ is only defined for $|M|\leq p-2$,
it transforms as
\bes
[J_\La, \EE_M(t)] &=&
\sum_{\scriptstyle \mu \atop \scriptstyle M_\mu\leq p-3} 
\sum_{\nu \atop \scriptstyle M_\nu\geq 1}
\hm \La_\mn M_\nu \phi_{M+\mu-\nu}(t),
\nle
{[}P_\vareps, \EE_M(t)] &=&
- \sum_{\scriptstyle \mu \atop \scriptstyle M_\mu\leq p-3} 
\vareps^\mu \phi_{M+\mu}(t).
\eens
If we postulate the corrsponding transformation law for the
antijet momentum,
\bes
[J_\La, \pi^*_M(t)] &=&
\sum_{\scriptstyle \mu \atop \scriptstyle M_\mu\leq p-3} 
\sum_{\scriptstyle \nu \atop \scriptstyle M_\nu\geq 1}
\hm \La_\mn (M_\mu+1) \pi^*_{M+\mu-\nu}(t),
\nle
{[}P_\vareps, \pi^*_M(t)] &=&
\sum_{\scriptstyle \mu \atop \scriptstyle M_\mu\geq 1} 
\vareps_\mu \pi^*_{M-\mu}(t),
\eens
the BRST charge $Q_D$ commutes with the Poincar\'e
generators. Analogously, we verify that the BRST charges $Q_T$ and
$Q_R$ associated with $\TT_M(t)$ and $\R_M(t)$ are Poincar\'e invariant;
the only difference is that the sums have to be truncated at different
values of $|M|$.

In contrast, the momentum constraint is not preserved by the full Poincar\'e
algebra. This is obvious since the actions of $P_\vareps$ in (\ref{JPpi})
and (\ref{Ppi}) differ.
Hence it is impossible to make the momentum BRST operator
$\QM = \sum_{k\in\KK} (\MM_k + \al_k) \gm_{-k}$ commute with $P_\vareps$.
The situation for Lorentz transformation is somewhat nicer. {F}rom
\be
[J_\La, \MM_M(t)] =
\sum_{\scriptstyle \mu \atop \scriptstyle M_\mu\leq p-1} 
\sum_{\scriptstyle \nu \atop \scriptstyle M_\nu\geq 1}
\hm \La_\mn (M_\mu+1) \MM_{M+\mu-\nu}(t),
\ee
it follows that
\be
[J_\La, \MM^\pp_k] = \La^\mu{}_\nu k_\mu {\d\/\d k_\nu} \MM^\pp_k,
\ee
as one expects from (\ref{JPreal}).

\subsection{ An alternative momentum constraint }

In the previous subsection we learned that the momentum constraint $\MM_k$
does not transform in a sensible way under translations. This is not
necessarily fatal; dynamics is fully specified by the dynamics and time
constraints, whereas the momentum constraint only relates canonical momenta
to velocities. Nevertheless, it might be worthwhile to consider an
alternative choice for the $\phi$ part of $\MM_k$, which makes Poincar\'e
invariance less badly broken.

Consider a plane wave defined by its Fourier transform
\be
\phi_k = \int d^d\!x\ \phi(x) \e^{ik\cdot x}.
\ee
Formally expanding $\phi(x)$ in the Taylor series (\ref{FreeTaylor}),
we obtain
\be
\phi_k = \e^{ik\cdot q} \sum_M (ik)^{-M-\one} \phi_M C_M,
\ee
where $\one = (1,1,...,1)$ is the unit multi-index and the constants
\be
C_M = {1\/M!} \int dy\ \e^{iy} y^M \equiv
\prod_\mu {1\/M_\mu!} \int dy^\mu\ \e^{iy^\mu} (y^\mu)^{M_\mu}.
\ee
This formula arises from the substitution $y = k\cdot(x-q)$, i.e.
$y^\mu = k_\mu (x^\mu - q^\mu)$. By integrating by parts
and discarding boundary terms, we obtain $C_M = i^M M!\, C_0$.
Hence we are led to define
\be
\phi_k(t) = (-i)^d\  \sum_M^p (-ik)^{-M-\one}
\int dt\ \e^{ik\cdot q(t)}\phi_M(t).
\label{planejet}
\ee
This expression clearly only makes sense if all $k_\mu \neq 0$.
The corresponding dynamics and time constraints read
\bes
\EE_k(t) &=& -(k^2-\w^2)\phi_k(t)
+ k^2 \sum_{|M|=0}^1 (-ik)^{-M-\one}\phi_M(t) \nl
&& -\ \w^2 \sum_{|M|=p-1}^p (-ik)^{-M-\one}\phi_M(t), \\
\TT_k(t) &=& \dot\phi_k(t) -ik_\mu\dot q^\mu(t)\phi_k(t)
- \sum_{|M|=p} (-ik)^{-M-\one}\dot\phi_M(t)
+ (-ik)^{-\one}\phi_M(t).
\eens
The linear combinations (\ref{planejet}) hence behave like plane waves,
apart from correction terms coming from the truncation to $p$-jets.
Define
\be
\pi_k(t) = \sum_M^p (-ik)^M \int dt\ \e^{ik\cdot q(t)}\pi_M(t).
\label{planemom}
\ee
The brackets read
\be
[\phi_k(t), \pi_{k'}(t')] = i \dlt_p(k,k') \dlt(t-t'),
\ee
where the definition of $\dlt_p(k,k')$ has been changed to
\be
\dlt_p(k,k') = (-1)^d \sum_M^p k^{-M-\one} {k'}^M.
\ee
In the limit $p \to \infty$
\be
\dlt_p(k,k') \to \prod_\mu {1\/k'_\mu + k_\mu}.
\label{dltp2}
\ee
If the sums over $M$ and $N$ would run over all $\ZZ^d$ rather than 
$\NN^d$, $\dlt_p(k,k')$ would approach $\dlt^d(k+k')$ in the 
$p\to\infty$ limit.

Given the linear combinations (\ref{planejet}) and (\ref{planemom}), we
can now give two alternative definitions for $\MM_k(t)$:
\be
\MM_k(t) =
\begin{cases}
\pi_k(t) - \dot \phi_k(t), \\
\pi_k(t) - ik_\mu \dot q^\mu(t)\phi_k(t).
\end{cases}
\label{MMkt2}
\ee
Their bracket becomes
\be
[\MM_k(t), \MM_{k'}(t')] = 
\begin{cases}
-i\dlt_p(k,k')\dot\dlt(t-t'), \\
(k_\mu - k_\mu') \dot q^\mu(t)\dlt_p(k,k') \dlt(t-t').
\end{cases}
\ee
Hence the integrals $\MM_k = \int dt\ \MM_k(t)$ satisfy CCR of the form
(\ref{MMkk}), with the definition of $\dlt_p(k,k')$ changed from
(\ref{dltp1}) to (\ref{dltp2}).

It is clear from the definition of $\phi_k(t)$ that the Poincar\'e
algebra acts like
\bes
[J_\La, \phi_k(t)] &=& -\La_\mn k^\mu {\d\/\d k_\nu} \phi_k(t) + ..., 
\nle
{[}P_\vareps, \phi_k(t)] &=& -\vareps^\mu k_\mu \phi_k(t) + ...,
\eens
where the ellipses stand for correction terms involving $\phi_M(t)$
with $|M|$ close to $p$ and zero. $\pi_k(t)$ and $\MM_k(t)$ transform 
in the same way. This definition of the momentum constraint thus 
appears to be manifestly covariant up to correction terms which can be
ignored in the limit $p\to\infty$. Note however that the variable
$k$ is restricted to the discrete index set $\KK$, and that a finite
Poincar\'e transformation in general takes us outside this set. $\KK$
becomes dense in the invariantly defined surface $k^2 = \w^2$ when 
$p\to\infty$, and one may argue that manifest covariance is recovered
in this limit.

That one must give up manifest Poincar\'e covariance is a
disappointment; as the name indicates, combining canonical quantization
with manifest covariance was a major goal with MCCQ. However, Poincar\'e
covariance is only broken by the momentum constraint. Since the dynamics
and time constraints are manifestly covariant (even general-covariant in
the gravity context), as is quantization in the sense of eliminating 
negative-frequency modes, dynamics and quantization are defined 
in a manifestly covariant way. Only the definition of canonical momenta
in terms of velocities break manifest covariance.

Therefore, we are led to a weaker definition of covariance in QJT($p$).
The history phase space $J^p\HP$ carries a well-defined representation
of the covariance algebra, and the BRST operator $Q_\TOT$ splits
into two terms, the dynamics part $Q_D+Q_T+Q_R+Q_q$ (plus
additional terms in the presence of gauge symmetries), and the momentum
part $\QM$. The former is manifestly covariant in the sense that this BRST
charge commutes with the covariance generators, but the latter is not.
Despite its weakened form, the requirement is still quite
strong. In particular, $H^\bullet(Q_D+Q_T+Q_R+Q_q)$ only carries a
well-defined (necessarily projective) action of diffeomorphism and
gauge algebra if
the history phase space is a space of trajectories in jet space; this
is clear from \cite{Lar98}.

\subsection{ Discussion }
\label{ssec:discjet}
To conclude the treatment of the free field within QJT, let us review
the most important lessons:
\begin{itemize}
\item
QJT($p$) gives a $p$-jet approximation $J^p\HH$ to the physical Hilbert
space, and the correct Hilbert space  $\HH$ should only arise in the
$p\to\infty$ limit. This is unlike the single harmonic oscillator,
where $J^p\HH = \HH$ for all $p$.
\item
QFT is recovered in the limit as QJT($\infty$), but only in the limit
that the observer's mass is infinite. The observer's position and
velocities are c-numbers only if $\Mobs=\infty$. Quantization is
hence taken more seriously in QJT then in QFT, because not only the
fields but also the observer's trajectory (and hence the operational
definition of time) are quantized.
\item
Repara\-metri\-za\-tions become problematic. Not only does the 
repara\-metrization
algebra acquire a central extension in QJT($p$), but the extension
diverges in the field limit $p\to\infty$.
\item
The Poincar\'e algebra does not act in a well-defined manner in QJT($p$).
The problems are located to the momentum constraint; dynamics and
quantization are implemented in a manifestly covariant way.
\end{itemize}

\section{ Free Maxwell field }

\subsection{ History formulation }

We now turn to the free Maxwell field, which is also described by a free
theory. The new feature here is that the model has a local $U(1)$ gauge
symmetry, which on the Lie algebra level is described by the Lie algebra
$\map(d, \uu(1))$ of maps from $d$-dimensional spacetime to $\uu(1)$.
This algebra has generators $\J_X$, where $X(x)$
is an arbitrary function on spacetime, and brackets $[\J_X, \J_Y] = 0$.

The dynamical degree of freedom is a gauge potential $A_\mu(x)$, which 
transforms under $\map(d, u(1))$ as
\be
[\J_X, A_\mu(x)] = -i\d_\mu X(x).
\label{JXA}
\ee
The canonical conjugate of $A_\mu(x)$ is denoted by $E^\mu(x)$; it will
eventually be identified with the electric field strength. The gauge
field $F_\mn(x) = \d_\mu A_\nu(x) - \d_\nu A_\mu(x)$ is clearly gauge
invariant, $[\J_X, F_\mn(x)] = 0$. In Fourier space, the gauge potential
has components $A_\mu(k)$ and the field strength is 
$F_\mn(k) = k_\mu A_\nu(k) - k_\nu A_\mu(k)$.
The dynamics constraint 
\be
\EE_\mu(k) = k^\nu F_{\nu\mu}(k) = k^2 A_\mu(k) - k_\mu k^\nu A_\nu(k),
\label{EEu1}
\ee
is subject to the redundancy
\be
k^\mu \EE_\mu(k) = k^\mu k^\nu F_{\mu\nu}(k) \equiv 0.
\ee
The gauge algebra is generated by the Gauss law constraint
\be
\J(k) = k_\mu E^\mu(k) \approx 0.
\ee
In Fourier space, the gauge generators can be written 
$\J_X = \intdk X(k) \J(-k)$.
The potential transforms as $[\J_X, A_\mu(k)] = ik_\mu X(k)$ and the field
strength is gauge invariance: $[\J_X, F_\mn(k)] = 0$.

We introduce standard polarization vectors $\eps^\ii_\mu(k)$,
$i = 0, 1, 2, 3$. If $k$ is the wave-vector for a photon travelling in
the $\mu=3$ direction, $k = (|k|,0,0,|k|)$, we have
$k^\mu\eps^{(0)}_\mu(k) = -k^\mu\eps^{(3)}_\mu(k) = |k|$.
It is convenient to introduce lightcone coordinates
\bes
\eps^\plus_\mu(k) &=&
{1\/\sqrt 2}(\eps^0_\mu(k) + \eps^{(3)}_\mu(k)) = {k_\mu\/|k|},
\nle
\eps^\minus_\mu(k) &=&
{1\/\sqrt 2}(\eps^0_\mu(k) - \eps^{(3)}_\mu(k)).
\eens
These polarization vectors satisfy the relations
\bes
k^\mu \eps^\ii_\mu(k) = 0, \hbox{ for } i = +,1,2, &\quad&
k^\mu \eps^\minus_\mu(k) = |k|,
\nlb{polar}
\eps^{\ii\mu}(k)\eps^\jj_\mu(k) = \eta^\ij, &&
\eta_\ij\eps^{\ii\mu}(k)\eps^\jj_\nu(k) = \delta^\mu_\nu,
\eens
where $\eta^{++} = \eta^{--} = 0$, $\eta^{+-} = \eta^{-+} = 1$.

Finally, we define the momentum constraint,
\be
\MM_\mu(k) = E_\mu(k) - F_{\mu0}(k)
= E_\mu(k) - k_\mu A_0(k) + k_0 A_\mu(k).
\label{MMu1}
\ee
Since $\MM_\mu(k)$ only depends on quantities which transform
homogeneously (even trivially) under gauge transformations, the momentum
constraint also has this property. Most components of the momentum
constraints don't commute with the dynamics constraint (\ref{EEu1});
rather
\be
[\MM_\mu(k), \EE_\nu(k')] = -i(k^2 \eta_\mn - k_\mu k_\nu)\dlt(k+k').
\ee
However, the combinations
\be
\MM^\ii(k) = \eta^\mn \eps^\ii_\mu(k) \MM_\nu(k)
\ee
do commute with the dynamics constraint, provided that $k^2=0$ and
$i=+,1,2$. These constraints are second class, since they do not commute
among themselves
\be
[\MM^\ii(k), \MM^\jj(k')] = 2ik_0\eta^\ij\dlt(k+k').
\ee
To fix this defect, we introduce oscillators $\al^\ii(k)$, also
defined for $k^2 = 0$ and $i=1,2$; no momentum constraint is needed for
the $i=+$ component because it is a gauge degree of freedom. 
The $\al$'s satisfy the CCR
\be
[\al^\ii(k), \al^\jj(k)] = - 2ik_0\eta^\ij\dlt(k+k').
\ee
Hence the improved momentum constraints $\MM^\ii(k) + \al^\ii(k)$
commute both with the dynamics constraints and among themselves.

\subsection{ Cohomological construction}

We now proceed to the cohomological construction of the physical phase
space $\PP$.
For all values of $k$, we introduce antifields and momenta according
to the following table.
\be
\barr{c|c|c|l}
\hbox{Field} & \hbox{Momentum} & \hbox{Parity} & \hbox{Constraint} \\
\hline
A_\mu(k) & E^\mu(-k) & B & \\
A^*_\mu(k) & E_*^\mu(-k) & F & \EE_\mu(k) \\
\zeta(k) & \xi(-k) & B & k^\mu A^*_\mu(k) \\
c(k) & b(-k) & F & k_\mu E^\mu(k) \\
\earr
\ee
Moreover, for $k^2 = 0$, $i = 1,2$, we introduce also the following
antifields and momenta:
\be
\barr{c|c|c|l}
\hbox{Field} & \hbox{Momentum} & \hbox{Parity} & \hbox{Constraint} \\
\hline
\theta^\ii(k) & \chi_\ii(-k) & B & \eps^{\ii\mu}(k) A^*_\mu(k)\\
\bt^\ii(k) & \gm_\ii(-k) & F & \MM^\ii(k) + \al^\ii(k)\\
c^*(k) & b_*(-k) & B & k_\mu E_*^\mu(k) \\
\al^\ii(k) & \al^\ii(-k) & B & -
\earr
\label{extraaf}
\ee
The BRST charge is $Q = Q_D + Q_S + Q_G + Q_A + Q_{GA} + \QM$, where
\bes
Q_D &=& \intdk k^\nu F_{\nu\mu}(k) E^\mu_*(-k), \nl
Q_S &=& \intdk k^\mu A^*_\mu(k) \xi(-k), \nl
Q_G &=& \intdk k_\mu E^\mu(k) c(-k), \nl
Q_{GA} &=& \intdk k_\mu E_*^\mu(k) c^*(-k)\dlt(k^2), \\
Q_A &=& \sum_{i=1}^2 \intdk
\eps^{\ii\mu}(k) A^*_\mu(k) \chi_\ii(-k) \dlt(k^2), \nl
\QM &=& \sum_{i=1}^2  \intdk \big( \eps^\ii_\mu(k)(E^\mu(k)
- F^{\mu0}(k)) + \al^\ii(k) \big) \gm_\ii(-k)\dlt(k^2).
\eens
Using the properties (\ref{polar}) of the polariation vectors, it
is straightforward to prove that $Q^2 = 0$.
The BRST operator acts on the fields and antifields as
\bes
\dlt c(k) &=& 0, \nl
\dlt A_\mu(k) &=& ik_\mu c(k)
- i\sum_{i=1}^2 \eps^\ii_\mu(-k) \gm_\ii(k) \dlt(k^2), \nl
\dlt A^*_\mu(k) &=& k^2 A_\mu(k) - k_\mu k^\nu A_\nu(k), \nl
\dlt \zeta(k) &=& ik^\mu A^*_\mu(k),
\label{dlt1u1}\\
\dlt \theta^\ii(k) &=& -i\eps^{\ii\mu}(k) A^*_\mu(k) \dlt(k^2), \nl
\dlt \bt^\ii(k) &=& \big( \eps^\ii_\mu(k)(E^\mu(k) - F^{\mu0}(k))
+ \al^\ii(k) \big) \dlt(k^2), \nl
\dlt \al^\ii(k) &=& 2ik_0\gm_\ii(k) \dlt(k^2).
\eens
where antifields with parenthesized indices $\ii$ are only defined for
$k^2 = 0$ and  $i= 1,2$. $Q$ acts on the conjugate momenta as
\bes
\dlt b(k) &=& -ik_\mu E^\mu(k), \nl
\dlt E^\mu(k) &=& i(k^2 E_*^\mu(k) - k^\mu k_\nu E_*^\nu(k))
- ik^0\eps^{\ii\mu}(k)\gm_\ii(k) \dlt(k^2), \nl
\dlt E_*^\mu(k) &=& -k^\mu \xi(k)
+ \sum_{i=1}^2 \eps^{\ii\mu}(k) \chi_\ii(-k) \dlt(k^2),
\label{dlt2u1}\\
\dlt \xi(k) &=& 0, \nl
\dlt \chi^\ii(k) &=& 0, \nl
\dlt \gm_\ii(k) &=& 0.
\eens

The cohomology is computed along the lines above. If $k^2 \neq 0$,
both the kernel and the image are generated by $c(k)$, $k^\mu A^*_\mu(k)$,
$\xi(k)$, and $k_\mu E_*^\mu(k)$, and by $\eps^\ii_\mu E_*^\mu(k)$ and
$\eps^{\ii\mu}(k)A_\mu(k)$, for $i = +,1,2$. Thus there is no
contribution to $H^\bullet_\cl(Q)$ for $k^2 \neq 0$.

For $k^2 = 0$, it is convenient to define
\bes
A^\ii(k) = \eps^{\ii\mu}(k) A_\mu(k), &\quad&
A_*^\ii(k) = \eps^{\ii\mu}(k) A^*_\mu(k), \nle
E^\ii(k) = \eps^\ii_\mu(k)E^\mu(k), &&
E_*^\ii(k) = \eps^\ii_\mu(k)E_*^\mu(k).
\eens
It follows that
\be
k^\mu A_\mu(k) = |k|A^\plus(k).
\ee
The relevant parts of (\ref{dlt1u1}) and (\ref{dlt2u1}) become for
$i=+,-$:
\bes
\dlt c(k) = 0, &\quad&
\dlt b(k) = |k| E^\plus(k), \nl
\dlt A^\minus(k) = i|k| c(k), &&
\dlt E^\minus(k) = -i |k|^2 E^\plus_*(k), \nl
\dlt A^\minus_*(k) = |k|^2 A^\plus(k), &&
\dlt E_*^\minus(k) = -|k| \xi(k), \\
\dlt \zeta(k) = i|k| A^\plus_*(k), &&
\dlt \xi(k) = 0, \nl
\dlt A^\plus(k) = \dlt A^\plus_*(k) = 0, &&
\dlt E^\plus(k) = \dlt E_*^\plus(k) = 0.
\eens
Hence both the kernel and the image consist of
$A^\plus(k)$, $A^\plus_*(k)$, $E^\plus(k)$, $E_*^\plus(k)$, $c(k)$ and
$\xi(k)$, and all fields vanish in cohomology; the longitudinal and
temporal parts of the gauge connection are unphysical.

In constrast, for $i=1,2$, the relevant parts of (\ref{dlt1u1}) and
(\ref{dlt2u1}) become
\bes
\dlt A^\ii(k) = -i \gm^\ii(k) \dlt(k^2), &&
\dlt E^\ii(k) = i\gm^\ii(k) \dlt(k^2), \nl
\dlt A_*^\ii(k) = 0, &&
\dlt E_*^\ii(k) = \chi^\ii(-k) \dlt(k^2), \nl
\dlt \theta^\ii(k) = -iA_*^\ii(k) \dlt(k^2), &&
\dlt \chi^\ii(k) = 0, \nl
\dlt \bt^\ii(k) = ( \MM^\ii(k)+ \al^\ii(k)) \dlt(k^2), &&
\dlt \gm^\ii(k) = 0, \nl
\dlt \al^\ii(k) = 2ik_0\gm^\ii(k) \dlt(k^2).
\ees
The kernel consists of $E^\ii(k) + k_0 A^\ii(k)$,
$\MM^\ii(k) + \al^\ii(k)$, $A_*^\ii(k)$ and $\gm^\ii(k)$, and all
combinations except for the first also belong to the image.
The cohomology $H^\bullet_\cl(Q)$ is thus generated by
\be
a^\ii(k) = E^\ii(k) + k_0 A^\ii(k) + x (\MM^\ii(k) + \al^\ii(k))
\ee
for all $k^2 = 0$ and $i = 1,2$, where $x$ is an arbitrary constant.
These are identified with creation
and annihilation operators for photons, depending on the sign of $k_0$.
We verify that they satisfy the CCR
\be
[a^\ii(k), a^\jj(k')] = 2ik_0 \eta^\ij\dlt(k+k'),
\ee
independent of the value of $x$.

The above analysis is slightly flawed, because there is additional
cohomology for $k=0$ (rather than just $k^2=0$), i.e. for soft photons.
This unwanted cohomology could be eliminated by introducing additional
antifields and ghosts, and adding new terms to the BRST operator.

We can now quantize by introducing a vacuum which is annihilated by
all oscillators with $k_0 < 0$; there is some ambiguity with modes
with $k_0 = 0$, which was irrelevant for the scalar field because
it has a mass gap. Otherwise, the treatment is completely analogous to
the non-covariant history quantization of the free scalar field in
section \ref{sec:scalar}.

\subsection{ Maxwell field in QJT }
\label{ssec:maxwell}

Expand the gauge field in a Taylor series
\be
A_\mu(x) = \sum_M^p {1\/M!} A_{\mu,M}(t) (x-q(t))^M.
\ee
The canonical momentum is defined by the non-zero CCR
\be
[A_{\mu,M}(t), E_{\nu,N}(t')] = i\eta_\mn \dlt(t-t').
\ee
The field strength corresponds to the jet
\be
F_{\mn,M}(t) \equiv F_{\mn,M}(A; t)
= A_{\nu,M+\mu}(t) - A_{\mu,M+\nu}(t).
\ee
The dynamics constraint,
\be
\EE_{\mu,M}(t) = \sum_\nu F_{\mn,M+\nu}(t)
= \sum_\nu A_{\nu,M+\mu + \nu}(t) - \sum_\nu \hn A_{\mu,M+2\nu}(t),
\ee
has the redundancy
\be
\sum_\mu \EE_{\mu,M+\mu}(t) = 0.
\ee
The corresponding gauge generators,
\be
\J_M(t) = \sum_\mu E^\mu_{M-\mu}(t),
\ee
commute with $E^\mu_M(t)$, $F_{\mn,M}(t)$ and the dynamics constraint
and acts as follows on the gauge field:
\be
[\J_M(t), A_{\nu,N}(t')] = -i\eta_{M-\nu,N} \dlt(t-t').
\label{JMA}
\ee
The gauge algebra takes the form
\be
[\J_M(t), \J_N(t')] = 0.
\ee
Alternatively, we could smear the generators with functions on
spacetime $X(x)$, $x \in \RR^4$:
\be
\J_X = \sum_M^p \int dt\ \d_M X(q(t)) \J_M(t),
\ee
where $\d_M$, $M = (M_0,M_1,...,M_{d-1})$, denotes the mixed partial
derivitives:
\be
\d_M X(x)
\equiv \underbrace{\d_0 .. \d_0}_{M_0}
\underbrace{\d_1 .. \d_1}_{M_1} ..
\underbrace{\d_{d-1} .. \d_{d-1}}_{M_{d-1}} X(x).
\ee
In this case, (\ref{JMA}) is replaced by
\be
[\J_X, A_{\mu,M}(t)] = - i\d_{M+\mu}X(q(t)).
\ee
For any jet $\phi_M(t)$ with reparametrization weight $\wt\phi_M(t) = 0$,
we define
\be
D\phi_M(t) \equiv \dot\phi_M(t) - \sum_\mu \dot q^\mu(t) \phi_{M+\mu}(t).
\label{Dphi}
\ee
The time constraint is defined by
\be
\TT_{\mu,M}(t) =  DA_{\mu,M}(t) \equiv \dot A_{\mu,M}(t)
- \sum_\nu \dot q^\nu(t) A_{\mu,M+\nu}(t).
\label{TA}
\ee
The time constraint does not commute with the gauge generators $\J_M(t)$;
instead,
\be
[\J_M(t), \TT_{\nu,N}(t')] = i\eta_{M,N+\nu}\dot\dlt(t-t')
+ i\sum_\rho \dot q^\rho(t) \eta_{M,N+\nu+\rho} \dlt(t-t').
\ee
However, the smeared generators do commute with the time constraint, 
\be
[\J_X, \TT_{\nu,N}(t')] = 0,
\ee
because $\dot X_{N+\nu}(q(t)) =
\sum_\rho \dot q^\rho(t) X_{N+\nu+\rho}(q(t))$.

Since the observer's trajectory points in the time
direction, we can use it instead of the fixed time direction ''$0$''.
Hence we replace $\dlt^\mu_0 \to \dot q^\mu(t)$ in all formulas.
New polarization vectors are defined by
\bes
\eps^\plus_\mu(k,\dot q) &=& k_\mu, \nl
\eps^\minus_\mu(k,\dot q) &=& k_\mu
- {2k_\nu \dot q^\nu\/\Mobs^2} \dot q_\mu, \\
\eps^\ii_\mu(k,\dot q) \eps^{\plus\mu}(k,\dot q) &=&
\eps^\ii_\mu(k,\dot q) \eps^{\minus\mu}(k,\dot q) = 0,
\quad i=1,2.
\eens
These polarization vectors satisfy the relations
\bes
k^\mu \eps^\ii_\mu(k,\dot q)= 0, \hbox{ for } i = +,1,2, &&
k^\mu \eps^\minus_\mu(k,\dot q) = k_\mu \dot q^\mu(t) \equiv |k|, \nl
\eps^{\ii\mu}(k)\eps^\jj_\mu(k) = \eta^\ij, \hbox{ for } i = 1,2, &&
\eps^\plus_\mu(k,\dot q) \eps^{\minus\mu}(k,\dot q) = 1, \\
\eps^\plus_\mu(k,\dot q) \eps^{\plus\mu}(k,\dot q) &=&
\eps^\minus_\mu(k,\dot q) \eps^{\minus\mu}(k,\dot q) = 0
\eens

We want to relate the electric field to the field strength by a relation
$E_\mu = F_{\mu0}$. We take the momentum constraint to be
\be
\MM_{\mu,M}(t) = E_{\mu,M}(t)
- {1\/M!}\sum_\nu \dot q^\nu(t) F_{\mu\nu,M}(t),
\ee
which is defined for $|M|\leq p-1$.
Modulo the time constraint, this is equivalent to the choice
\be
\MM_{\mu,M}(t) = E_{\mu,M}(t) + {1\/M!}\dot A_{\mu,M}(t)
- {1\/M!}\sum_\nu \dot q^\nu(t) A_{\nu,M+\mu}(t).
\ee
The momentum constraint clearly commutes with the gauge algebra,
\break $[\J_M(t), \MM_{\nu,N}(t')] = 0$, since $\J_M(t)$ commutes with 
the field strength. Moreover, the combinations
\be
\MM^\ii(k) = \sum_M^{p-1} (-ik)^M \int dt\ \e^{ik\cdot q(t)}
\eps^{\ii\mu}(k,\dot q(t)) \MM_{\mu,M}(t),
\label{MAik}
\ee
where $k^2 = 0$ and $i = +,1,2$, also commutes with the dynamics and
time constraints.
We verify that these generators satisfy the algebra
\bes
&&[\MM^\ii(k), \MM^jj(k')] = \si^\ij(k,k') = -\si^{ji}(k',k)\nl
&\equiv& \sum_M^{p-1} (-ik)^M (-ik')^M \times \\
&&\times\ \int dt\ \e^{ik\cdot q(t)} 
\eps^\ii_\mu(k,\dot q(t)) \eps^\jj_\nu(k',\dot q(t))
(\dot q^\nu(t)k'_\mu - \dot q^\mu(t)k_\nu).
\eens

We take the BRST operator to be
$Q_A + Q_D + Q_S + Q_T + Q_{GT} + Q_{DT} + Q_{ST} + \QM$, where
\bes
Q_A &=& \sum_M^{p} \int dt\
c_{M+\mu}(t) E^\mu_M(t), \nl
Q_D &=& \sum_M^{p-2} \int dt\
\sum_\nu F_{\nu\mu,M+\nu}(A(t)) E^{*\mu}_M(t), \nl
Q_S &=& \sum_M^{p-3} \int dt\
\sum_\mu A^*_{\mu,M+\mu}(t) \xi_M(t), \nl
Q_{GT} &=& \sum_M^p \int dt\
Dc_M(t) \bar b_M(t),
\label{Qmaxwell}\\
Q_{AT} &=& \sum_M^{p-1} \int dt\ ( DA_{\mu,M}(t)
-\bar c_{M+\mu}(t)) \bar E^\mu_M(t), \nl
Q_{DT} &=& \sum_M^{p-3} \int dt\ ( DA^*_{\mu,M}(t)
-\sum_\nu \hn F_{\nu\mu,M+\nu}(\bar A(t))) \bar E^{*\mu}_M(t), \nl
Q_{ST} &=& \sum_M^{p-4} \int dt\ ( D\zeta_\M(t)
- \sum_\mu \hm\bar A^*_{\mu,M+\mu} (t)) \bar\xi_M(t), \nl
\QM &=& \sum_{i=1}^2 \sum_{k\in\KK} (\MM^\ii(k) + \al^\ii(k)) \gm_\ii(k),
\eens
It is straightforward to verify that $Q^2 = 0$, using e.g.
\bes
\{Q_A, Q_D\} &=& \sum_M \int \sum_\nu
(\hn c_{M+\mu+2\nu}(t) - \hn c_{M+\nu+\mu+\nu}(t)) E^{*\mu}_M(t), \nl
\{Q_{GT}, Q_{AT}\} &=& -\{Q_A, Q_{AT}\}
= i\int Dc_{M+\mu}(t) \bar E^\mu_M(t), \\
\{Q_{AT}, Q_{DT}\} &=& -\{Q_D, Q_{DT}\}
= i\int \sum_\nu F_{\nu\mu,M+\nu}(DA(t)) \bar E^{*\mu}_M(t).
\eens

\subsection{ Solution of cohomology}

We first consider the unbarred jets and antijets. Ignoring the
momentum constraint, the BRST operator acts as
\bes
\dlt c_M(t) &=& 0, \nl
\dlt A_{\mu,M}(t) &=& -ic_{M+\mu}(t),
\label{dltAm}\\
\dlt A^*_{\mu,M}(t) &=& \sum_\nu \hn F_{\nu\mu,M+\nu}(t) =
\sum_\nu \hn (A_{\mu,M+2\nu}(t) - A_{\nu,M+\mu+\nu}(t)), \nl
\dlt\zeta_M(t) &=& -i\sum_\mu \hm A^*_{\mu,M+\mu}(t)
\equiv -i\div A^*_M(t).
\eens
The kernel is generated by $c_M(t)$, $\div A^*_M(t)$, and
gauge invariant functionals of $A_{\mu,M}(t)$,
i.e. functionals of $F_{\mn,M}(t)$.
The image is generated by $c_M(t)$, $\div A^*_M(t)$, and
$\EE_{\mu,M}(t) = \sum_\nu \hn F_{\nu\mu,M+\nu}(t)$.
Hence the cohomology is generated by the equivalence classes
\be
F_{\mn,M}(t) + x^\rho_\mn \EE_{\rho,M}(t),
\label{Fcohom}
\ee
where $x^\rho_\mn$ are arbitary.

Introduce the linear combinations
\bes
c_k(t) = \sum_M^{p+1} (ik)^{-M} c_M(t), &&
A^*_{\mu,k}(t) = \sum_M^{p-2} (ik)^{-M} A^*_{\mu,M}(t), \nl
A_{\mu,k}(t) = \sum_M^{p} (ik)^{-M} A_{\mu,M}(t), &&
\zeta_k(t) = \sum_M^{p-3} (ik)^{-M} \zeta_M(t).
\ees
Equation (\ref{dltAm}) then takes the form
to
\bes
\dlt c_k(t) &=& 0, \nl
\dlt A_{\mu,k}(t) &=& k_\mu c_k(t), \nle
\dlt A^*_{\mu,k}(t) &=& (k^2 \dlt^\nu_\mu - k_\mu k^\nu) A_{\nu,k}(t), \nl
\dlt \zeta_k(t) &=& k^\mu A^*_{\mu,k}(t),
\eens
which we recognize as (\ref{dlt1u1}). Hence the transverse photons
\be
A^\ii_k(t) = \eps^{\ii\mu}(k, \dot q(t)) A_{\mu,k}(t), \quad
\hbox{where $k^2 = 0$, $i = 1,2$},
\label{Aik}
\ee
belong to the physical operators. As usual, these expressions are not
all linearly independent, because the $A^\ii_k(t)$ are linear combinations
of the $(d-2){d+p \choose d}$ modes $A_{\mu,M}(t)$, where we generalized
to $d$ dimensions. Therefore, the
cohomology is generated by $A^\ii_k(t)$, where $k$, $k^2 = 0$, belongs to
an index set $\KK$. The $k\in\KK$ must be linearly independent, but
$\KK$ is otherwise arbitrary. Since $\dim\KK = r_{p,d}$, the number of
independent modes equals $(d-2)r_{p,d}$.

We next consider the momenta corresponding to the jets in (\ref{dltAm}).
The BRST charge acts as
\bes
\dlt b_M(t) &=& \sum_\mu \hm E^\mu_{M-\mu}(t) \equiv \div E^\mu_M(t), \nl
\dlt E^\mu_M(t) &=& \sum_\nu i
(\hn E^{*\mu}_{M-2\nu}(t) - E^{*\nu}_{M-\mu-\nu}(t))
\equiv i F^\mn_{\nu,M}(E^*(t)), \nl
\dlt E^{*\mu}_M(t) &=& \xi_{M-\mu}(t), \\
\dlt \xi_M(t) &=& 0.
\eens
Introducing the linear combinations
\bes
b_k(t) = \sum_M^{p+1} (-ik)^M b_M(t), &&
E^{*\mu}_k(t) = \sum_M^{p-2} (-ik)^M E^{*\mu}_M(t), \nl
E^\mu_k(t) = \sum_M^{p} (-ik)^M E^\mu_M(t), &&
\xi_k(t) = \sum_M^{p-3} (-ik)^M \xi_M(t),
\ees
we obtain
\bes
\dlt b_k(t) &=& -ik_\mu E^\mu_k(t), \nl
\dlt E^\mu_k(t) &=& i(k^2\dlt^\mu_\nu - k^\mu k_\nu) E^{*\nu}_k(t), \nl
\dlt E^{*\mu}_k(t) &=& -i k^\mu \xi_k(t), \\
\dlt \xi_k(t) &=& 0.
\eens
Again, this is recognized as (\ref{dlt2u1}), and the cohomology is
generated by the transverse electric field
\be
E^\ii_k(t) = \eps^\ii_\mu(k, \dot q(t)) E^\mu_k(t), \quad
\hbox{where $k\in\KK$, $i = 1,2$}.
\ee

We now turn to the barred antijets. The BRST operator (\ref{Qmaxwell})
acts like
\bes
\dlt\bar c_M(t) &=& -iDc_M(t), \nl
\dlt\bar A_{\mu,M}(t) &=& DA_{\mu,M}(t) -\bar c_{M+\mu}(t), \\
\dlt\bar A^*_{\mu,M}(t) &=& -i(DA^*_{\mu,M}(t)
-\sum_\nu \hn F_{\nu\mu,M+\nu}(\bar A(t))), \nl
\dlt\bar \zeta_M(t) &=& D\zeta_M(t) - \sum_\mu \hm\bar A^*_{\mu,M+\mu}(t).
\eens
This is completely analogous to how the time constraint eliminates the
$t$ dependence for the free scalar field. Morally,
the barred antifields implement the constraint $DA_{\mu,M}(t) \approx 0$,
i.e. $\dot A_{\mu,M}(t) \approx \sum_\nu \dot q^\nu(t) A_{\mu,M}(t)$ for
the physical components of the $A$-field. In order to implement this
constraint consistently, we also need to impose constraints on the
other unbarred antijets.
Analogously, the barred antijet momenta are eliminated through
\bes
\dlt\bar b_M(t) &=& \sum_\mu \hm \bar E^\mu_{M-\mu}(t), \nl
\dlt\bar E^\mu_M(t) &=& i\sum_\nu (\hn
\bar E^{*\mu}_{M-2\nu}(t) - \bar E^{*\nu}_{M-\mu-\nu}(t)), \nl
\dlt\bar E^{*\mu}_M(t) &=& \bar \xi_{M-\mu}(t), \\
\dlt\bar \xi_M(t) &=& 0.
\eens

So far, we have calculated the cohomology
$H^\bullet_\cl(Q_A + Q_D + Q_S + Q_T + Q_{GT} + Q_{DT} + Q_{ST})$,
and found that $H^0$ is generated by transverse
modes $E^\ii_k$ and $A^\ii_k$, where $i = 1,2$ and $k\in\KK$.
To construct the $p$-jet approximation to the true physical phase
space $J^p\PP$, we finally need to identify the velocities 
$\sum_\nu \dot q^\nu(t) F^\mn_k(t)$, where
\be
F^\mn_k(t) = \sum_M^{p} (ik)^{-M} F^\mn_M(t),
\ee
with the momenta
$E^\mu_k(t)$. This task is accomplished by the momentum constraint
(\ref{MAik}), exactly as for the free scalar field. $C(J^p\PP)$ is
hence generated by $p$-jet approximations to transverse photons,
\be
a^\ii(k) \propto \sum_\mu\int dt\ \eps^\ii_\mu(k,\dot q(t))
(E^\mu_k(t) - \sum_\nu\dot q_\nu(t) F^\mn_k(t)),
\ee
for $i=1,2$ and $k\in\KK$.

\subsection{ Hamiltonian and quantization}

The physical Hamiltonian in $\HP^*$ is given by $\H_A = \H^0_A + \H^1_A$,
where
\bes
\H^0_A &=&  \int dt\ \bigg(
- i \sum_M^{p+1} \dot c_M(t) b_M(t)
+ \sum_M^p \dot A_{\mu,M}(t) E^\mu_M(t) \nl
&&-\  i \sum_M^{p-2}\dot A^*_{\mu,M}(t) E^{*\mu}_M(t)
+ \sum_M^{p-3} \dot \zeta_M(t) \xi_M(t) \\
&&+\  \sum_M^p \ddt \bar c_M(t) \bar b_M(t)
-i\sum_M^{p-1} \ddt \bar A_{\mu,M}(t) \bar E^\mu_M(t) \nl
&&+\  \sum_M^{p-3} \ddt \bar A^*_{\mu,M}(t) \bar E^{*\mu}_M(t)
-i \sum_M^{p-4} \ddt \bar \zeta_M(t) \bar \xi_M(t) \bigg).
\eens
\bes
\H^1_A &=& \int dt\ \sum_\nu q_*^\nu(t) \bigg(
- i \sum_M^{p+1} c_{M+\nu}(t) \bar b_M(t)
+ \sum_M^p \int dt\ A_{\mu,M+\nu}(t) \bar E^\mu_M(t) \nl
&&-\  i \sum_M^{p-2} A^*_{\mu,M+\nu}(t) \bar E^{*\mu}_M(t)
+ \sum_M^{p-3} \zeta_{M+\nu}(t) \bar \xi_M(t) \bigg).
\ees
Unlike the total Hamiltonian $\H_\TOT = \H^0_A + \H_\phi$, which
translates both the fields and the observer's trajectory in time, the
physical Hamiltonian translates the fields relative to the observer.
We have e.g.
$[H_A, A_{\mu,k}(t)] = -i\dot A_{\mu,k}(t)$ and
$[H_A, E^\mu_k(t)] = -i\dot E^\mu_k(t)$, and hence
\be
[H_A,a^\ii(k)] =
k_\mu\int dt\ \dot q^\mu(t) \e^{ik\cdot q(t)} a^\ii_k(t)
= k_\mu u^\mu a^\ii(k),
\ee
where we used that the solution to the observer's dynamics constraint is
$q^\mu(t) \approx u^\mu t + s^\mu$. The physical Hamiltonian thus picks out
the projection of $k$ along the observer's trajectory, which is the
physical Hamiltonian of a photon with wave-vector $k$.

To quantize the theory, we introduce a vacuum state $\ket0$,
annihilated by all negative frequency modes in $J^p\HP^*$. Under the
ansatz (\ref{assumption}), this means that all modes with negative $m$
are assumed to annihilate the vacuum. The state cohomology
$J^p\HH = H_\st^0(Q)$ is then a $p$-jet approximation to the photon
Fock space.

\subsection{ Yang-Mills theory and gauge anomalies}
\label{ssec:anomaly}

As noted in subsection \ref{ssec:maxwell} above, the gauge generators
\be
\J_X = \sum_M^p \int dt\ \sum_\mu X_{M+\mu}(q(t)) E^\mu_M(t)
\label{JXabel}
\ee
satisfy the abelian algebra of gauge transformations, $[\J_X,\J_Y]=0$,
which is identified with the algebra $\map(d,\uu(1))$ of maps
from $d$-dimensional
space to $\uu(1)$. In the free Maxwell case, there are no contributions
to $\J_X$ from the antijets, because they all transform trivially
under $\uu(1)$. However, interacting theories can be treated
with similar methods. As sketched in the next section, interactions
bring in some new problems, but the key phenomena can be understood
by analogy with the Maxwell field.

As the simplest example of an interacting gauge theory, consider a
pure non-abelian gauge theory with gauge symmetry $\map(d, \g)$, where
$\g$ is some finite-dimensional Lie algebra. All
fields and antijets transform in the adjoint representation, except
$A^a_{\mu,M}$ which transforms as a connection; $a$ is a $\g$
index. There are now several contributions to the gauge generators,
\be
\J_X = \J_X^G + \J_X^A + \J_X^D + \J_X^S + \J_X^{GT} + \J_X^{AT}
+ \J_X^{DT} + \J_X^{ST},
\ee
where each term is a bilinear in a field jet and the corresponding
canonical momentum, according to the following table:
\[
\barr{cccccccc}
\J_X^G & \J_X^A & \J_X^D & \J_X^S &
\J_X^{GT} & \J_X^{AT} & \J_X^{DT} & \J_X^{ST} \\
c^a_M & A^a_{\mu,M} & A^{*a}_{\mu,M} & \zeta^a_M &
\bar c^a_M & \bar A^a_{\mu,M} & \bar A^{*a}_{\mu,M} & \bar\zeta^a_M\\
\earr
\]
E.g., the part acting on the antijet $A^{*a}_{\mu,M}(t)$ is
\be
\J_X^D = \sum_M^{p-2} if^{abc} \int dt\
X^a(q(t)) \no{ A^{*b}_{\mu,M}(t) E^{*c\mu}_M(t) },
\label{JXD}
\ee
where $f^{abc}$ are the totally antisymmetric structure constants of
$\g$ and double dots indicate normal ordering w.r.t. frequency.

It was shown in \cite{Lar98} that such a bilinear combination acquires
an anomaly, which turns $\map(d,\g)$ into a higher-dimensional
generalization of the affine Kac-Moody algebra $\hat\g$. Similarly,
in a general-covariant theory, the diffeomorphism generators $\L_\xi$,
where $\xi = \xi^\mu(x)\d_\mu$,
satisfy a multi-dimensional generalization of the Virasoro algebra, i.e.
an abelian but non-central extension of the algebra of vector fields in
$d$ dimensions, $\vect(d)$.
The abelian extension of $\vect(d)\ltimes\map(d,\g)$
is defined by the brackets
\bes
[\L_\xi,\L_\eta] &=& \L_{[\xi,\eta]}
+ {1\/2\pi i}\int dt\ \dot q^\rho(t)
\Big\{ c_1 \d_\rho\d_\nu\xi^\mu(q(t))\d_\mu\eta^\nu(q(t)) +\nl
&&+\ c_2 \d_\rho\d_\mu\xi^\mu(q(t))\d_\nu\eta^\nu(q(t)) \Big\}, \nl
{[}\L_\xi, \J_X] &=& \J_{\xi X},
\label{VirKM}\\
{[}\J_X, \J_Y] &=& \J_{[X,Y]} - {k\/2\pi i}\dlt^{ab}
\int dt\ \dot q^\rho(t)\d_\rho X_a(q(t))Y_b(q(t)), \nl
{[}\L_\xi,q^\mu(t)] &=& \xi^\mu(q(t)), \nl
{[}\J_X, q^\mu(t)] &=& {[}q^\mu(t), q^\nu(t')] = 0,
\eens
To see that this Lie algebra generalizes the Virasoro and affine algebras
to several dimensions, define for each $m = (m_\mu) \in \ZZ^d$:
$L_\mu(m) = \L_\xi$ for $\xi^\nu(x) = -i\exp(im\cdot x)\dlt^\nu_\mu$,
$J^a(m) = \J_X$ for $X^a(x) = \exp(im\cdot x)J^a$, and
$S^\mu(m) = {1\/2\pi i}\int dt\ \dot q^\mu(t) \exp(im\cdot x)$.
The algebra (\ref{VirKM}) then acquires the form
\bes
[L_\mu(m), L_\nu(n)] &=& n_\nu L_\mu(m+n) - m_\mu L_\nu(m+n) +\nl
&&-\ ( c_1 m_\nu n_\mu + c_2 m_\mu n_\nu)
m_\rho S^\rho(m+n), \nl
{[}L_\mu(m), J^a(n)] &=& n_\mu J^a(m+n),
\label{VirKMTor}\\
{[}J^a(m), J^b(n)] &=& if^{abc} J^c(m+n) - k m_\mu S^\mu(m+n), \nl
m_\mu S^\mu(m) &\equiv& 0, \qquad \hbox{for all}\ m.
\eens
The last equation expresses that the integral of a total derivative 
vanishes.
Since the equation $m S(m) = 0$ implies that $S(m)$ is proportional to
a Kronecker delta, (\ref{VirKMTor}) reduces to $Vir\ltimes\hat\g$ when
$d=1$.

The parameters $c_1 - c_4$ and $k$ were also computed in that paper.
In particular, the gauge anomaly for a $p$-jet is given by
\be
k = \pm y {d+p \choose d},
\ee
where $y$ is value of the second Casimir operator (here in the adjoint
representation), $d$ is the number of dimensions, $p$ is the truncation
order, and the sign depends on Grassmann parity.
If we now truncate the jets as indicated in the definition of the
BRST operator (\ref{Qmaxwell}), i.e.
\[
\barr{l|cccccccc}
\hbox{Jet} & c^a_M & A^a_{\mu,M} & A^{*a}_{\mu,M} & \zeta^a_M &
\bar c^a_M & \bar A^a_{\mu,M} & \bar A^{*a}_{\mu,M} & \bar\zeta^a_M\\
\hline
\hbox{Order} & p+1 & p & p-2 & p-3 & p & p-1 & p-3 & p-4 \\
\hbox{Components} & 1 & d & d & 1 & 1 & d & d & 1 \\
\hbox{Parity} & F & B & F & B & B & F & B & F \\
\earr
\]
the total abelian charge becomes
\bes
k(p) &=& y\bigg( - {d+p+1 \choose d} + d{d+p \choose d}
- d{d+p-2 \choose d} + {d+p-3 \choose d} \nl
&+& {d+p \choose d} - d{d+p-1 \choose d}
+ d{d+p-3 \choose d} - {d+p-4 \choose d} \bigg) 
\label{kp}\\
&=& y\bigg( - {d+p \choose d-1} + d{d+p-1 \choose d-1}
- d{d+p-3 \choose d-1} + {d+p-4 \choose d-1} \bigg).
\eens
This result result immediately leads to two observations. First, and
unlike the abelian case, we can not factor out the gauge transformations
due to the anomaly. Second, $k(p) \to \infty$ when $p \to \infty$,
which indicates that the field theory limit is problematic.

Since the gauge algebra is anomalous, we can not impose the gauge
symmetry as a constraint, and thus it would appear that the $Q_G$
and $Q_{GT}$ terms must be discarded from the BRST operator. This
would be disturbing, because we want QJT to correspond to
QFT in the limit $p\to\infty$, and in QFT there is no anomaly. However,
it might be possible to save part of the gauge invariance. Truncate the
sums in $\J_X^G$, $\J_X^A$, $\J_X^D$ and $\J_X^S$ at $|M|\leq p-3$, and
the  sums in $\J_X^{GT}$, $\J_X^{AT}$, $\J_X^{DT}$ and $\J_X^{ST}$
at $|M|\leq p-4$. Hence we replace e.g. (\ref{JXD}) by
\be
\J_X^{D(p-3)} = \sum_M^{p-3} if^{abc} \int dt\
X^a(q(t)) \no{ A^{*b}_{\mu,M}(t) E^{*c\mu}_M(t) }.
\ee
Then we see that the modified gauge generators
\bes
\J_X^{(p-3)} &=&
\J_X^{G(p-3)} + \J_X^{A(p-3)} + \J_X^{D(p-3)} + \J_X^{S(p-3)} \\
&&+\ \J_X^{GT(p-4)} + \J_X^{AT(p-4)} + \J_X^{DT(p-4)} + \J_X^{ST(p-4)}
\eens
do indeed satisfy $\map(d, \g)$ without an anomaly, because the
abelian charge is now
\bes
k(p) &=& y\bigg( (-1 + d - d + 1) {d+p-3 \choose d}
+ (1 - d + d - 1) {d+p-4 \choose d} \bigg) \nl
&=& 0.
\ees
The $\map(d,\uu(1))$ symmetry acts on full history phase space
$J^p\HP$, but it acts as a gauge symmetry only on $J^{p-3}\HP$; 
the action on the space of jet components with $p-2\leq |M|\leq p$
is anomalous. The origin of this anomaly is that we make a regularization,
which can not preserve all properties of the unregularized theory.
In this case, the regularization consists of truncating the jets; we
can identify $\HP \cong J^\infty\HP$, but replacing $J^\infty\HP$ with
$J^p\HP$ amounts to a regularization. In this process, some properties
of the original theory is lost; e.g., the jet components with
$p-1\leq|M|\leq p$ are not subject to a dynamics constraint.

Thus we can not implement the full symmetry $\J_X$ as a constraint,
but only the truncated gauge symmetry $\J_X^{(p-3)}$. Returning to the
abelian case, we should hence replace $Q_A$ and $Q_{AT}$ in
(\ref{Qmaxwell}) by
\bes
Q_A &=& \sum_M^{p-3} \int dt\
c_{M+\mu}(t) E^\mu_M(t), \nle
Q_{AT} &=& \sum_M^{p-4} \int dt\ ( DA_{\mu,M}(t)
-\bar c_{M+\mu}(t) ) \bar E^\mu_M(t).
\eens
However, since the gauge field transforms as a connection, we need
the ghost jets $c_M$ and $\bar c_M$ up to one order higher, i.e.
for $|M|\leq p-2$ and $|M|\leq p-3$, respectively; cf. the appearence
of $c_{M+\mu}$ above. Finally, we must modify one
more term in the BRST operator, namely
\be
Q_{GT} &=& \sum_M^{p-2} \int dt\ Dc_M(t) \bar b_M(t).
\ee
This means that the total anomaly (\ref{kp}) is replaced by
\bes
&& k(p) = y\bigg( - {d+p-2 \choose d} + d{d+p \choose d}
- d{d+p-2 \choose d} + {d+p-3 \choose d} \nl
&&\ +\ {d+p-3 \choose d} - d{d+p-1 \choose d}
+ d{d+p-3 \choose d} - {d+p-4 \choose d} \bigg) \nl
&&\ =  y\bigg( - {d+p-3 \choose d-1} + d{d+p-1 \choose d-1}
- d{d+p-3 \choose d-1} + {d+p-4 \choose d-1} \bigg) \nl
&&\ =  y\bigg( d{d+p-1 \choose d-1}
- d{d+p-3 \choose d-1} - {d+p-4 \choose d-2} \bigg).
\ees
This expression is still nonzero, and diverges with $p$ whenever
$d > 1$. However, the last term only diverges when $d > 2$, which is
a clear improvement. Let us study the relevant terms more closely.
If we truncate the ghost $c_M$ at order $p+1$, the contributions from
$c_M$, $\zeta_M$ and their barred antijets sum up to
\be
- {d+p \choose d-1} + {d+p-4 \choose d-1},
\ee
which is infinite for $d > 1$. However, if we instead truncate the
ghost at order $p-2$, the relevant sum becomes
\be
- {d+p-3 \choose d-1} + {d+p-4 \choose d-1} = - {d+p-4 \choose d-2},
\ee
which is finite also when $d=2$.

Of course, we don't want to avoid infinite anomalies only when $d \leq 2$,
but also in the physical case $d = 4$. It was noticed in \cite{Lar02}
that we can arrange the field content to cancel the infinite parts of
the anomalies, and this procedure naturally singles out $d = 4$
dimensions. This follows from the properties of the binomial coefficients
and the fact that we must introduce jets ranging from order $p$
($A_{\mu,M}$) to order $p-4$ ($\bar\zeta_M$).
Compared to the calculation of $k(p)$ in this subsection,
we also need to add fermions to cancel more infinities.

This prescription produced a qualitatively
correct field content, but there was a glaring discrepancy: fermionic
antijets truncated at order $p-2$ were also needed, which at the
time was interpreted as some kind of supersymmetry. It was
then observed in \cite{Lar05a} that the right kind of cancellation
would occur, if we reinterpreted these antijets as ghosts. However,
why the ghosts should be truncated at order $p-2$ rather than at order
$p+1$ remained a mystery, until now: the ghosts should be truncated at
order $p-2$ because the gauge symmetry is truncated at order $p-3$.

We emphasize that the anomalies in (\ref{VirKM}) is a distinguishing
feature of QJT. It is clear that the relevant cocycles are functionals
of the observer's trajectory $q^\mu(t)$. Hence these anomalies can not
arise in QFT, where the observer has not been explicitly introduced.

\section{ Interactions }
\label{sec:interactions}

\subsection{ General structure }
\label{ssec:general}

The detailed study of interacting theories within the present framework
is postponed to another paper. However, we briefly sketch what
modifications are involved. In this subsection, we use the standard
abbreviated notation, where a single generalized index $\al$ can stands
either for a discrete index, a continuous time variable $t$, or a
continuous spacetime coordinate $x$, or a combination of all.
Contraction of generalized indices involves summation over repeated
discrete indices and integration over repeated continuous variables.

Assume that we have a dynamical system described by
some bosonic degrees of freedom $\fa$ and an action functional $S$. We
introduce history momenta $\pa$, subject to the nonzero Poisson brackets
\be
[\fa, \pb] = i\dlt^\al_\bt.
\label{fapb}
\ee
The dynamics constraint reads
\be
\Ea = \da S \equiv {\dlt S\/\dlt \fa} = i[\pa, S] \approx 0.
\label{Ea}
\ee
The second derivative of the action is symmetric:
\be
\da\Eb = \db\Ea = {\dlt^2 S\/\dlt \fa \dlt \fb}.
\label{daEb}
\ee
We assume that there exist functions $\ua_\w = \ua_\w(\phi)$, such
that\footnote{ Henneaux and Teitelboim do not make this assumption in
chapter 17 of \cite{HT92}, eq. (17.4). Since they only consider fields
and antifields and not their conjugate momenta, this does not affect
$H^0_\cl(Q) = C(\PP) = C(\HP^*)/\II$, but it makes the higher cohomology
groups nonempty. Note that the redundancies (\ref{uE}) are present
already for the harmonic oscillator (\ref{Ekharm})
($\EE_\w = \EE_{-\w} \equiv 0$).}
\be
\ua_\w \Ea &\equiv& 0.
\label{uE}
\ee
The index $\w$ labels solutions to the Euler-Lagrange equations.
The {\em solution set} $\W$ is the set of such labels $\w \in \W$.
Such functions $\ua_\w$ clearly exist if $\Ea = \Ea^0 = L_\ab\fb$ is
linear, and if $\Ea = \Ea^0 + \vareps\Ea^1$, we can
prove the existence of such functions to all orders in $\vareps$. One
makes the ansatz
\be
\ua_\w = \sum_m \ua_{\w\mm} \vareps^m,
\ee
and constructs the functions $\ua_{\w\mm}$ order by order in
perturbation theory.

Physically, (\ref{uE}) simply expresses that the
Euler-Lagrange equations have solutions. Namely, in a cohomological
formulation, we start from an equal number of field and antifield histories,
$\fa$ and $\fsa$. Hence the net dimension $\dim H^\bullet_\cl(Q) = 0$,
where fermionic degrees of freedom count negative. However, the
cohomology should be a resolution of $C(\PP)$, i.e. $H^0_\cl(Q) = C(\PP)$
and $H^n_\cl(Q) = 0$, $n \neq 0$. Unless the physical phase space is empty,
this implies $\dim H^\bullet_\cl(Q) = \dim H^0_\cl(Q) > 0$, a contradiction.
This paradox is resolved by introducing further bosonic antifields
corresponding to the redundancies (\ref{uE}), giving a net surplus
of bosonic degrees of freedom. Of course, if the number of fields
exceed the number of antifields in $\HP^*$, as is the case in QJT,
the redundancies (\ref{uE}) may be absent.

Define
\be
\pi_\w = \ua_\w \pa.
\label{pw}
\ee
It follows from (\ref{Ea}) and (\ref{uE}) that all $\pi_\w$ commute
with the action, $[\pi_\w, S] = 0$. By the Jacobi identity,
$[[\pi_\w, \pi_\ww], S] = 0$. By assumption, it must therefore be
possible to expand $[\pi_\w, \pi_\ww]$ in terms of $\pi_\ups \in \W$:
\be
[\pi_\w, \pi_\ww] = if_{\w\ww}{}^\ups \pi_\ups,
\label{pipi}
\ee
whereas explicit calculation yields
\be
f_{\w\ww}{}^\ups \ua_\ups &=& \ub_\ww\db\ua_\w - \ub_\w\db\ua_\ww.
\label{fuu}
\ee
The brackets (\ref{pipi}) define the {\em momentum algebra}
$\m$. Note that the structure functions $\ua_\w(\phi)$ and
$f_{\w\ww}{}^\ups(\phi)$ depend on the field $\fa$ in general. The Jacobi
identities yield
\bes
&&f_{\w\ww}{}^\varsi f_{\ups\varsi}{}^\la  +
f_{\ww\ups}{}^\varsi f_{\w\varsi}{}^\la  +
f_{\ups\w}{}^\varsi f_{\ww\varsi}{}^\la =
\nlb{Jacobi}
&&\ua_\w\da f_{\ww\ups}{}^\la +
\ua_\ww\da f_{\ups\w}{}^\la +
\ua_\ups\da f_{\w\w}{}^\la.
\eens

We now construct the BRST operator which eliminates the dynamics
constraint (\ref{Ea}) and the antifield constraint implied by
(\ref{uE}). Introduce antifields and antifield momenta according to
the following table
\be
\barr{c|c|c|l}
\hbox{Field} & \hbox{Momentum} & \hbox{Parity} & \hbox{Constraint} \\
\hline
\fa & \pa & B & \\
\fsa & \psa & F & \Ea \\
\theta_\w & \chi^\w & B & \ua_\w \fsa \\
\earr
\ee
The nonzero Poisson brackets are, in addition to (\ref{fapb}),
\be
\{\fsa, \psb\} = \dlt^\bt_\al, \qquad
[\theta_\w, \chi^\ww] = i\dlt^\ww_\w.
\label{fsapsb}
\ee
The BRST charge reads $Q = Q_D + Q_A$, where
\be
Q_D = \Ea \psa, \qquad
Q_A = \ua_\w \fsa \chi^\w.
\label{QDA}
\ee
$Q$ acts on $C(\HP^*)$ as
\bes
\dlt \fa = 0, &&
\dlt \pa = i\da\Eb\psb + i\da\ub_\w \fsb\chi^\w, \nl
\dlt \fsa = \Ea, &&
\dlt \psa = \ua_\w \chi^\w, \\
\dlt \theta_\w = -i\ua_\w \fsa, &&
\dlt \chi^\w = 0.
\eens
By construction, the field part of the cohomology equals
$H^0_\cl(Q) = C(\phi)/\II$, where $C(\phi)$ is the space of functions over 
$\fa$ and $\II$ is the ideal generated by $\Ea$, and
$H^n_\cl(Q) = 0$ if $n\neq 0$.
In view of (\ref{uE}), this space is generated by independent elements
$\fw(\phi)$ labelled by $\w\in\W$, where
\be
\fa = \fw \ua_\w.
\ee
Let us now consider the momentum part.
$\Ea\psa = \dlt( \fsa\psa - i \theta_\w\chi^\w)$
and $\chi^\w$ belong both to the kernel and to the image,
and thus vanish in cohomology. We also note that
$\dlt \pa = i\da \dlt(\fsb\psb)$.
Define
\be
\tpi_\w = \ua_\w \pa + i\da\ub_\w\fsb\psa
- f_{\w\ww}{}^\ups \theta_\ups\chi^\ww.
\label{tpiw}
\ee
Using (\ref{daEb}), (\ref{uE}), (\ref{fuu}) and (\ref{Jacobi}), it
is straightforward to verify that $\tpi_\w$ is BRST closed,
$\dlt\tpi_\w = 0$.
The space of all $\tpi_\w$, $\w\in\W$, is not closed under the
bracket in $\HP^*$. However, it is closed up to a BRST exact term:
\be
[\tpi_\w, \tpi_\ww] = if_{\w\ww}{}^\ups \tpi_\ups
+ \dlt G_{\w\ww},
\label{tpitpi}
\ee
where
\be
G_{\w\ww} = -i\da f_{\w\ww}{}^\ups \theta_\ups \psa.
\ee
When restricted to $H^\bullet_\cl(Q_D+Q_A)$, the $\tpi_\w$ define a proper
Lie algebra isomorphic to the momentum algebra $\m$ (\ref{pipi})
generated by $\pi_\w$.

So far, we have found that $H^0_\cl(Q_D+Q_A)$ is generated by $\fw$
and $\tpi_\w$. The implementation of the momentum constraint is
complicated by the fact that the momentum algebra $\m$ is no longer
abelian. Therefore, we need to pass from the variables $(\fw, \tpi_\w)$
to new variables $(q^\w, p_\w)$, in which the momentum constraint
is abelian. 
For simplicity, we first assume that the structure constants
$f_{\w\ww}{}^\ups$ do not depend on $\phi$. The bracket between
$\fw$ and $\tpi_\ww$ must obey the Jacobi identities, which leads to
\be
[\fw, \tpi_\ww]	= -if_{\ups\ww}{}^\w \phi^\ups + c^\w_\ww,
\label{fp}
\ee
where the structure constants $c^\w_\ww$ obeys the cocycle condition
\be
f_{\w\varsi}{}^\ups c^\varsi_\ww - f_{\ww\varsi}{}^\ups c^\varsi_\w
= f_{\w\ww}{}^\varsi c^\ups_\varsi.
\ee
Moreover, $[\fw, \phi^\ww] = 0$ because
$\fw$ depends on the fields $\fa$ only.	The bracket (\ref{fp}) thus
defines an abelian extension of the coadjoint representation of the
momentum algebra $\m$.
Define new variables $(q^\w, p_\w)$, with Poisson brackets
\bes
[q^\w, p_\ww] &=& i\dlt^\w_\ww, \nle
[q^\w, q^\ww] &=& [p_\w, p_\ww]  =  0,
\eens
by
\bes
\tpi_\w &=& f_{\w\ww}{}^\ups q^\ww p_\ups - i p_\ww c^\ww_\w,
\nlb{abel1}
\fw &=& q^\w.
\eens
It is straightforward to verify that these expressions satisfy
(\ref{pipi}) and (\ref{fp}).

More generally, assume that there are functions $v^\ww_\w(\phi)$,
which depend on $\fw$ with $\w \in \W$ only (and not on general
histories $\fa$), satisfying
\be
f_{\w\ww}{}^\ups v^\varrho_\ups &=&
v^\varsi_\ww\d_\varsi v^\varrho_\w
- v^\varsi_\w\d_\varsi v^\varrho_\ww.
\label{fvv}
\ee
This relation is essentially the same as (\ref{fuu}), except that
it only depends on elements in the solution set $\W$.
Then we replace (\ref{abel1}) by
\bes
\tpi_\w &=& v^\ww_\w(q) p_\ww,
\nlb{abel2}
\fw &=& q^\w.
\eens
One checks that the bracket $[\tpi_\w, \tpi_\ww]$ is still given by
the momentum algebra (\ref{pipi}), and that the $\fw$ transform as
\be
[\fw, \tpi_\ww] = i v^\w_\ww(\phi).
\ee

We now invert the coordinate transformations (\ref{abel1}) or
(\ref{abel2}), and obtain a map $(\fw, \pw) \to (q^\w, p_\w)$.
Introduce a set of
antisymmetric constants $c_{\w\ww} = -c_{\ww\w}$, $\w, \ww \in \W$;
$c_{\w\ww}q^\w$ can be thought of as the velocity.
Define the momentum constraint by
\be
\MM_\w = p_\w - c_{\w\ww} q^\ww.
\label{Mpq}
\ee
It satisfies the brackets
\be
[\MM_\w, \MM_\ww] = -2ic_{\w\ww}.
\ee
Introduce new antifields $\al_\w$ satisfying the same algebra with
opposite sign:
\be
[\al_\w, \al_\ww] = 2ic_{\w\ww}.
\ee
Further introduce fermionic antifields $\bt_\w$ with canonical momentum
$\gm^\w$; $\{\bt_\w, \gm^\ww\} = \dlt^\ww_\w$. The momentum part of the
BRST operator is defined by
\be
\QM = (p_\w - c_{\w\ww} q^\ww + \al_\w)\gm^\w.
\ee
It acts on $H^\bullet_\cl(Q_D+Q_A)$ as
\bes
\dlt q^\w = -i\gm^\w ,&&
\dlt p_\w = i c_{\w\ww} \gm^\ww, \nl
\dlt \bt_\w = \MM_\w + \al_\w, &&
\dlt \gm^\w = 0, \\
\dlt \al_\w = -2ic_{\w\ww}\gm^\ww.
\eens
The kernel is generated by $p_\w + c_{\w\ww} q^\ww$,
$p_\w - c_{\w\ww} q^\ww + \al_\w$, and $\gm^\w$, and the image by
$\MM_\w+\al_\w$ and $\gm^\w$. The cohomology is thus generated by
\be
a_\w = 	p_\w + c_{\w\ww} q^\ww + x^\ww_\w(\MM_\ww+\al_\ww),
\ee
for arbitrary constants $x^\ww_\w$. It is easily verified that the
bracket
\be
[a_\w, a_\ww] = 2ic_{\w\ww}
\label{osc}
\ee
is independent of these constants. Hence the physical phase space
$\PP = C(a_\w)$ is identified with the space of functions over $a_\w$.

It might appear that we have transformed the original interacting theory
to a free theory defined by the oscillators (\ref{osc}), but this is not
the case. Since the Hamiltonian is the generator of rigid time
translations, it has a very simple form in $\HP$:
\be
\H(\fa,\pa) = h^\bt_\al \fa \pb,
\label{HHP}
\ee
where $h^\bt_\al$ is a constant matrix. That the Hamiltonian has this
simple bilinear form is a major advantage of a history formulation.
However, when expressed in terms of the cohomology variables, the
Hamiltonian
\be
\H = \H(\fw, \pw) = \H(q^\w,p_\w) = \H(a_\w)
\ee
becomes a complicated, nonlinear function. Hence the theory is not a
free theory; we have merely expressed the bracket in Darboux variables.

\subsection{Gauge transformations}

In the previous subsection, we assumed that all degrees of freedom were
physical. In a gauge theory, some degrees of freedom are redundant,
leading to identities of the form
\be
\ra_a \Ea \equiv 0,
\label{rE}
\ee
where $\ra_a = \ra_a(\phi)$ may depend on the fields $\fa$.
Equation (\ref{rE}) can be written as $[J_a, S] = 0$, where the gauge
generators are
\be
J_a = \ra_a \pa.
\label{Ja}
\ee
There is a striking similarity between the redundancy related to the
existence of solutions (\ref{uE}) and the gauge redundancy (\ref{rE}),
and between the the momentum generators $\pi_\w$ in (\ref{pw}) and
the gauge generators $J_a$ in (\ref{Ja}). In the present abbreviated
formalism, there appears to be no difference at all, apart from a
change in lettering. However, in the fuller formalism of the next
subsection, where we explicitly keep track of the time variable, there is
a difference: the momentum generators are time independent, but the gauge
generators depend on time. This amounts to replacing
$\al \to (\al, t)$, $a \to (a, t)$, but $\w \to \w$, in all formulas.
Since a well-posed Cauchy problem can not depend on arbitrary functions
of time, the gauge degrees of freedom should be eliminated, at least
classically.

Both gauge and momentum generators preserve the action,
$[J_a, S] = [\pi_\w, S] = 0$. If the $J_a$ and $\pi_\w$ exhaust the
linear functions of $\pa$ with this property, it must be possible to
expand the brackets as
\bes
[\pi_\w, \pi_\ww] &=& if_{\w\ww}{}^\ups \pi_\ups + if_{\w\ww}{}^a J_a, \nl
{[}\pi_\w, J_a] &=& if_{\w a}{}^\ww \pi_\ww + if_{\w a}{}^b J_b, \\
{[}J_a, J_b] &=& if_{ab}{}^\w \pi_\w + if_{ab}{}^c J_c.
\eens
Now we make the simplifying assumption that this algebra is in fact a
direct sum $\m \oplus \g$ of the momentum algebra $\m$ and the gauge
algebra $\g$, i.e. $f_{\w\ww}{}^a = f_{\w a}{}^\ww = f_{\w a}{}^b
= f_{ab}{}^\w = 0$. This means that the gauge generators $J_a$ are closed
under the bracket:
\be
[J_a, J_b] = if_{ab}{}^c J_c,
\label{JJ}
\ee
where
\be
f_{ab}{}^c \ra_c = \rb_b \db \ra_a - \rb_a \db \ra_b.
\ee
The Jacobi identities imply
\bes
f_{ab}{}^d f_{cd}{}^e +
f_{bc}{}^d f_{ad}{}^e +
f_{ca}{}^d f_{bd}{}^e =
\ra_a \da f_{bc}{}^e +
\ra_b \da f_{ca}{}^e +
\ra_c \da f_{ab}{}^e.
\eens
Moreover, from $f_{a\w}{}^b \ra_b = f_{a\w}{}^\ww \ua_\ww = 0$ follows
the identity
\be
\ub_\w \db \ra_a - \rb_a \db \ua_\w = 0,
\label{ur}
\ee
which will be used extensively below.

The algebra (\ref{JJ}) is an open algebra, where the structure functions
$f_{ab}{}^c(\phi)$ depend on the field $\fa$. Since our purpose is to
give a rapid sketch on how gauge symmetries fit into the picture, rather
than dealing with the most general case, we make the simplifying
assumption that the $f_{ab}{}^c$ are in fact $\phi$-independent constants;
this amounts to
\be
\da f_{ab}{}^c = 0.
\ee

A profound difference between the momentum redundancy (\ref{uE}) and the
gauge redundancy (\ref{rE}) is that the former is used to eliminate half
of the variables, because the momentum constraint identifies momenta and
velocities, whereas the gauge variables are truly redundant and should
be eliminated altogether. The antifields $\theta_\w$ do not affect
$H^0_\cl(Q) = C(\HP)/\II$, where $\II$ is the ideal generated by the dynamics,
but only ensure that the higher cohomology groups
vanish. In contrast, a gauge symmetry should change the physical phase
space to $H^0_\cl(Q) = C'(\HP)/\II$, where $C'(\HP)$ is the space of
gauge-invariant functions over $\HP$.

To enforce gauge invariance in cohomology, we introduce fermionic ghosts
$c^a$ with canonical momenta $b_a$, as well as bosonic second-order
antifields $\zeta_a$ with momenta $\xi^a$; the nonzero Poisson brackets
read
\be
\{c^a, b_b\} = \dlt^a_b, \qquad
[\zeta_a, \xi^b] = i\dlt^b_a.
\ee
The total gauge generator is $J^\TOT_a = J^\phi_a + J^\gh_a$, where
\bes
J^\phi_a &=& \ra_a \pa + i\da \rb_a \fs_\bt\psa
- f_{ab}{}^c \zeta_c \xi^b, \nle
J^\gh_a &=& - if_{ab}{}^c c^b b_c.
\eens
We verify that $J^\phi_a$ and $J^\gh_a$ satisfy the brackets (\ref{JJ})
separately, and that they commute; hence $J^\TOT_a$ also satisfies the
same algebra $\g$.

The BRST operator, before the momentum constraint is taken into account,
has the form $Q = Q_G + Q_D + Q_A + Q_S$, where
\bes
Q_G &=& (J^\phi_a + \half J^\gh_a) c^a \nl
&=& \ra_a c^a\pi_\al + i\da\rb_a c^a \fs_\bt \psa
- f_{ab}{}^c c^a\zeta_c \xi^b - {i\/2}f_{ab}{}^c c^a c^b b_c, \nl
Q_D &=& \Ea\psa, \\
Q_A &=& \ua_\w \fsa \chi^\w, \nl
Q_S &=& \ra_a \fsa \xi^a.
\eens
$Q$ acts on the fields as
\bes
\dlt c^a &=& - {i\/2}f_{bc}{}^a c^b c^c, \nl
\dlt \fa &=& -i\ra_a c^a, \nl
\dlt \fsa &=& \Ea + i\da \rb_a c^a\fsb, \\
\dlt \theta_\w &=& -i\ua_\w \fsa, \nl
\dlt \zeta_a &=& -i\ra_a\fsa - if_{ab}{}^c c^b\zeta_c.
\eens
It is straightforward although quite tedious to verify nilpotency on
all fields, e.g. $\dlt^2 \theta_\w = 0$ and $\dlt^2\zeta_a = 0$, using
(\ref{ur}) and antisymmetry of the product $c^a c^b$. The cohomology 
is given by $H^0_\cl(Q) = C'(\phi)/\II$, $H^n_\cl(Q) = 0$ for
all $n>0$, where $C'(\phi)$ denotes the space of gauge-invariant
functions of $\fa$, and $\II$ is the ideal generated by $\Ea$. A basis
for this space is functions $\fw$ defined by $\fa = \fw\ua_\w$.

$Q$ acts on the conjugate momenta as
\bes
\dlt b_a &=& J^\phi_a + J^\gh_a = J^\TOT_a, \nl
\dlt \pa &=& i\da\rb_a c^a \pb - \da\db\rc_a c^a \fsc \psb \nl
&&+\ i\da\Eb\psb + i\da\ub_\w \fsb\chi^\w + i\da\rb_a\fsb\xi^a, \nl
\dlt \psa &=& - i\db\ra_a c^a\psb + \ua_\w\chi^\w + \ra_a\xi^a, \\
\dlt \chi^a &=& 0, \nl
\dlt \xi^a &=& - if_{bc}{}^a c^b \xi^c.
\eens
The cohomology is still generated by $\tpi_\w$ given by (\ref{tpiw}).
The reason is that the gauge algebra $\g$ acts trivially on the $\m$
indices $\w$, because we assumed that $[\g,\m]=0$. Without this
assumption, the gauge algebra acts on the solution space $\W$ as well,
and the formulas become more involved.

Hence $H^\bullet_\cl(Q_G+Q_D+Q_A+Q_S)$ is generated by $\fw$ and $\pw$.
Since these variables are gauge invariant, the momentum constraint
is imposed exactly as in the previous subsection.

\subsection{ Time dependence }

Let us now single out one of the indices $\al$ as the time variable.
The formalism in the previous subsections then applies with the following
modifications:
In quantum mechanics, $\fa \to \fa(t)$, $\al$ a discrete index;
in field theory, $\fa \to \fa(\xx,t)$, $\xx\in\RR^3$ the spatial
coordinate;
in QJT, $\fa \to (q^\mu(t), \fa_M(t))$, $M$ a multi-index and $q^\mu(t)$
the observer's trajectory.

In general, we assume a dynamical system described by some degrees of
freedom $\fa(t)$ and an action functional $S$. Introduce history
momenta $\pa(t)$, subject to the nonzero Poisson brackets
\be
[\fa(t), \pb(t')] = i\dlt^\al_\bt\dlt(t-t').
\ee
The dynamics constraint reads
\be
\Ea(t) = {\dlt S\/\dlt \fa(t)} = i[\pa(t), S] \approx 0.
\ee
The second derivative of the action is symmetric:
\be
{\dlt \Eb(t')\/\dlt \fa(t)} ={\dlt \Ea(t)\/\dlt \fb(t')}
= {\dlt^2 S\/\dlt \fa(t) \dlt \fb(t')}.
\ee
Because the Euler-Langrange equations $\Ea(t) = 0$ have solutions, there
are functions $\ua_\w(t)$, such that
\be
\int dt\ \ua_\w(t) \Ea(t) &\equiv& 0.
\label{uawt}
\ee
The momentum operators
\be
\pi_\w = \int dt\ \ua_\w(t) \pa(t).
\ee
satisfy an algebra which formally is given by the brackets (\ref{pipi}),
where instead of (\ref{fuu}) the structure constants are now defined by
\be
f_{\w\ww}{}^\ups \ua_\ups(t) &=&  \int dt' \bigg(
\ub_\ww(t'){\dlt\ua_\w(t)\/\dlt \fb(t')}
- \ub_\w(t'){\dlt\ua_\ww(t)\/\dlt \fb(t')} \bigg).
\ee

In contrast, a gauge index is assumed to include time, so we replace
$a \to (a,t)$ in the formulas from the previous subsection.
Hence assume that there exist functions $\ra_a(t,t')$, such that
\be
\int dt'\ \ra_a(t,t') \Ea(t') &\equiv& 0.
\ee
The gauge generators
\be
J_a(t) =  \int dt'\ \ra_a(t,t') \pa(t')
\ee
satisfy an algebra
\be
[J_a(t), J_b(t')] = i \int dt''\ f_{ab}{}^c(t,t';t'') J_c(t''),
\ee
where the structure functions $f_{ab}{}^c(t,t';t'')$ are defined by
\[
\int ds\ f_{ab}{}^c(t,t';s) \ra_c(s,t'')
= \int ds\ \bigg( \rb_b(t',s){\dlt\ra_a(t,t'')\/\dlt\fb(s)}
- \rb_a(t,s){\dlt\ra_b(t',t'')\/\dlt\fb(s)} \bigg).
\]
Moreover, we assume that the momentum algebra $\m$ and the gauge algebra
$\g$ commute, which leads to the identities (\ref{ur}):
\be
\int ds\ \bigg( \ub_\w(s) {\dlt \ra_a(t,t')\/\dlt \fb(s)} -
\rb_a(t,s) {\dlt \ua_\w(t')\/\dlt \fb(s)} \bigg) = 0.
\ee

We introduce antifields with nonzero brackets:
\be
\{\fsa(t), \psb(t)\} = \dlt^\bt_\al \dlt(t-t'), \qquad
[\theta_\w, \chi^\ww] = i\dlt^\ww_\w.
\ee
The BRST charge reads $Q = Q_G + Q_D + Q_A + Q_S$, where
\bes
Q_G &=& \int dt\ (J^\phi_a(t) + \half J^\gh_a(t)) c_a(t), \nl
Q_D &=& \int dt\ \Ea(t) \psa(t), \\
Q_A &=& \int dt\ \ua_\w(t) \fsa(t) \chi^\w, \nl
Q_S &=& \iint dt\ ds\ \ra_a(t,s) \fsa(s) \xi^a(t).
\eens
$H^\bullet_\cl(Q)$ is generated by $\fw$, where $\fa(t) = \fw \ua_\w(t)$, and
\[
\tpi_\w = \int dt\ \ua_\w(t) \pa(t) +
i \iint dt\ dt'\ {\dlt\ub_\w(t')\/\fa(t)}\fsb(t')\psa(t)
- f_{\w\ww}{}^\ups \theta_\ups\chi^\ww.
\]
The momentum algebra satisfied by $\tpi_\w$ takes the same form
(\ref{pipi}), up to a BRST exact term, since the solution set $\W$
contains no explicit time dependence. Hence the momentum constraint
is treated exactly as in the previous subsection.

The Hamiltonian in $\HP^*$ is $\H = \H^0 + \H^1$, where
\[
\H^0 = \int dt\ \bigg( -i\dot c^a(t)b_a(t) + \dot\fa(t)\pa(t)
-i\dot\phi^*_\al(t)\psa(t) + \dot\zeta_a(t)\xi^a(t) \bigg),
\]
and $\H^1$ involves the time-independent antifields. In the quantum
theory, energy is bounded from below. The vacuum $\ket0$ is therefore
annihilated by all negative-frequency modes, where frequency is
the Fourier transform variable.

\subsection{ Free theory }

Let us specialize the construction in the subsection \ref{ssec:general}
to free theories, e.g. the harmonic oscillator or the free scalar field
considered earlier in this paper. The action is then a quadratic form
\be
S = \half L_\ab\fa\fb,
\ee
where $L_\ab = L_\ba$ is a symmetric matrix. The dynamics
constraint (\ref{Ea}) becomes
\be
\Ea = L_\ab\fb \approx 0,
\ee
and the redundancy (\ref{uE})
\be
\ua_\w L_\ab = L_\ab \ub_\w = 0.
\ee
In other words, the constant vector $\ua_\w$ is a solution to the
equations of motion, whereas the solution set $\W$ ($\w\in\W$) labels
such solutions. The dynamics and antifield constraints give rise to the
BRST operators
\be
Q_D + Q_A = L_\ab\fa\psb + \fsa\ua_\w\chi^\w,
\ee
and $H^\bullet_\cl(Q_D + Q_A)$ is generated by $\pi_\w = \ua_\w \pa$ and
$\fw$, where $\fa = \fw\ua_\w$. Since $\ua_\w$ now is a constant matrix,
is has a one-sided inverse $k_\al^\w$, defined by
\be
k_\al^\w \ua_\ww = \dlt^\w_\ww.
\label{inverse}
\ee
Note that $\ua_\w$ and $k_\al^\w$ are not square matrices, since the
history index $\al$ runs over a much larger set of values than the
solution index $\w$; hence the inverse is one-sided. We can use
(\ref{inverse}) to solve for $\fw$:
\be
\fw = k_\al^\w \fa,
\ee
and we verify that (\ref{inverse}) leads to the expected Poisson bracket
$[\fw,\pi_\ww] = i\dlt^\w_\ww$.
Since $\da\ub_\w = 0$, the momentum algebra $\m$ is
abelian, and $f_{\w\ww}{}^\ups = 0$. Hence there is no need to
abelianize $\m$ by passing to new coordinate; we can take $q^\w = \fw$
and $p_\w = \pw$.
The momentum constraint (\ref{Mpq}) becomes simply
\be
\MM_\w = \pw - c_{\w\ww} \phi^\ww.
\ee
and the cohomology is generated by
\be
a_\w = 	\pw + c_{\w\ww} \phi^\ww + x^\ww_\w(\MM_\ww+\al_\ww),
\ee

In particular, we recover the treatment of the harmonic oscillator
in section \ref{sec:harmosc} by making the substitutions
\bes
&&\fa \to \phi(x) = \fw \e^{i\w x}, \qquad
\pa \to \pi(x) \nl
&&L_\ab \to \dlt''(x-x') + \w^2\dlt(x-x'), \nl
&&\ua_\w \to \e^{i\w x}, \qquad
k_\al^\w \to {1\/2\pi i} \e^{-i\w x}, \qquad
k_\al^\w \ua_\ww \to \dlt_{\w+\ww}, \\
&& \fw = {1\/2\pi i} \int dx\ \e^{-i\w x}\phi(x), \quad
\pw = \int dx\ \e^{i\w x} \pi(x), \nl
&&\phi_\w = \phi^{-\w}, \quad
c_{\w\ww} = 2\pi i\dlt_{\w+\ww}.
\eens

The Hamiltonian in the history phase space is given by (\ref{HHP}):
$\H = h^\bt_\al \fa\pb$. We assume that the constants $\ua_\w$ are
eigenvectors of the matrix $h^\al_\bt$, i.e.
\be
h^\al_\bt \ub_\w = \eps_\w \ua_\w.
\ee
The Hamiltonian in the extended history phase space $\HP^*$,
\be
\H &=& h^\bt_\al (\fa \pb - i\fsb \psa)
+ \sum_\w \eps_\w (-i \theta_\w \chi^\w + \bt_\w\gm^\w
+ \half \al_\w \al_{-\w} ),
\eens
acts e.g. as follows
\bes
[\H, \fa] = -ih^\al_\bt \fb, &&
[\H, \pa] = ih_\al^\bt \pb, \nl
{[}\H, \fw] = -\eps_\w \fw, &&
[\H, \pw] = \eps_\w \pw, \\
{[}\H, c_{\w\ww}] = (\eps_\w + \eps_\ww)c_{\w\ww}, &&
[\H, \phi_\w] = \eps_\w \phi_\w,
\eens
and analogously on other antifields, as indicated by the index structure.
It follows that $\H$ commute with a pair of two contracted indices, and
in particular $[\H, Q] = 0$, because in the BRST charge $Q$ consistes of
terms with all indices contracted.

Again, we write down the substitutions which apply to the harmonic
oscillator.
\bes
&&h^\al_\bt \to -\dlt'(x-x'), \quad
h^\al_\bt \ub_\w \to \w \e^{i\w x}, \quad
\eps_\w \to \w, \nl
&&\H = h^\bt_\al \fa \pb \to \int dx\ \phi'(x)\pi(x), \\
&&{[}\H, \phi(x)] = -i\phi'(x), \quad
[\H, \fw] = -\w\fw, \quad
[\H, a_\w] = \w a_\w.
\eens
The physical Hilbert space by the creation operators $a_\w$ with
energy $\w$.

\section{ Discussion }
\label{sec:discussion}

\subsection{ Ultralocality }
\label{ssec:ultra}

QJT amounts to two inventions: a formulation and cohomological treatment
of dynamics as a constraint in the history phase space, and the
replacement of fields by jets. We argued in the introduction that the
latter is physically significant, but the former is merely a technical
reformulation. Why is it needed?

The answer is that QJT is an ultralocal theory. Not only do
interactions have to be local in spacetime, but everything must
be formulated in terms of data on the observer's trajectory. In particular,
non-local integrals like the action functional or the Hamiltonian do not
make sense within QJT. Consider e.g. a system described by an action
functional $S[\phi]$. In QJT, we must expand the Lagrangian density in a
Taylor series around $q^\mu(t)$, \viz
\be
\L(\phi(x),\d_\mu\phi(x)) = \sum_M^p {1\/M!} \L_M(t) (x-q(t))^M.
\ee
Substituting this series into the action functional, we obtain
\bes
S[\phi] &=& \int d^4x\ \L(\phi(x),\d_\mu\phi(x)) \nle
&=& \sum_M^p {1\/M!} \L_M(t) \int d^4x\ (x-q(t))^M.
\eens
Clearly, the terms $\int d^4x\ (x-q(t))^M$ are ill defined, and hence
we can not base a formulation on non-local integrals like $S[\phi]$.
In QJT we must specify dynamics in terms of local data, such as the
Euler-Lagrange equations $\EE(x)$, which lead to well-defined Taylor
data $\EE_M(t)$.

\subsection{ The limits of manifest covariance }
\label{ssec:mancov}

The main novelty in the present work,
compared to previous papers \cite{Lar04,Lar05a,Lar05b,Lar05c}, is the
introduction of the momentum constraint. Unfortunately, this breaks
manifest covariance, provided that we insist on a cohomological
formulation; this was shown in subsection \ref{ssec:Poincare}.
To see that this is inevitable, it suffices to count the
degrees of freedom, in analogy with what was done for the harmonic
oscillator in (\ref{harmcount}).

For a field theory in $d$-dimensional spacetime, histories are labelled
by an $d$-dimensional index set. For a free theory, there is
one history for each $k \in \RR^d$; we write $\dim\HP = \RR^d$.
However, the physical phase space $\PP$ is the space of solutions to
the Euler-Lagrange equations, labelled by Cauchy data; hence
$\dim\PP = \RR^{d-1}$. We can not construct $\PP$ using history antifields
alone, because either all $k \in \RR^d$ cancel exactly, and
$\dim \PP = 0$, or $\dim\PP \sim \RR^d$, which is too large. To construct
the correct phase space with $\dim\PP = \RR^{d-1}$, we also
need antifields labelled by solutions to the Euler-Lagrange equations.

The counting is similar for QJT($p$), as discussed in subsection
\ref{ssec:Poincare}. There are two types of degrees of freedom:
histories in $p$-jet space, and momentum antifields associated with the
finite index set $\KK$. The former have infinitely many components (one
for each value of $t$), and the latter have only finitely many.
Since QJT($p$) is a $p$-jet regularization of QFT, it must only depend on
finitely many degrees of freedom, which can not achieved by combining
$p$-jet trajectories alone. Hence we must introduce also some antifields
labelled by the index set $\KK$. This inevitably breaks manifest
covariance because $\KK$ is not Poincar\'e invariant.
The total BRST operator is $Q = Q_D + \QM$, where the dynamics
part $Q_D$ is associated with histories and the momentum part $\QM$ with
the index set $\KK$.

Nevertheless, QJT maintains a stronger version of covariance
than QFT does in canonical formulations, because the introduction of the
observer's trajectory defines a local time direction in a covariant
way. Dynamics and canonical quantization are done in a manifestly
covariant fashion, but the definition of the canonical momentum
necessarily depends on the non-covariant index set $\KK$.
In the limit $p\to\infty$, $\KK$ becomes dense on the invariantly
defined surface $k^2 = \w^2$, and one may argue that the theory
is manifestly covariant.

\subsection{ The need for gauge anomalies }
\label{ssec:needanom}

Gauge anomalies are usually considered as a sign of inconsistency, which
must be avoided at all cost. However, this is factually incorrect. There
exist theories with a classical gauge symmetry, which cease to be gauge
theories after quantization, due to gauge anomalies. The best known
example is the free subcritical string. According to the no-ghost theorem,
clearly stated in section 2 of \cite{GSW87}, the free bosonic string in
$d$ dimensions can be quantized with a ghost-free spectrum for all $d
\leq 26$. The anomaly turns the classical conformal gauge symmetry into a
quantum global symmetry, which acts on the Hilbert space instead of
reducing it. This is not fatal, because unitarity is maintained.

It is true that the anomaly violates unitarity when $d > 26$, and that
the interacting string is inconsistent also when $d < 26$. However, this
is irrelevant if we specifically consider the free string as an example
of a classical gauge theory which exhibits a consistent gauge anomaly.
Moreover, the free bosonic string can be considered as a useful toy
model of quantum gravity coupled to scalar fields in two dimensions
\cite{NPZ05}.

If the correct quantum theory has a global infinite-dimensional symmetry,
there is no way to distinguish it from a gauge theory by looking at the
classical limit alone. Classically, we can always write down a nilpotent
BRST operator, and thus reduce the theory, both if the original quantum
symmetry was global or gauge.

As was emphasized in \cite{Lar06}, it is also important which version of
the gauge algebra we consider. For Laurent and Fourier polynomials, an
anomaly is necessary for nonzero charge (for conformal symmetry, $L_0
\neq 0$ implies that all $L_m \neq 0$ because $[L_m, L_{-m}] = 2m L_0$),
whereas for compact support and ordinary polynomials ($L_m$ with
$m \geq -1$), there exists no anomaly. It is easy to see that this is true
for all kinds of gauge symmetries, e.g. conformal, diffeomorphisms, or
Yang-Mills. Since we know that nonzero charge exists, and string theory
tells us that Laurent polynomials are admissible, gauge anomalies are
in fact inevitable.

\subsection{ QJT is not QFT }

$p$-jets are an approximation to $\infty$-jets, which may be identified
with fields, to the extent that infinite Taylor series can be identified
with the functions to which they converge. Hence one may view QJT($p$)
as a regularization of QFT. This regularization has the unique property
of preserving manifest covariance (before the momentum constraint is
taken into account), but it is a regularization nonetheless, and we
might na\"\i vely expect to recover QFT results in the limit
$p \to \infty$. However, this fails due to the presence of anomalies.

Gauge and diff anomalies in QFT were classified long ago \cite{Bon86}.
In particular, in four dimensions there are gauge anomalies proportional
to the third Casimir, and no diff anomalies. In QJT there are
other types of anomalies, described by the extensions in (\ref{VirKM}):
a gauge anomaly proportional to the second Casimir, and diff anomalies
in all spacetime dimensions including four. The presence of new
anomalies clearly shows that QJT is not merely a regularization of QFT,
but something genuinely new. This is a good thing, because it is well
known that QFT is incompatible with gravity, and this no-go theorem
does hence not apply to QJT.

At first sight, the inequivalence between QFT and QJT($p$) in the
$p\to\infty$ limit might be surprising, because classically a $p$-jet
is simply an approximation to a classical field.
The origin of this discrepancy is that a Taylor
expansion introduces a new datum: the expansion point (or rather the
family of expansion points referred to as the observer's trajectory). This
difference is significant, because the cocycles in (\ref{VirKM})
are functionals of $q^\mu(t)$. If we deal directly with the fields
themselves, we do not introduce the observer's trajectory, and the
relevant cocycles can not even be formulated.

Let us rephrase this non-equivalence slightly differently.
Quantization and the field theory ($p\to\infty$) limit do not commute.
The relation can be illustrated by a diagram. Let
CFT and CJT($p$) denote classical field theory and $p$-jet theory, and
QFT and QJT($p$) their quantum counterparts. We have the following
situation:
\[
\begin{array}{ccccccc}
CJT(p)&&& \stackrel{Quantization}{\longrightarrow} &&& QJT(p) \\
\Big\downarrow\vcenter{\rlap{$\scriptstyle{p\to\infty}$}} &&&&&&
\Big\downarrow\vcenter{\rlap{$\scriptstyle{p\to\infty}$}} \\
CJT(\infty) &=& CFT &\stackrel{Quantization}{\longrightarrow} &
QFT &\neq& QJT(\infty)
\end{array}
\]

\subsection{ Possible experimental signatures }

QJT has clearly not reached the stage where it is possible to make
predictions about experiments, although the Hilbert space for the
harmonic oscillator is correctly reproduced. However, it is possible that
one signature of QJT has already been observed; dark energy and perhaps
also dark matter could be manifestations of diffeomorphism anomalies,
whose existence (\ref{VirKM}) in four dimensions is the most striking
departure from QFT.

The argument relies on low-dimensional quantum gravity, where the
cosmological constant is related to a conformal anomaly as
\[
c \sim {\ell_{cc}\/\ell_{Planck}} = 10^{30} = (10^{120})^{1/4}.
\]
Here the experimental value of the cosmological constant in four
dimensions has been used. Since both diffeomorphism and conformal
anomalies are extensions of infinite-dimensional spacetime algebras,
it would be natural if they manifest themselves in a similar way --
as a cosmological constant.

\subsection{ The observer and quantum gravity }
\label{ssec:obsQG}

Although the anomaly argument shows that QJT is inequivalent from
conventional QFT, even in the $p\to\infty$ limit, there must some limit
in which QFT is recovered. We argued in subsection \ref{ssec:QFTlim} that
this limit amounts to replacing the operator-valued curve $q^\mu(t)$
by a classical curve. Physically, this means that the observer is
assumed macroscopic and classical, so her position and velocity are
unaffected by the act of observation. This is clearly an idealization.
Every physical observer is built from a finite, albeit perhaps large,
number of elementary particles, and is thus subject to the laws
of quantum mechanics.

Hence we may say that QJT amounts to a quantization of the observer and
her clock, which is the operational definition of time.
This clearly has implications for the interpretation of quantum
mechanics; in particular, QJT disfavors the Copenhagen interpretation,
which contains an assumption about a classical observer.

We can now intuitively understand why conventional quantization of
gravity fails. In order that the observer be unaffected by the act of
observation, we must assume that she is macroscopic and thus infinitely
massive. Conventional formulations of quantum theory, and QFT in
particular, thus secretly introduce an infinitely massive observer into
the scene. This is usually a good approximation if we ignore gravity, but
it will wreck havoc in a theory of gravity which couples to this infinite
mass. E.g., in the presence of gravity, an infinitely massive observer
will immediately create a black hole. This is, in my opinion, the
physical reason underlying the difficulties with quantum gravity.

\subsection{Principles of quantum gravity}

One may expect that the fundamental principles underlying quantum gravity
are the same as those underlying general relativity and QFT, namely
\begin{itemize}
\item
Background independence.
\item
Locality.
\item
Quantum theory with a separable Hilbert space and energy bounded
from below.
\end{itemize}
Background independence and locality are incompatible within the
framework of QFT, because the only unitary proper lowest-energy
representation of the spacetime diffeomorphism algebra is the trivial
one. However, this no-go theorem does not apply to QJT, because there are
new diff anomalies described by the algebra (\ref{VirKM}), which allow
non-trivial, projective representations. This is analogous to the
ordinary Virasoro algebra, which only has one unitary irrep with $c=0$
(the trivial one), but many unitary irreps with $c>0$.

Hence QJT is the unique way to combine all three fundamental principles
of quantum gravity.

\subsection{Conceptual problems of quantum gravity}

As is well known, quantum gravity leads to numerous conceptual problems,
known as the problem of time in various guises; for a nice list, see the
discussion in section 3 of Carlip's review \cite{Car01}. That author
notices that all these problems can be sidestepped by defining time as
``the reading of a clock'', but dismisses this option for two reasons:
\begin{enumerate}
\item
A clock only measures time along its worldline.
\item
Any physical clock has a finite probability of sometimes running backwards.
\end{enumerate}
These objections are evaded in QJT, in the following ways:
\begin{enumerate}
\item
In QJT, everything is formulated in terms of data living on the clock's
worldline $q^\mu(t)$, namely the Taylor coefficients $\phi_M(t)$.
The magic of analyticity makes it possible to reconstruct the non-local
field data $\phi(x)$ from the local Taylor data $\phi_M(t)$. In this
sense, QJT only needs to make statements about events on the observer's
trajectory.
\item
Both the Taylor data $\phi_M(t)$ and proper time  $\tau(t)$ are functions
of an underlying c-number parameter $t$, which determines the causal
structure; in particular, proper time is defined by
$\dot \tau^2(t) = g_{\mn,0}(t)\dot q^\mu(t) \dot q^\nu(t)$,
where $g_{\mn,0}(t)$ is the zero-jet of the metric.
\end{enumerate}

\subsection{ Conclusion }

In this paper, the major flaws in \cite{Lar04,Lar05a,Lar05b,Lar05c} were
corrected, and a history-oriented quantization scheme was developed,
yielding perfect agreement with conventional canonical quantization.
We then performed the passage to QJT. This gives the correct results
for the harmonic oscillator, and interesting discrepancies (quantization 
of observer, reparametrization anomalies) for the free scalar field.

There are several obvious directions for future research. First of all,
the formalism must be extended to interacting theories, along the lines
sketched in section \ref{sec:interactions}. Of particular interest is to
understand the relation between renormalization and QJT. Another
long-standing open problem, first introduced in \cite{Lar02}, is to
understand the field limit. Generically, the new gauge and diff anomalies
in QJT($p$) diverge when $p\to\infty$, which appears inconsistent.
As discussed in subsection \ref{ssec:anomaly} and in \cite{Lar02,Lar04},
the divergent parts can be
cancelled. This works best in four dimensions, which is encouraging,
but the details do not seem to work out.


\begin{thebibliography}{99}

\bibitem{Bon86} L. Bonora, P. Pasti and M. Tonin,
  {\it The anomaly structure of theories with external gravity},
  J. Math. Phys. {\bf 27} (1986) 2259--2270.

\bibitem{Car01} S. Carlip,
  {\it Quantum gravity: a progress report},
  Rept. Prog. Phys. {\bf64} (2001) 885,
  {\tt gr-qc/0108040}

\bibitem{GSW87} M.B. Green, J.H. Schwarz and E. Witten,
   {\it Superstring theory, volume I: Introduction},
   Cambridge Univ. Press (1987).

\bibitem{HT92} M. Henneaux, and C. Teitelboim,
  {\em Quantization of gauge systems},
  Princeton Univ. Press (1992)

\bibitem{Ish95} C. J. Isham and N. Linden,
  {\it Continuous histories and the history group in generalised
  quantum theory},
  J. Math. Phys. {\bf 36} (1995) 5392.

\bibitem{Jac95} R. Jackiw,
  {\it Two lectures on Two-Dimensional Gravity},
  {\tt gr-qc/9511048} (1995)

\bibitem{Lar98} T.A. Larsson,
  {\it Extended diffeomorphism algebras and trajectories in jet space},
  Comm. Math. Phys. {\bf 214} (2000) 469--491,
  {\tt math-ph/9810003}

\bibitem{Lar02} T.A. Larsson,
  {\it Koszul-Tate cohomology as lowest-energy modules of non-centrally
  extended diffeomorphism algebras},
  {\tt math-ph/0210023} (2002)

\bibitem{Lar04} T.A. Larsson,
  {\it Manifestly covariant canonical quantization I: the free scalar
  field},
  {\tt hep-th/0411028} (2004)

\bibitem{Lar05a} T.A. Larsson,
  {\it Manifestly covariant canonical quantization II: Gauge theory
  and anomalies},
  {\tt hep-th/0501043} (2005)

\bibitem{Lar05b} T.A. Larsson,
  {\it Manifestly covariant canonical quantization III: Gravity,
  locality, and diffeomorphism anomalies in four dimensions},
  {\tt hep-th/0504020} (2005)

\bibitem{Lar05c} T.A. Larsson,
  {\it Manifestly covariant canonical quantization of gravity
  and diffeomorphism anomalies in four dimensions},
  in Focus on Quantum Gravity Research, ed. David C. Moore,
  pp 261-310, Nova Publishers (2006)

\bibitem{Lar06} T.A. Larsson,
  {\it Local, global, divergent -- which gauge symmetries are redundant?},
  {\tt math-ph/0603024} (2006)

\bibitem{NPZ05}  H. Nicolai, K. Peeters and M. Zamaklar,
  {\it Loop quantum gravity: an outside view},
  Class. Quant. Grav. {\bf22} (2005) R193,
  {\tt hep-th/0501114}.

\bibitem{RM94} S.E. Rao and R.V. Moody,
  {\it Vertex representations for $N$-toroidal Lie algebras and a
  generalization of the Virasoro algebra},
  Comm. Math. Phys. {\bf 159} (1994) 239--264.

\bibitem{Sard02} G. Sardanashvily,
  {\it Ten lectures on jet manifolds in classical and quantum field
  theory},
  {\tt math-ph/0203040} (2002)

\bibitem{Sav98} K. Savvidou,
  {\it The action operator for continuous-time histories},
  J.Math.Phys. {\bf 40} (1999) 5657-5674,
  {\tt  gr-qc/9811078}

\end{thebibliography}
\end{document}